\documentclass{article}

\usepackage{arxiv}

\usepackage[utf8]{inputenc} % allow utf-8 input
\usepackage[T1]{fontenc}    % use 8-bit T1 fonts
\usepackage{hyperref}       % hyperlinks
\usepackage{url}            % simple URL typesetting
\usepackage{booktabs}       % professional-quality tables
\usepackage{amsfonts}       % blackboard math symbols
\usepackage{nicefrac}       % compact symbols for 1/2, etc.
\usepackage{microtype}      % microtypography
\usepackage{lipsum}		% Can be removed after putting your text content
\usepackage{graphicx}
\usepackage{natbib}
\usepackage{doi}
\usepackage{amsmath}
\usepackage{amssymb}
\usepackage{multirow}
\usepackage{makecell}
\usepackage{multicol}

\title{Differentiation methods as a systematic uncertainty source in equation discovery}

\author{ \href{https://orcid.org/0000-0000-0000-0000}{\includegraphics[scale=0.06]{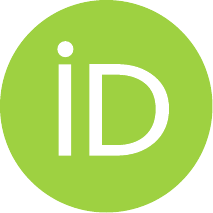}\hspace{1mm}M.D.~Khilchuk}\thanks{Use footnote for providing further
		information about author (webpage, alternative
		address)---\emph{not} for acknowledging funding agencies.} \\
	NSS Lab\\
	ITMO University\\
	Saint-Petersburg, 197101, Russia \\
	\texttt{mdkhilchuk@itmo.ru} \\
	\And
	\href{https://orcid.org/0000-0000-0000-0000}{\includegraphics[scale=0.06]{orcid.pdf}\hspace{1mm}I.O.~Markov} \\
	NSS Lab\\
	ITMO University\\
	Saint-Petersburg, 197101, Russia \\
	\texttt{iomarkov@itmo.ru} \\
    \And
	\href{https://orcid.org/0000-0000-0000-0000}{\includegraphics[scale=0.06]{orcid.pdf}\hspace{1mm}A.A.~Hvatov} \\
	NSS Lab\\
	ITMO University\\
	Saint-Petersburg, 197101, Russia \\
	\texttt{alex\_hvatov@itmo.ru} \\
}

\renewcommand{\shorttitle}{Differentiation as an uncertainty source in equation discovery}

\hypersetup{
pdftitle={Differentiation methods as a systematic uncertainty source in equation discovery},
pdfsubject={q-bio.NC, q-bio.QM},
pdfauthor={M.D.Khilchuk, I.O.Markov, A.A.Hvatov},
pdfkeywords={Equation discovery, Differentiation methods, Feature engineering, Sensitivity analysis},
}

\begin{document}
\maketitle

\begin{abstract}
	In differential equation discovery algorithms, numerical differentiation is usually a fixed preliminary step. Current methods improve robustness with data subsampling and sparsity but often ignore the variability from the differentiation method itself. We show that this choice systematically introduces uncertainty, affecting both equation form and parameter estimates. Our study indicates that high-resolution schemes can magnify measurement noise, while heavily regularized methods may mask real physical variations, which leads to method-dependent findings. By evaluating six differentiation techniques on various partial differential equations under diverse noise levels using SINDy and EPDE frameworks, we consistently notice methodological biases in the determined models. This underscores the importance of selecting differentiation methods as a key modeling choice and highlights a path to enhance ensemble-based discovery by diversifying methodologies.
\end{abstract}

% keywords can be removed

\section{Introduction}
Four critical components define the framework of any machine learning model: architecture, parameters, features, and the objective function. Similarly, modern approaches to differential equation discovery treat differential equations as machine learning models. This perspective raises key questions about how to assess the quality of the discovered DE and the associated uncertainties, leveraging established evaluation techniques from machine learning and sensitivity analysis (SA). Although uncertainty assessment is important in classical ML, it is particularly vital in differential equation discovery to form ensembles.

Uncertainty can be assessed for each component of either a classical ML model or a differential equation as an ML model:

\texttt{Architectural uncertainty} is typically assessed through techniques such as pruning \cite{blalock2020state} or ensemble methods \cite{lakshminarayanan2017simple}, which quantify the robustness to structural variations.

\texttt{Parameter uncertainty} is often analyzed via sensitivity analysis. Local SA methods, such as the one-at-a-time (OAT) approach \cite{hamby1994review}, assess the effect of small perturbations in individual inputs and keep others constant. However, these methods capture non-linearities and interactions poorly. In contrast, global SA methods, such as Sobol indices \cite{sobol1993sensitivity}, evaluate the impact of variations across the parameter space and account for input interactions.

\texttt{Feature uncertainty} is managed through preprocessing, data augmentation, sampling, feature engineering, and strategies to handle noisy inputs, such as robust normalization techniques \cite{werner2023imbalanced}.

\texttt{The objective function}, although often defined by design, can introduce uncertainties when there is misalignment between the target and the model's capacity \cite{gonzalez2020improved}. 

Recent differential equation discovery methods allow us to treat differential equations as machine learning methods. Therefore, we could also find analogs to machine learning components. Every equation discovery method aims to identify the equation structure and terms likely to appear in the governing equation for the data; this structure is closely related to a neural network architecture, as it describes how features and layers are interconnected.

The second step is to identify the parameters. The parameters are the coefficients within the differential equation that frequently correspond to physical properties. They can be referred to as neural network weights (and are essentially coefficients of a specialized type of linear regression).

Advancements in differential equation discovery techniques have refined the assessment of uncertainty for these components. For example, parameter uncertainty has been addressed using ensemble-based approaches, such as E-SINDy \cite{fasel2022ensemble}, which employs term library ensembling to handle parameter robustness . Structural uncertainty has been explored using methods like multi-objective evolutionary optimization combined with Bayesian networks, as demonstrated by \cite{hvatov2023towards}. These approaches enable researchers to more accurately quantify structural robustness and align the DE identified with physical phenomena.

Unlike traditional machine learning, the objective function in DE discovery is more constrained. It is often defined as the discrepancy of the equation, evaluated either in a strong form (e.g., term-by-term residuals) or in a weak form (e.g., weak formulations such as wSINDy \cite{messenger2021weak}). Solver-based methods are also employed to minimize discrepancies between observed data and solutions generated by the identified DE, such as physics-informed criterion (PIC) and others.

The critical distinction between differential equation discovery and machine learning lies in the treatment of features. The sole feature in differential equation discovery is based on observational data; it could be a time series or a field (a multidimensional tensor that contains time as one axis). However, to build an equation, we require differentials with respect to every axis up to the given order. Differentials of the input data are not provided in most cases and must therefore be computed numerically. Thus, from a machine learning perspective, the features are engineered within the algorithm.

Noisy measurements pose a challenge to numerical differentiation, leading to errors in derivative estimates. Stable numerical differentiation techniques (for example, finite differences, polynomial interpolation, or methods based on machine learning \cite{chartrand2011numerical}) have been proposed to address these problems. However, the choice of differentiation method can significantly impact the quality of the discovered DE model. Variations in derivative computation propagate uncertainty into both the estimated parameters and the structure of the resulting differential equation.

Despite progress in addressing parameter and structural uncertainties in DE discovery, the impact of differentiation methods on feature uncertainty remains underexplored. This paper aims to systematically assess how differentiation techniques influence the quality of discovered models, with a particular focus on parameter and structural accuracy under varying levels of data uncertainty. We describe differentiation as a ``feature engineering'' source of uncertainty in differential equation discovery. Additionally, we experimentally prove the apparent fact that better differentiation quality leads to better discovery, but also the non-obvious fact that different methods should be used for noisy and clean data to achieve better performance.

\section{Differential equation discovery background}
\label{sec:discovery}
As noted above, in the context of a differential equation as a machine learning model, we can distinguish the components of such a model: structure/architecture, parameters, features, and objective function. 

For differential equations discovery, as input, we have the data placed on a discrete grid $X=\newline\left\{x^{(i)}=\left(x_1^{(i)}, \ldots x_{\operatorname{dim}}^{(i)}\right)\right\}_{i=1}^{i=N}$, where $N$ is the number of observations and dim is the dimensionality of the problem. We mention a particular case of time series, for which $\operatorname{dim}=1$ and $X=\left\{t_j\right\}_{i=1}^{i=N}$.

It is also assumed that for each point on the grid, there is an associated set of observations $U=\left\{u^{(i)}=\left(u_1^{(i)}, \ldots, u_L^{(i)}\right)\right\}_{i=1}^N$ to define a grid map $u$ : $X \subset \mathbb{R}^{\operatorname{dim}} \rightarrow U \subset \mathbb{R}^L$. This grid and observations can be used as input data or features in the machine learning model.

From differential equation theory, we expect $u$ to represent not only a function but also a jet, which is essentially differential up to a given order $r$ in form:

\begin{equation}
    J^r=(x_1,...,x_{\text{dim}};u;D_1u;D_2u;...;D_ru)
\end{equation}

,where $D_r=\bigcup \limits_{|\alpha|=r} \{\frac{\partial^ru}{\partial x_1^{\alpha_1}...\partial x_r^{\alpha_r}}\}$ is the set of all partial differentials of order $r$ and $\alpha =\{\alpha_1,...\alpha_{\text{dim}}\}, |\alpha|=\sum \limits_{i=1}^{i=\text{dim}}\alpha_i$ is just a differential multi-index. Since we usually have a single observation set $u$ we omit it from the notation $J^r(u)$

Any differential equation of an order $r$ is just a surface in a jet space $J^r$. Let $\mathcal{T}$ be a set of basis functions (monomials, compositions) acting on
$\bar{J}^r$. Then $S \subset \mathcal{T}$ represents selected terms, and $P$ is the set of admissible coefficients. It could be a function of independent coordinates or just constants. Then the surface has the following form:

\begin{equation}
    M(S,P) = \sum_{s \in S} p_s \cdot s(\bar{J}^r)=0
    \label{eq:diffeq_surface}
\end{equation}

Having an analytical jet is the best possible case for the discovery of \textit{differential equations}. We look for the surface within the jet space, nothing more. The real-case scenario significantly differs from the ideal "continuous" case, namely: (a) in most cases the jets are restored just from observation data $U$ with numerical differentiation,(b) we cannot look for any surface, we restrict the surface search space, and (c) if we want to find a "governing" law, we need to go beyond simple symbolic regression: the equation may not have unique solution, we may overlook some terms, terms presented in theoretical equation could have small magnitude and appear as numerical noise. In general, we require one to assess the uncertainty in both the structure and the coefficients. In most cases, it is done using ensembles.

\textbf{(a) On jets and numerical differentiation} Returning to the discrete setup, we have an approximate jet $\bar{J^r} = \{(x^{(i)},u^{(i)},D_hu^{(i)},...(D_h)^ru^{(i)})\}_{i=1}^{i=N}$ with $D_h$ denoting an arbitrary numerical differentiation operator. Here is much uncertainty. Generally, we do not know how the discrete observation set $U = \{u^{(i)}\}_{i=1}^{i=N}$ is connected with a true function $u$. For the selected numerical differentiation method $D_h$, we usually only know the order of approximation for a first-order differential, and we apply it several times to have higher-order differentials without any guarantee of higher-order differential approximation.

We expect that if the differentiation method is correct, then in a point-wise manner $\bar{J^r} \xrightarrow[N \to +\infty]{} J^r$. There is an ambiguity. In a real-world scenario, we cannot obtain more observational data. From the other side, for a finite computation starting from a certain number of points, the process becomes ill-posed. We discuss how to approach assessing it below. 

\textbf{(b) Equation discovery problem statement} Let us assume that the discrete jet $\bar{J^r}$ is already computed for the observation data $U$. Therefore, we can treat items in $\bar{J^r}$ as symbols and formulate a symbolic regression problem.

As stated above, we have to find an explicit surface \eqref{eq:diffeq_surface}. Any machine learning requires restricting the search space to a finite one. As the first step, we define the loss function $L(M(S,P))$ and formulate the optimization problem.

\begin{equation}
S^*,P^*=\underset{S \in \Sigma,P \in \Pi}{\operatorname{argmin}}L(M(S,P))
\label{eq:symbolic regression}
\end{equation}

The methods of equation discovery differ in the way they determine $L(\cdot)$, the parameterization of the model $M(S,P)$, and the set of restrictions of the structure $\Sigma$, as well as the set of parameters $\Pi$. The structure and parameters are analogous to those of a machine learning algorithm's model architecture. Unlike machine learning $\Sigma$ and $\Pi$ in differential equation discovery, the optimization algorithm determines the results. Below, we briefly outline the main groups of methods.

In equation discovery, in the first place, we care about how $\Sigma$ is built. As a classical algorithm in the area, we consider another algorithm, Sparse Identification of Nonlinear Dynamics (SINDy) \cite{brunton2016discovering}. 

For the SINDy case, we manually determine the longest sentence $\Sigma_{\text{long}}$ possible and fix it. The optimization is performed only by $P$, which is essentially a vector of the numerical coefficients near each word of $\Sigma_{\text{long}}$. We need to make $P$ as sparse as possible, which is done with classical LASSO regression. In SINDy, we compute the loss function by using the discrepancy over the discrete grid. 

\begin{equation}
P^*=\underset{P \in \Pi}{\operatorname{argmin}} ||M(\Sigma_{\text{long}},P)||_2+\alpha ||P||_1
\label{eq:SINDy_loss}
\end{equation}

In \eqref{eq:SINDy_loss} we denote by $||\cdot||_2$ the mean discrepancy in the computation grid $X$ and by $||\cdot||_1$ is the $l_1$ norm. Since SINDy usually works with constant coefficients, we could use the $l_1$ norm to determine the sparsity of the set of parameters $P$. In some sense, it is a measure of the complexity of the surface in terms of the number of symbols needed to describe it.

Evolutionary approaches and reinforcement learning have their own rules to construct $S$ for a model. Every equation $S_i$ appearing within the optimization process is evaluated using the SINDy approach \eqref{eq:SINDy_loss} with discrepancy or, as is done in EPDE, by constructing the Pareto frontier over the discrepancy and complexity criteria. Both discrepancy computation and Patero frontier forming are done as part of the fitness function computation or to form a reward for the reinforcement learning agent.

There are also more robust measures. For a given surface $M(S,P)$, we try to restore the continuous function $u$ that exactly generates the surface and then compare it with observations $U$. It, of course, requires the solution of the equation. We note that in this case, we do not need to consider jets $J^r$; instead, we begin working with the fibers $u$ and no longer need to consider the differentials $D_r$. In that case, all surfaces are single-connected, i.e., the solution of the equation is unique, which is, of course, a limitation, but it is more robust than a discrepancy measure.

There are also some intermediate cases, such as PIC. Here we spatially handle jets, but temporally restore continuous paths. It could be considered as jet factorization and partial fiber projection.

Ultimately, after optimization of equation \eqref{eq:symbolic regression}, we obtain a single symbolic expression that represents a relation found in the discrete jet $\bar{J^r}$. We need to assess sensitivity to the data subsampling method, noise, and differentiation.

\textbf{(c) Ensemble sensitivity analysis} To assess sensitivity and pick stable appearing equations, we need to form ensembles. To briefly mention, E-SINDy is the first algorithm for the equation discovery method \cite{fasel2022ensemble} that addresses this problem. We also have different ensembling methods \cite{hvatov2023towards}.

Unlike what is usually stated in the literature, we do not need to have a single stable equation, but an ensemble that has a common part "in mean". Additionally, we typically consider the sensitivity and optimization stability of data subsampling. However, the sensitivity of the differentiation method, although it is an important part as shown above, is usually omitted.

\section{Data differentiation problem statement and proposed methods}
\label{sec:denoising}
As seen above, the problem of using derivatives as features and overall differentiation methods is rarely mentioned in equation discovery. The differentials are used as symbols to form a discrete jet. Strictly speaking, discrete jets differ greatly based on the differentiation method used.

In particular, differentials here are considered handcrafted features within symbolic regression from a machine learning perspective. We are interested in the contribution the differentiation algorithm makes to equation discovery and the importance of the differentiation error. The ad hoc solution is that the error should not be too large for most problem statements.

To illustrate the problem, let us consider a straightforward numerical differentiation problem statement. The input to any equation discovery algorithm typically consists of noisy measurements. Denoting the true (noise‐free) state by $\overline{u}(t,\mathbf{x})$ and the observed data by

\begin{equation}
   u(t,\mathbf{x}) = \overline{u}(t,\mathbf{x}) + \epsilon(t,\mathbf{x}) 
   \label{eq:noised_data}
\end{equation}

Typically, one assumes that $\epsilon(t,\mathbf{x})$ arises from additive white Gaussian noise (AWGN). In particular, each measurement is modeled as

\begin{equation}
    u(t,\mathbf{x}) \sim \mathcal{N}\bigl(\overline{u}(t,\mathbf{x}),\,\sigma^2(t,\mathbf{x})\bigr),
\quad \sigma(t,\mathbf{x}) = \kappa\,\bigl|\overline{u}(t,\mathbf{x})\bigr|
\label{eq:noise_in_data_mechanism}
\end{equation}

for some proportionality constant $\kappa$. Let us consider the simplest case of the central difference differentiation of a single-variable function:

\begin{equation}
   u'(x) \approx \frac{u(x + h) - u(x - h)}{2\,h} 
\end{equation}

Substituting into the central difference noised data \eqref{eq:noised_data} yields

\begin{multline}
   \tilde{u}'(x)
=
\frac{\bigl[u(x+h) + \epsilon(x+h)\bigr] - \bigl[u(x-h) + \epsilon(x-h)\bigr]}{2\,h}
= \\=
\underbrace{\frac{u(x+h) - u(x-h)}{2\,h}}_{\displaystyle\text{(deterministic, truncation part)}}
\;+\;
\underbrace{\frac{\epsilon(x+h) - \epsilon(x-h)}{2\,h}}_{\displaystyle \eta(x)} 
\end{multline}

Because $\epsilon(x+h)$ and $\epsilon(x-h)$ are independent Gaussian random variables with variance $\kappa^2\,\bigl|\overline{u}(t,\mathbf{x})\bigr|^2$, it follows that

\begin{equation}
    \mathrm{Var}\bigl[\eta(x)\bigr] 
\;=\; 
\mathrm{Var}\Bigl[\tfrac{\epsilon(x+h) - \epsilon(x-h)}{2\,h}\Bigr]
\;=\; 
\frac{2\,\kappa^2\,\bigl|\overline{u}(t,\mathbf{x})\bigr|^2}{4\,h^2}
\;=\; 
\frac{\kappa^2\,\bigl|\overline{u}(t,\mathbf{x})\bigr|^2}{2\,h^2},
\end{equation}

hence, the noise-induced standard deviation in the derivative estimate scales like $\kappa\,|u(x)|/(\sqrt{2}\,h)$.

On the other hand, a standard Taylor remainder analysis shows that the deterministic error (truncation) of the central difference approximation is $\mathcal{O}(h^2)$. Denoting by $k$ the constant that bounds the second derivative term in the remainder, one may write:

\begin{equation}
    E_{\mathrm{trunc}}(h) \approx k\,h^2,\qquad
E_{\mathrm{noise}}(h) \approx \frac{\kappa\,|u(x)|}{\sqrt{2}\;h}.
\end{equation}

Therefore, the total error in the central-difference derivative for noisy data can be viewed as

\begin{equation}
    E_{\mathrm{total}}(h) \sim k\,h^2 \;+\; \frac{\kappa\,|u(x)|}{\sqrt{2}\,h}.
    \label{eq:error_expansion}
\end{equation}

Minimizing $E_{\mathrm{total}}(h)$ with respect to $h$ yields an optimal grid spacing and the corresponding minimal error scales as

\begin{equation}
    h^* \;=\; \Bigl(\tfrac{c}{2\,k}\Bigr)^{1/3}, 
\qquad
c = \frac{\kappa\,|u(x)|}{\sqrt{2}},
\qquad
 E_{\mathrm{min}} \sim c^{2/3}\,k^{1/3}\bigl(2^{-2/3} + 2^{1/3}\bigr).
\end{equation}

Thus, although reducing $h$ decreases the truncation error, it amplifies the noise. It implies that an intermediate (non-zero) $h^*$ balances both contributions. In symbolic regression, where derivatives are treated as features, choosing $h$ too small will cause the noise amplification term to dominate, leading to spurious high-frequency artifacts. Choosing $h$ too large will smear genuine gradients and obscure critical dynamics. The differentiation algorithm should be correct to draw a general conclusion for noisy data, but increasing differentiation precision may lead to worse results.

For a general differential equation discovery algorithm, we have only a fixed differentiation method as a hyperparameter and a fixed grid, but the equation is not fixed. So, in some sense, we could only manipulate $k$ from the above equations. We could only assess the variance of the discrepancy distribution for a known equation-" answer ". However, we do not have any guarantees that none of the other equations could eventually achieve a lower discrepancy.

The discrepancy measure is used as the optimization criterion in most equation discovery methods. Since the discrepancy is averaged over the whole grid, we already use the weak form. However, as shown in the paper, the differentiation algorithm and error play a significant role in this process.

\section{Experiments}

We will conduct complex numerical experiments to investigate the influence of the above differentiation methods in DE discovery. DE discovery will be carried out using sparse regression \cite{brunton2016discovering} and an evolutionary approach \cite{maslyaev2021partial}.

All experiments were carried out using six differentiation methods in total, which are referred to as \texttt{Gradient}, \texttt{Adaptive}, \texttt{Polynomial}, \texttt{Spectral}, and \texttt{Total\_var}. The detailed description of the methods used is in Appendix~\ref{app:differentiation_formulation}. For the particular algorithms' parameters and realization, please refer to the code.

\subsection{Experimental setup}
Second-order ODE and several types of partial differential equations were chosen, each with different solutions: analytical (KdV), numerical (Burgers, wave, Laplace). Additionally, we use a data-driven model of ocean behavior \cite{ross2023benchmarking} where the exact equation is not known a priori (referred to as pyqg below).

The workflow includes selecting and generating data; as noted earlier, it is either obtaining an analytical solution in the form of a matrix of values or finding a solution matrix using numerical methods, setting boundary and initial conditions, where necessary, and choosing constants. After that, all the data obtained are differentiated by the described methods, while the derivatives sought are those that, as is known in advance, occur in the equation. These may be derivatives of the form $\frac{\partial u(t, \mathbf{x})}{\partial x}$, $\frac{\partial u(t, \mathbf{x})}{\partial t}$,$\frac{\partial^2 u(t, \mathbf{x})}{\partial x^2}$, etc. 

Then, an evolutionary algorithm is applied using the EPDE framework. Data were loaded with the grid and all derivatives, and then we chose a multi-objective mode. The population size is 7 for all equations, and the number of training epochs ranges from 30 to 80, depending on the equation's complexity; the maximum number of terms in each equation is 8. This is done to obtain greater variability in the equations. Then, the algorithm is run; one run yields approximately 5-7 equations per the Pareto frontier. We perform only 50 runs for each equation to minimize variation in the data and more accurately estimate the average approximation for all coefficients.

For each data set, a series of experiments was conducted, resulting in box plots (refer to the appendices) that show the distribution of coefficients preceding the correct terms. The difference between the obtained equations and the true ones was also analyzed using the Structural Hamming Distance (SHD) metric.

For each experiment, noise was added to the data as \eqref{eq:noise_in_data_mechanism}, with $\kappa=\{0,0.5,1\}$\%, which is referred to as \texttt{ noise}. However, we provide results for $\text{noise}=1$ for the particular equations. Other cases are discussed in the corresponding section.

\section{Results}

\subsection{Ordinary differential equation}
As a simple example, we consider a second-order ODE in the form   $mu'' + qu' + ku = 0$ with parameters $m = 1$, $q = 0.25$, $k = 3$, and initial conditions $u(0) = 1, u'(0) = 0$.The results visualized as boxplots are given in Appendix~\ref{app:ODE_res} with a further discussion in Section~\ref{sec:discussion}.

\begin{table}[ht!]
\caption{Coefficients values calculated with EPDE, noise = 0\%}
\centering
\begin{tabular}{|c|c|c|c|}
\hline
Methods/Terms & u                & u'               & u''              \\ \hline
Gradient      & 0.0219 ± 0.0082  & -0.2156 ± 0.0048 & -1               \\ \hline
Adaptive      & 0.0197 ± 0.0077  & -0.2086 ± 0.0053 & -1               \\ \hline
Polynomial    & 0.0541 ± 0.0452  & -0.2348 ± 0.0002 & -1               \\ \hline
Spectral      & -0.0312 ± 0.0412 & -0.3008 ± 0.0392 & -0.8997 ± 0.0460 \\ \hline
Inverse       & -0.0142 ± 0.0282 & -0.5213 ± 0.0632 & -0.5069 ± 0.1026 \\ \hline
Total\_var         & -0.3901 ± 0.0032 & -0.9952 ± 0.0068 & -0.9353 ± 0.0003 \\ \hline
Ground truth  & 3                & 0.25             & 1                \\ \hline
\end{tabular}
\end{table}

\begin{table}[ht!]
\caption{Coefficients values calculated with EPDE, noise = 0.5\%}
\centering
\begin{tabular}{|c|c|c|c|}
\hline
Methods/Terms & u                & u'               & u''              \\ \hline
Gradient     & 0.0205 ± 0.0000  & -0.2222 ± 0.0043 & -1               \\ \hline
Adaptive     & 0.0152 ± 0.0112  & -0.2173 ± 0.0055 & -1               \\ \hline
Polynomial   & 0.0363 ± 0.0331  & -0.2412 ± 0.0058 & -1               \\ \hline
Spectral     & -0.0288 ± 0.0241 & -0.2694 ± 0.0375 & -0.9048 ± 0.0480 \\ \hline
Inverse      & -0.0255 ± 0.0177 & -0.2840 ± 0.0533 & -0.1348 ± 0.0804 \\ \hline
Total\_var        & -0.3937 ± 0.0027 & -1               & -0.9387          \\ \hline
Ground truth & 3                & 0.25             & 1                \\ \hline
\end{tabular}
\end{table}

\begin{table}[ht!]
\caption{Coefficients values calculated with EPDE, noise = 1\%}
\centering
\begin{tabular}{|c|c|c|c|}
\hline
Methods/Terms & u                & u'               & u''              \\ \hline
Gradient     & 0.0098 ± 0.0251  & -0.2248 ± 0.0053 & -1               \\ \hline
Adaptive     & 0.0444 ± 0.0523  & -0.2149 ± 0.0040 & -1               \\ \hline
Polynomial   & 0.0983 ± 0.0748  & -0.2455 ± 0.0082 & -1               \\ \hline
Spectral     & -0.0651 ± 0.0643 & -0.3254 ± 0.0391 & -0.9172 ± 0.0400 \\ \hline
Inverse      & 0.0639 ± 0.0191  & -0.1889 ± 0.0410 & 0.0789 ± 0.0746  \\ \hline
Total\_var        & -0.3949 ± 0.0027 & -1.0002 ± 0.0003 & -0.9371 ± 0.0003 \\ \hline
Ground truth & 3                & 0.25             & 1                \\ \hline
\end{tabular}
\end{table}

\begin{table}[ht!]
\caption{Coefficients values calculated with SINDy at different noise levels}
\centering
\begin{tabular}{|c|ccc|ccc|ccc|}
\hline
\multirow{2}{*}{Methods} & \multicolumn{3}{c|}{Noise 0 \%} & \multicolumn{3}{c|}{Noise 0.5 \%} & \multicolumn{3}{c|}{Noise 1 \%} \\
\cline{2-10}
& u & u' & u'' & u & u' & u'' & u & u' & u'' \\ \hline
Gradient & 2.845 & 0.208 & 1 & 2.824 & 0.212 & 1 & 2.754 & 0.212 & 1 \\ \hline
Adaptive & 2.385 & 0.249 & 1 & 2.368 & 0.253 & 1 & 2.353 & 0.256 & 1 \\ \hline
Polynomial & 2.874 & 0.193 & 1 & 2.854 & 0.206 & 1 & 2.785 & 0.176 & 1 \\ \hline
Spectral & 3.199 & - & 1 & 3.180 & - & 1 & 3.197 & - & 1 \\ \hline
Inverse & 2.732 & 0.264 & 1 & 3.697 & 0.376 & 1 & 3.240 & 0.268 & 1 \\ \hline
Total\_var & 0.413 & 1.070 & 1 & 0.409 & 1.066 & 1 & 0.414 & 1.072 & 1 \\ \hline
Ground truth & 3 & 0.25 & 1 & 3 & 0.25 & 1 & 3 & 0.25 & 1 \\ \hline
\end{tabular}
\end{table}

%\clearpage
%\newpage

\subsection{Korteweg -- de Vries equation}

The Korteweg-de Vries equation is a partial differential equation $u'_{t} + u'"_{xxx} + 6uu'_{x} = 0$, which is one of the few that has analytical one-soliton and two-soliton solutions.

We will study its single-soliton solution, presented in the following form $u(x,t) = \frac{2 (k^2)}{ch^2(-k (\mathbf{x} - 4(k^2)t))}$ ,where $k = 0.7$ is the constant that determines the velocity of the soliton $4k^2$ and the amplitude $2k^2$. The results visualized as boxplots are given in Appendix~\ref{app:KdV_res} with a further discussion in Section~\ref{sec:discussion}.

\begin{table}[ht!]
\caption{Coefficients values calculated with EPDE, noise =0\%}
\centering
\begin{tabular}{|c|c|c|c|}
\hline
Methods/Terms & du/dt            & d\textasciicircum{}3u/dx\textasciicircum{}3 & u*du/dx          \\ \hline
Gradient      & -0.4565 ± 0.2045 & 0.0008 ± 0.0038                             & -1.3444 ± 0.1839 \\ \hline
Adaptive      & -0.5102 & -                                           & -1.9143 ± 0.0645 \\ \hline
Polynomial    & -0.5045 ± 0.4228 & -                                           & -0.0303 ± 0.0011 \\ \hline
Spectral      & 0.0202 ± 0.0401  & 0.0002 ± 0.0000                             & -0.2297 ± 0.1165 \\ \hline
Inverse       & 0.0142 ± 0.0007  & -                                           & -0.2412 ± 0.0004 \\ \hline
Total\_var         & -0.8334 ± 0.4282 & 1.1503                                      & -0.9770 ± 0.0231 \\ \hline
Ground truth  & 1                & 1                                           & 6                \\ \hline
\end{tabular}
\end{table}

\begin{table}[ht!]
\caption{Coefficients values calculated with EPDE, noise =0.5\%}
\centering
\begin{tabular}{|c|c|c|c|}
\hline
Methods/Terms & du/dt            & d\textasciicircum{}3u/dx\textasciicircum{}3 & u*du/dx          \\ \hline
Gradient     & -0.1973 ± 0.2523 & -0.0001          & -1.5692 ± 0.1447 \\ \hline
Adaptive     & -0.5081 ± 0.0035 & -0.0801 ± 0.0554                & -0.7084 ± 0.2506 \\ \hline
Polynomial   & -0.9822 ± 0.0248 & -0.0022          & -0.8874 ± 0.2106 \\ \hline
Spectral     & 0.0191 ± 0.0614  & -0.0000 ± 0.0003 & -0.1706 ± 0.1092 \\ \hline
Inverse      & -0.4698 ± 0.1770 & 0.0001 ± 0.0003  & 0.0419 ± 0.0954  \\ \hline
Total\_var        & -0.8164 ± 0.1561 & 1.1569 ± 0.1997  & -0.9282 ± 0.0466 \\ \hline
Ground truth & 1                & 1                & 6                \\ \hline
\end{tabular}
\end{table}

\begin{table}[ht!]
\caption{Coefficients values calculated with EPDE, noise =1\%}
\centering
\begin{tabular}{|c|c|c|c|}
\hline
Methods/Terms & du/dt            & d\textasciicircum{}3u/dx\textasciicircum{}3 & u*du/dx          \\ \hline
Gradient     & -0.3711 ± 0.1267 & -0.0819 ± 0.0647 & -0.7199 ± 0.3075 \\ \hline
Adaptive     & -0.4008 ± 0.0431 & -0.0553 ± 0.0512   & -0.3934 ± 0.1230 \\ \hline
Polynomial   & -0.6898 ± 0.1512 & -0.1111 ± 1.4098 & -0.9766 ± 0.1917 \\ \hline
Spectral     & 0.0320 ± 0.0667  & -0.0001 ± 0.0006 & -0.1036 ± 0.0779 \\ \hline
Inverse      & -0.1240 ± 0.0939 & 0.1313 ± 0.1499  & 0.0565 ± 0.0776  \\ \hline
Total\_var        & -0.8394 ± 0.1006 & -                & -0.6780 ± 0.1103 \\ \hline
Ground truth & 1                & 1                & 6                \\ \hline
\end{tabular}
\end{table}

%\clearpage
%\newpage

\begin{table}[ht!]
\caption{Coefficients values calculated with SINDy at different noise levels}
\centering
\begin{tabular}{|c|ccc|ccc|ccc|}
\hline
\multirow{2}{*}{Methods} & \multicolumn{3}{c|}{Noise 0 \%} & \multicolumn{3}{c|}{Noise 0.5 \%} & \multicolumn{3}{c|}{Noise 1 \%} \\
\cline{2-10}
& du/dt & d\textasciicircum{}3u/dx\textasciicircum{}3 & u*du/dx & du/dt & d\textasciicircum{}3u/dx\textasciicircum{}3 & u*du/dx & du/dt & d\textasciicircum{}3u/dx\textasciicircum{}3 & u*du/dx \\ \hline
Gradient & 1 & -0.009 & 0.077 & 1 & 0.158 & 1.295 & 1 & 0.052 & 0.443 \\ \hline
Adaptive & 1 & - & 0.195 & 1 & 0.146 & 1.320 & 1 & 0.056 & 0.681 \\ \hline
Polynomial & 1 & - & 0.595 & 1 & - & 0.472 & 1 & - & 0.841 \\ \hline
Spectral & 1 & -0.067 & 2.530 & 1 & -0.066 & 2.530 & 1 & -0.067 & 2.523 \\ \hline
Inverse & 1 & 0.072 & 0.025 & 1 & - & - & 1 & - & - \\ \hline
Total\_var & 1 & -4.011 & -0.599 & 1 & -4.010 & -0.639 & 1 & -4.015 & -0.731 \\ \hline
Ground truth & 1 & 1 & 6 & 1 & 1 & 6 & 1 & 1 & 6 \\ \hline
\end{tabular}
\end{table}

\subsection{Burger's equation}   Burger's equation has the form  $u'_{t} + uu'_{x} = vu''_{xx}$ ,where $v = 0.05$ is the diffusion coefficient.

The solution was obtained using an implicit numerical scheme for the diffusion term and an explicit numerical scheme for the convective term. An initial condition was set, and the right and left boundaries were fixed at zero. The results visualized as boxplots are given in Appendix~\ref{app:Burgers_res} with a further discussion in Section~\ref{sec:discussion}.

\begin{table}[ht!]
\caption{Coefficients values calculated with EPDE, noise =0\%}
\centering
\begin{tabular}{|c|c|c|c|}
\hline
Methods/Terms & du/dt            & d\textasciicircum{}2u/dx\textasciicircum{}2 & u*du/dx          \\ \hline
Gradient      & -0.9454 ± 0.0301 & 0.0346 ± 0.0133                             & -0.8945 ± 0.0327 \\ \hline
Adaptive      & -0.4548 ± 0.6341 & 0.0402 ± 0.026                             & -0.7677 ± 0.4413 \\ \hline
Polynomial    & -0.9283 ± 0.0332 & 0.0439 ± 0.0188                             & -0.8931 ± 0.0556 \\ \hline
Spectral      & -0.4025 ± 0.0549 & 0.0032 ± 0.0171                             & -0.3732 ± 0.0849 \\ \hline
Inverse       & -0.2118 ± 0.1229 & -0.0239 ± 0.1173                            & 0.0773 ± 0.1174  \\ \hline
Total\_var         & -0.4373 ± 0.2731 & -1.2180 ± 0.2525                            & -0.1324 ± 0.1249 \\ \hline
Ground truth  & 1                & -0.05                                       & 1                \\ \hline
\end{tabular}
\end{table}

\begin{table}[ht!]
\caption{Coefficients values calculated with EPDE, noise =0.5\%}
\centering
\begin{tabular}{|c|c|c|c|}
\hline
Methods/Terms & du/dt            & d\textasciicircum{}2u/dx\textasciicircum{}2 & u*du/dx          \\ \hline
Gradient     & -0.8727 ± 0.0398 & 0.0428 ± 0.0035  & -0.8520 ± 0.0479 \\ \hline
Adaptive     & -0.5185 ± 0.0164 & 0.0069 ± 0.0031  & -0.0718 ± 0.0337 \\ \hline
Polynomial   & -0.9428 ± 0.0150 & 0.0378 ± 0.0202  & -0.9510 ± 0.0382 \\ \hline
Spectral     & -0.3064 ± 0.0521 & 0.0138 ± 0.0033  & -0.3226 ± 0.0810 \\ \hline
Inverse      & -0.3064 ± 0.0521 & 0.0138 ± 0.0033  & -0.3226 ± 0.0810 \\ \hline
Total\_var        & 0.0008 ± 0.0011 & -0.0041 ± 0.0115 & -0.5039 ± 0.0086 \\ \hline
Ground truth & 1                & -0.05            & 1                \\ \hline
\end{tabular}
\end{table}

%\clearpage
%\newpage

\begin{table}[ht!]
\caption{Coefficients values calculated with EPDE, noise =1\%}
\centering
\begin{tabular}{|c|c|c|c|}
\hline
Methods/Terms & du/dt            & d\textasciicircum{}2u/dx\textasciicircum{}2 & u*du/dx          \\ \hline
Gradient     & -0.4088 ± 0.0448 & 0.0051 ± 0.0269  & -0.5262 ± 0.0840 \\ \hline
Adaptive     & 0.5498 ± 0.0191 & 0.0033 ± 0.0026 & 0.0577 ± 0.0274 \\ \hline
Polynomial   & -0.8245 ± 0.0360 & 0.0384 ± 0.0208  & -0.9395 ± 0.0375 \\ \hline
Spectral     & -0.3414 ± 0.0472 & 0.0049 ± 0.0207  & -0.3910 ± 0.0798 \\ \hline
Inverse      & -0.1569 ± 0.0575 & 0.0238 ± 0.0420  & -0.0533 ± 0.0453 \\ \hline
Total\_var        & -0.4989 ± 0.1903 & -0.3595 ± 0.2082 & -0.0269 ± 0.0465 \\ \hline
Ground truth & 1                & -0.05            & 1                \\ \hline
\end{tabular}
\end{table}

\begin{table}[ht!]
\caption{Coefficients values calculated with SINDy at different noise levels}
\centering
\begin{tabular}{|c|ccc|ccc|ccc|}
\hline
\multirow{2}{*}{Methods} & \multicolumn{3}{c|}{Noise 0 \%} & \multicolumn{3}{c|}{Noise 0.5 \%} & \multicolumn{3}{c|}{Noise 1 \%} \\
\cline{2-10}
& du/dt & d\textasciicircum{}2u/dx\textasciicircum{}2 & u*du/dx & du/dt & d\textasciicircum{}2u/dx\textasciicircum{}2 & u*du/dx & du/dt & d\textasciicircum{}2u/dx\textasciicircum{}2 & u*du/dx \\ \hline
Gradient & 1 & -0.044 & 0.952 & 1 & -0.044 & 0.955 & 1 & - & 0.661 \\ \hline
Adaptive & 1 & -0.045 & 0.951 & 1 & -0.045 & 0.951 & 1 & - & 0.661 \\ \hline
Polynomial & 1 & -0.058 & 1.057 & 1 & -0.055 & 1.039 & 1 & -0.050 & 1.004 \\ \hline
Spectral & 1 & - & 0.273 & 1 & - & 0.277 & 1 & - & 0.271 \\ \hline
Inverse & 1 & -0.134 & 0.205 & 1 & - & 0.188 & 1 & -0.129 & 0.202 \\ \hline
Total\_var & 1 & 1.765 & - & 1 & 1.763 & - & 1 & 1.736 & -0.014 \\ \hline
Ground truth & 1 & -0.05 & 1 & 1 & -0.05 & 1 & 1 & -0.05 & 1 \\ \hline
\end{tabular}
\end{table}

\subsection{Wave equation}   Wave equation has the form $u''_{xx} = c^2u''_{tt}$, where $c = 0.25$ is the propagation speed of the wave. The initial conditions were set as a sinusoidal function, and the boundary conditions were fixed at zero. The finite difference method was then used to solve the problem. The results visualized as boxplots are given in Appendix~\ref{app:Wave_res} with a further discussion in Section~\ref{sec:discussion}.

\begin{table}[ht!]
\caption{Coefficients values calculated with EPDE at different noise levels}
\centering
\small
\begin{tabular}{|c|cc|cc|cc|}
\hline
\multirow{2}{*}{Methods} & \multicolumn{2}{c|}{\footnotesize Noise 0\%} & \multicolumn{2}{c|}{\footnotesize Noise 0.5\%} & \multicolumn{2}{c|}{\footnotesize Noise 1\%} \\
\cline{2-7}
& \footnotesize $u_{xx}$ & \footnotesize $u_{tt}$ & \footnotesize $u_{xx}$ & \footnotesize $u_{tt}$ & \footnotesize $u_{xx}$ & \footnotesize $u_{tt}$ \\ \hline
Gradient & -1 & -0.0005\textsuperscript{$\pm$.0338} & 0.0037\textsuperscript{$\pm$.0075} & 0.0688\textsuperscript{$\pm$.0199} & -0.0041\textsuperscript{$\pm$.0068} & 0.0064\textsuperscript{$\pm$.0070} \\ \hline
Adaptive & -1 & -0.0579\textsuperscript{$\pm$.0408} & 0.0008\textsuperscript{$\pm$.0061} & 0.0972\textsuperscript{$\pm$.0131} & 0.0004\textsuperscript{$\pm$.0003} & 0.0108\textsuperscript{$\pm$.0056} \\ \hline
Polynomial & -1 & -0.0827\textsuperscript{$\pm$.0430} & 0.0018\textsuperscript{$\pm$.0030} & 0.1539\textsuperscript{$\pm$.0345} & -0.0001\textsuperscript{$\pm$.0005} & 0.1041\textsuperscript{$\pm$.0191} \\ \hline
Spectral & - & - & 0.1549\textsuperscript{$\pm$.0041} & - & 0.1533\textsuperscript{$\pm$.0039} & - \\ \hline
Inverse & -0.9486\textsuperscript{$\pm$.0502} & 0.0038\textsuperscript{$\pm$.0162} & -0.8171\textsuperscript{$\pm$.1025} & -0.5975\textsuperscript{$\pm$.1638} & -0.8946\textsuperscript{$\pm$.0792} & -0.6434\textsuperscript{$\pm$.1901} \\ \hline
Total\_var & -1.0049\textsuperscript{$\pm$.0097} & -0.9904\textsuperscript{$\pm$.0191} & -1 & -1 & -0.3441\textsuperscript{$\pm$.1927} & -1.0066\textsuperscript{$\pm$.0092} \\ \hline
Ground truth & 1 & -0.0625 & 1 & -0.0625 & 1 & -0.0625 \\ \hline
\end{tabular}
\end{table}

\begin{table}[ht!]
\caption{Coefficients values calculated with SINDy at different noise levels}
\centering
\begin{tabular}{|c|cc|cc|cc|}
\hline
\multirow{2}{*}{Methods} & \multicolumn{2}{c|}{Noise 0 \%} & \multicolumn{2}{c|}{Noise 0.5 \%} & \multicolumn{2}{c|}{Noise 1 \%} \\
\cline{2-7}
& d\textasciicircum{}2u/dx\textasciicircum{}2 & d\textasciicircum{}2u/dt\textasciicircum{}2 & d\textasciicircum{}2u/dx\textasciicircum{}2 & d\textasciicircum{}2u/dt\textasciicircum{}2 & d\textasciicircum{}2u/dx\textasciicircum{}2 & d\textasciicircum{}2u/dt\textasciicircum{}2 \\ \hline
Gradient & - & - & 1 & -0.193 & 1 & -0.395 \\ \hline
Adaptive & 1 & -0.055 & 1 & -0.163 & 1 & -0.332 \\ \hline
Polynomial & 1 & -0.063 & 1 & -0.049 & 1 & -0.1 \\ \hline
Spectral & 1 & - & 1 & - & 1 & - \\ \hline
Inverse & 1 & -0.008 & 1 & -0.221 & 1 & -4.586 \\ \hline
Total\_var & 1 & -0.007 & 1 & -0.027 & 1 & -0.079 \\ \hline
Ground truth & 1 & -0.0625 & 1 & -0.0625 & 1 & -0.0625 \\ \hline
\end{tabular}
\end{table}

\subsection{Laplace equation} $u''_{xx} + u''_{yy} = 0$ . The Dirichlet boundary conditions were set, and the problem was solved using the finite difference method. The results visualized as boxplots are given in Appendix~\ref{app:Laplace_res} with a further discussion in Section~\ref{sec:discussion}.

\begin{table}[ht!]
\caption{Coefficients values calculated with EPDE at different noise levels}
\centering
\small
\begin{tabular}{|c|cc|cc|cc|}
\hline
\multirow{2}{*}{Methods} & \multicolumn{2}{c|}{\footnotesize Noise 0\%} & \multicolumn{2}{c|}{\footnotesize Noise 0.5\%} & \multicolumn{2}{c|}{\footnotesize Noise 1\%} \\
\cline{2-7}
& \footnotesize $u_{xx}$ & \footnotesize $u_{yy}$ & \footnotesize $u_{xx}$ & \footnotesize $u_{yy}$ & \footnotesize $u_{xx}$ & \footnotesize $u_{yy}$ \\ \hline
Gradient & -1 & -0.997 & 0.2072\textsuperscript{$\pm$.1540} & 0.5129\textsuperscript{$\pm$.0095} & 0.4476\textsuperscript{$\pm$.1022} & 0.4962\textsuperscript{$\pm$.0192} \\ \hline
Adaptive & -1 & -0.997 & 0.4755\textsuperscript{$\pm$.1158} & 0.5150\textsuperscript{$\pm$.0065} & 0.4047\textsuperscript{$\pm$.0796} & 0.5056\textsuperscript{$\pm$.0164} \\ \hline
Polynomial & -1 & -0.9964 & 0.5139\textsuperscript{$\pm$.0077} & 0.3348\textsuperscript{$\pm$.0365} & 0.5100\textsuperscript{$\pm$.0140} & 0.2119\textsuperscript{$\pm$.0521} \\ \hline
Spectral & -1 & -1 & 0.4728\textsuperscript{$\pm$.0783} & 0.5393\textsuperscript{$\pm$.0137} & 0.3388\textsuperscript{$\pm$.1160} & 0.5333\textsuperscript{$\pm$.0117} \\ \hline
Inverse & -0.9985\textsuperscript{$\pm$.0007} & -0.9955\textsuperscript{$\pm$.0013} & -0.0000\textsuperscript{$\pm$.0025} & 0.5579\textsuperscript{$\pm$.0239} & 0.5868\textsuperscript{$\pm$.0061} & 0.3661\textsuperscript{$\pm$.0714} \\ \hline
Total\_var & -1 & -1 & 0.3411\textsuperscript{$\pm$.1787} & 0.1874\textsuperscript{$\pm$.0382} & 0.3295\textsuperscript{$\pm$.0528} & 0.2150\textsuperscript{$\pm$.0566} \\ \hline
Ground truth & 1 & 1 & 1 & 1 & 1 & 1 \\ \hline
\end{tabular}
\end{table}

\begin{table}[ht!]
\caption{Coefficients values calculated with SINDy at different noise levels}
\centering
\begin{tabular}{|c|cc|cc|cc|}
\hline
\multirow{2}{*}{Methods} & \multicolumn{2}{c|}{Noise 0 \%} & \multicolumn{2}{c|}{Noise 0.5 \%} & \multicolumn{2}{c|}{Noise 1 \%} \\
\cline{2-7}
& d\textasciicircum{}2u/dx\textasciicircum{}2 & d\textasciicircum{}2u/dy\textasciicircum{}2 & d\textasciicircum{}2u/dx\textasciicircum{}2 & d\textasciicircum{}2u/dy\textasciicircum{}2 & d\textasciicircum{}2u/dx\textasciicircum{}2 & d\textasciicircum{}2u/dy\textasciicircum{}2 \\ \hline
Gradient & 1 & 1.028 & 1 & - & 1 & - \\ \hline
Adaptive & 1 & 1.129 & 1 & - & 1 & -0.457 \\ \hline
Polynomial & 1 & 1.009 & 1 & - & 1 & - \\ \hline
Spectral & 1 & - & 1 & - & 1 & - \\ \hline
Inverse & 1 & 0.62 & 1 & - & 1 & - \\ \hline
Total\_var & 1 & - & 1 & - & 1 & - \\ \hline
Ground truth & 1 & 1 & 1 & 1 & 1 & 1 \\ \hline
\end{tabular}
\end{table}

For every experiment we also record structural Hamming distance placed in Appendix~\ref{app:SHD} and differentiation error in Appendix~\ref{app:diff_error}.

%\clearpage
%\newpage

\subsection{Quasigeostrophic potential vorticity} Original data were obtained using the pyqg framework \footnote{\url{http://github.com/pyqg/pyqg}} for quasi-geostrophic modeling. The maximum number of terms was extended to 15 to capture complex dynamics. Since the exact governing equations are unknown, we evaluate the discovered equations by comparing the discrepancy between the original data and numerical solutions from a Physics-Informed Neural Networks (PINNs) solver. PINNs are necessary due to the high non-linearity that renders conventional FEM inadequate. The general form of the governing equation is $\mathbf{V_g} \cdot \nabla q = 0$, where $V_g$ represents geostrophic velocity and $q$ denotes potential vorticity.

The equations presented were derived using Savitzky-Golay (SG) filtering and spectral domain differentiation methods, respectively, as alternative approaches failed to capture the eddy-driven structure of the derivatives, resulting in suboptimal preprocessing. The solutions to these equations similarly exhibit a lack of regions with pronounced eddy behavior, which may indicate a tendency toward identifying broader-scale features in the data. Equations discovered:

via spectral domain differentiation
\begin{equation}
\begin{aligned}
&3.2602 \times 10^{-6} u_{xx} + 0.0067028 u + 0.7095 u_y \\
&- 0.6485 u_y \cos(1.7965 y) + 2.5201 \times 10^{-5} u u_{yy}\\
&- 0.01010 u_x u - 1.2018 \times 10^{-6} u_{xx} u_{yy} \\
&+ 3.2084 \times 10^{-5} y u_{yy} + 2.4292 \times 10^{-5} u_{xx} u_y\\
&  + 0.1998 u_y \sin(2.7363 y) - 0.000223 - y u_y = 0
\end{aligned}
\end{equation}

via SG filtering
\begin{equation}
\begin{aligned}
    & 0.044994162 \, u_x - 5.34527 \times 10^{-5} \, u_{xx} \\
    &- 0.000760196 \, u_x u_{yy} + 0.001192827 - u_x u = 0 \\
\end{aligned}
\end{equation}

Visual representations of the original data, numerical solutions, and error maps are provided in Appendix~\ref{app:pyqg}. These results demonstrate that, in real-world cases, we cannot consistently achieve results for unknown equations and that we require ensembles that include both data subsampling and differentiation uncertainty.

\section{Discussion}
\label{sec:discussion}

Our experiments reveal a counterintuitive reality: numerically precise differentiation is not a remedy for equation discovery. As shown in Tab.~\ref{tab:integral}, methods such as Spectral achieve a minimal differentiation error (e.g., $9.988 \cdot 10^{-6}$ at 0\% noise), but produce poor structural precision (SHD = $4 \pm 0.13$). In contrast, despite the high differentiation error (1.963 for 1\% noise level). Polynomial consistently delivers superior structural recovery (SHD = 3 ± 0.13). This paradox arises because noise amplification from high-precision methods introduces misleading high-frequency artifacts. As we illustrate in a concrete example (see \eqref{eq:error_expansion}), optimal discovery requires strategic smoothing rather than maximal precision within a single algorithm run.

%\clearpage
%\newpage

\begin{table}[ht!]
\caption{Differentiation Method Performance Analysis}
\resizebox{\columnwidth}{!}{
\begin{tabular}{llllll}
                    & Noise Level & D1/D2/D3 error & Coeff. Error (Mean ± SD) & SHD (Mean ± SD) & Key Insight                            \\
\multirow{3}{*}{\tiny{\rotatebox[origin=c]{90}{Gradient}}}   & 0\%         & 0.0002/0.00027/0.00050             & 0.7309 ± 0.0515          & 2 ± 0.0782      & Precise but noisy; good for clean data \\
                            & 0.5\%       & 0.0017/0.0024/0.0170             & 0.9179 ± 0.0683          & 3 ± 0.1153      & Noise amplifies rapidly                \\
                            & 1\%         & 0.0065/0.0036/0.0502             & 1.0068 ± 0.2449          & 4 ± 0.1482      & Avoid for noisy PDEs                   \\
\multirow{3}{*}{\tiny{\rotatebox[origin=c]{90}{Adaptive}}}   & 0\%         & 0.0016/0.2023/0.0121             & 0.7309 ± 0.0515          & 2 ± 0.0823      & Precise but noisy; good for clean data \\
                            & 0.5\%       & 0.0025/0.2079/0.0221             & 0.9179 ± 0.0683          & 3 ± 0.0723      & Noise amplifies rapidly                \\
                            & 1\%         & 0.0341/0.6116/0.0432             & 1.0068 ± 0.2449          & 4 ± 0.0971      & Avoid for noisy PDEs                   \\
\multirow{3}{*}{\tiny{\rotatebox[origin=c]{90}{Polynomial}}} & 0\%         & 0.0193/0.3344/0.0350             & 0.8971 ± 0.0517          & 2 ± 0.1300      & \textbf{Low SHD despite high diff. error}  \\
                            & 0.5\%       & 0.0236/0.6421/0.0366            & 0.9611 ± 0.0390          & 3 ± 0.1260      & Robust structure recovery              \\
                            & 1\%         & 0.0302/1.9630/0.0390            & 0.9148 ± 0.1551          & 3 ± 0.1323      & Best SHD-noise tradeoff                \\
\multirow{3}{*}{\tiny{\rotatebox[origin=c]{90}{Spectral}}}   & 0\%         & 0.0683/11.7856/14.1910            & 1.1074 ± 0.0461          & 4 ± 0.1309      & \textbf{High precision, poor SHD}           \\
                            & 0.5\%       & 0.0716/13.0997/14.1966           & 1.1737 ± 0.0462          & 5 ± 0.1558      & Boundary artifacts dominate            \\
                            & 1\%         & 0.0800/17.8008/14.2271           & 1.1924 ± 0.0466          & 4 ± 0.1498      & Unreliable under noise                 \\
\multirow{3}{*}{\tiny{\rotatebox[origin=c]{90}{Inverse}}}    & 0\%         & 1.5583/52.5442/0.4601            & 1.0482 ± 0.0518          & 4 ± 0.1929      & Moderate SHD, high coeff. variance     \\
                            & 0.5\%       & 1.5655/71.5591/0.4617            & 1.2806 ± 0.0656          & 4 ± 0.1763      & \textbf{Worst coeff. error at low noise}    \\
                            & 1\%         & 1.5676/76.2041/0.4931           & 1.3120 ± 0.0816          & 5 ± 0.1537      & Avoid for high-order terms      \\ 
\multirow{3}{*}{\tiny{\rotatebox[origin=c]{90}{Total\_var}}}    & 0\%         & 1.6218/52.9117/0.3273            & 1.0482 ± 0.0518          & 3 ± 0.1055      & Smoothing improves structural recovery    \\
                            & 0.5\%       & 1.6292/54.4074/0.3275            & 1.2806 ± 0.0656          & 3 ± 0.1049      & Consistent performance across noise levels    \\
                            & 1\%         & 1.6296/58.7889/0.3275           & 1.3120 ± 0.0816          & 3 ± 0.1065      & \textbf{Good structure despite high error}      \\ 
\end{tabular}
}
\label{tab:integral}
\end{table}

\begin{figure}[ht!]
    \centering
    \includegraphics[width=0.48\textwidth]{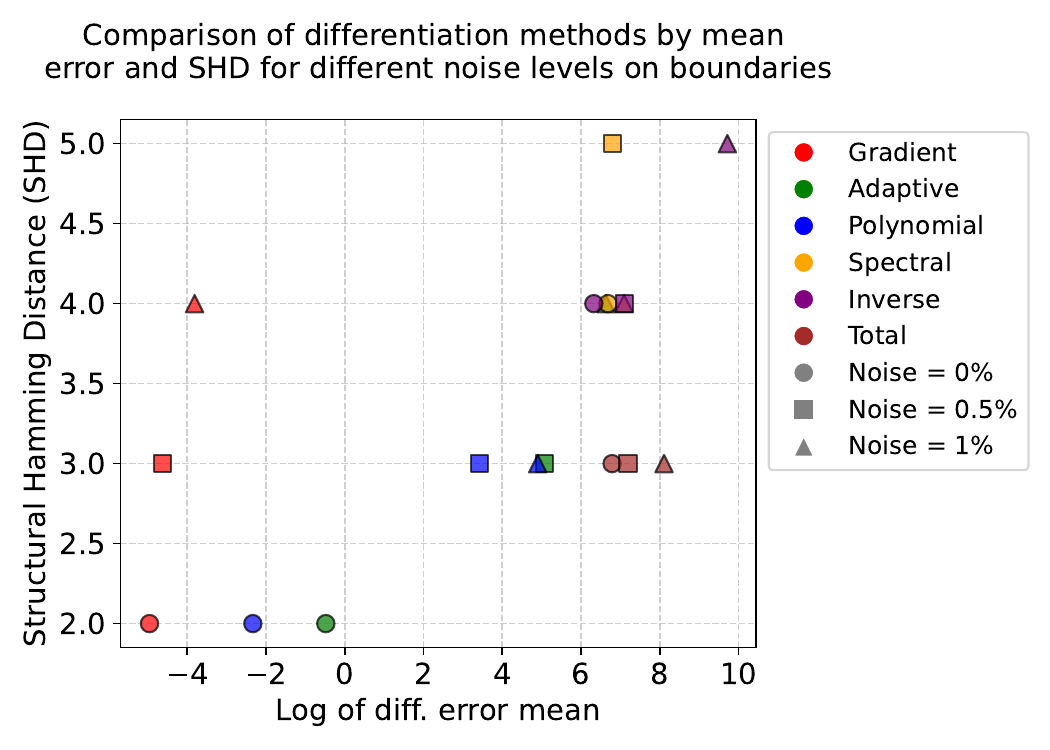}\hfill
    \includegraphics[width=0.48\textwidth]{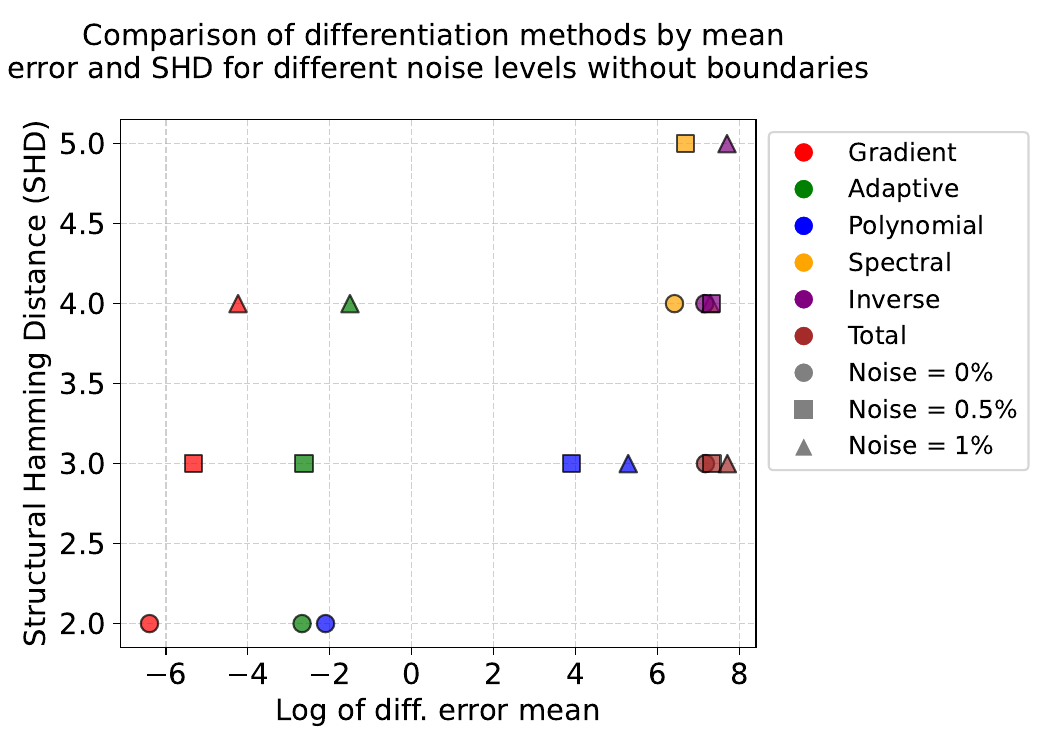}
    \caption{Comparison of differentiation methods by mean differentiation error and SHD for different noise levels on boundaries (left) and without boundaries (right)}
    \label{fig:rel_scatter}
\end{figure}

To illustrate the general dependencies, we plot the SHD and error values for different methods on a scatter plot, as shown in Fig.~\ref{fig:rel_scatter}.

\section{Conclusion}

The paper considers another aspect of differential equation discovery as a machine learning method. 

The error of the differentiation algorithm, as the "feature engineering" method plays a role in the general uncertainty, is often left out of the scope.

The main results are as follows.

\begin{itemize}
    \item The differentiation is an important part of every differential equation discovery method
    \item The differentiation is a reliable uncertainty source and may be used in ensembles to get the models on a different process scale
    \item Absolute value of differentiation error is less important -- very precise methods give poor discovery results in some cases
    \item Significant differentiation error is allowed in the derivative, and in some cases, it is necessary to accept it to work with noisy data
\end{itemize}

We also note that the conclusion remains the same regardless of the method used, whether it is LASSO regression-based SINDy or evolutionary EPDE.

\bibliographystyle{unsrtnat}
\bibliography{references}  %%% Uncomment this line and comment out the ``thebibliography'' section below to use the external .bib file (using bibtex) .





% \end{thebibliography}

\newpage

\appendix

\section{Differentiation approach formulation}
\label{app:differentiation_formulation}

\subsection{Savitzky-Golay filtering}

%One of the approaches, considered in this work, involves approximation of the input data with the fully-connected artificial neural network (ANN). Since the modeled function can be considered as a continuous function, sampled from a subset of a Euclidean space ($\Omega \in \mathbb{R}^{n}$), mapping into the Euclidean space of $\mathbb{R}^{m}$, where $m$ is the number of dependent variables, it can be represented as the ANN. One of the valuable properties of the artificial neural network, is that the low-frequency signal in data is learned first, while the further training approximates the high-frequency components \cite{rahaman2019spectral}. Thus, by training an ANN representation of the process, we can obtain its low-frequency approximation, which can be further differentiated with decreased noise component.

Savitzky-Golay (SG) filtering, developed in \cite{savitzky1964smoothing}, is a commonly used approach to signal or data filtering, coupled with an opportunity to compute derivatives, involves a least squares-based local fitting of the polynomials to represent the data. To the set of data samples along an axis, we introduce the window of (commonly, odd) length $N = 2M + 1$, allowing the construction of series of polynomials $P_{0}(x), P_{1}(x), \; ...$ up to (even) order $n$, $n < N$ to approximate the data in the interior of our domain. With the selection of appropriate window size, from which the function values are used for the approximation, and polynomial order, the overdetermined system is constructed. Its solution provides the polynomial coefficients that represent the smoothed signal, without oscillations, caused by the random error. Even though the boundaries of length $M$ can be processed in a separate way, with the finite-difference schema or by a shifted approximation, the quality of results tend to decrease, thus for the equation discovery only the domain interior shall be used.

During the calculation of the partial derivative $u'_{j}$ for the sample $u(x_i)$, matching the $x_i$ grid node along the $j$-th axis, we select samples $\mathbf{u}_i = (u_{i - M}, u_{i - M + 1}, \; ... \; , u_{i}, \; ... \;, u_{i + M})$ in the aforementioned window. Using the corresponding coordinates $\mathbf{y}_i = (x_{i - M}, \; ... \; , x_{i}, \; ... \;, x_{i + M})$, we introduce the least-square problem of detecting coefficient vector $\mathbf{\alpha} = (\alpha_{0}, \; ... \;, \alpha_{n-1})$ for the series $P_{0}, \; ... \; , P_{n-1}$. The representation of data samples is as follows:

\begin{equation}
\label{eq:representation_poly}
u_i = \sum_{k = 0}^{n-1} \alpha_{k} P_{k}(x_i).
\end{equation}

\begin{equation}
\label{eq:representation_lstsq}
\mathbf{\alpha} = \arg \min_{\alpha'} \vert \mathbf{u}_i - P \mathbf{y}_i \vert,
\end{equation}

where matrix $P$ contains values of the polynomials in the grid nodes.

In our case, we utilize orthogonal Chebyshev polynomials of the first kind, where by $C_{m}^{2k}$ we denote the number of combination of $2k$ elements from the set of cardinality $m$:

\begin{equation}
\label{eq:cheb_poly}
T_{m}(x) = \sum_{k=0}^{\lfloor m/2 \rfloor}C_{m}^{2k} (x^2 - 1)^{k} x^{m - 2k}%\frac{(x + \sqrt{x^{2} - 1})^{n} + (x - \sqrt{x^{2} - 1})^{n}}{2}
\end{equation}

Having a series of Chebyshev polynomials with calculated coefficients, differentiation can be held analytically. Using the representation of data as series in \ref{eq:representation_poly}, we get the derivative as $u'_i = \sum_{k = 0}^{n-1} \alpha_{k} U_{k}(x_i)$, where $U_{k}$ is a Chebyshev polynomial of the second kind.

\begin{equation}
\label{eq:cheb_der_poly}
U_{m}(x) = \sum_{k=0}^{\lfloor m/2 \rfloor}C_{m + 1}^{2k + 1} (x^2 - 1)^{k} x^{m - 2k}%\frac{(x + \sqrt{x^{2} - 1})^{n} + (x - \sqrt{x^{2} - 1})^{n}}{2}
\end{equation}

Although the provided approach is capable of filtering the data and stably calculating the derivatives, work \cite{schmid2022and} suggests that modification of Savitzky-Golay filtering by adding fitting weights or by implementing other filters, such as Whittaker-Henderson filter, can lead to better results in noise suppression.

\subsection{Spectral domain differentiation}

Although the process of differentiation in the spatial domain can be complicated for the data, described with an arbitrary function, in the Fourier domain the derivatives can be estimated in term-to-term basis \cite{johnson2011notes}. In general, the series of the derivatives, taken on a term-to-term basis may not converge. However, if we assume that the data represents continuous piecewise smooth function that has piecewise differentiable derivatives, the data can be differentiated term-to-term. 

A discrete Fourier transform (DFT) is the basis for our implementation of spectral domain differentiation. Let us examine a case of one-dimensional data, even though the algorithm can operate on multi-dimensional data, with the canonical discrete Fourier transform algorithm replaced by n-dimensional DFT. In data-driven equation discovery problems, one-dimensional data $u(t)$ is viewed from the point of view of samples $u_n = u(n T/N), n = 0, 1, \; ...\; , N-1$, where $T$ is the length of time interval and $N$ - the number of samples, and the corresponding coordinates will be $t_n = n T/N, n = 0, 1, \; ...\; , N-1$. The Fourier coefficients are denoted as $\hat{u}_k$, and they are calculated as:

\begin{equation}
\hat{u}_k = \frac{1}{N} \sum_{n = 0}^{N-1} u_n exp(-2 \pi i \frac{n k}{N}).
\label{eq:discrete_Fourier_transform}
\end{equation}

In many cases, the data are provided on the regular (even multi-dimensional) grid, thus to improve the algorithm performance a fast Fourier transform can be used. Due to the lower computational complexity, the increase in performance is substantial. The process of data reconstruction, using the obtained Fourier coefficients, is held with an inverse discrete Fourier transform:

\begin{equation}
u_n = \sum_{k = 0}^{N-1} \hat{u}_k exp(2 \pi i \frac{n k}{N}).
\label{eq:inv_discrete_Fourier_transform}
\end{equation}

Full term-by-term differentiation is performed in the Fourier domain, and the derivatives values are computed by the inverse DFT. For example, an expression for the first-order derivative has form, as in Eq.~\ref{eq:inv_discrete_Fourier_transform_diff}.

\begin{equation}
u'(t_k) = \sum_{0 < k < \frac{N-1}{2}}  \frac{2 \pi i}{T} k \left( \hat{u}_n exp(2 \pi i \frac{n k}{N}) - \hat{u}_{N - k} exp(-2 \pi i \frac{n k}{N}) \right).
\label{eq:inv_discrete_Fourier_transform_diff}
\end{equation}

Filtering with the desired properties can be done with low-pass filters that pass signals with lower frequencies, while dampen the high-frequency ones. Butterworth filter is a representative of such tools, and is flat for the passband (the frequencies that we do not want to penalize). The latter property prevents distortion of the modeled process by introducing factors, close to $1$, to the low-frequency Fourier components. The penalizing factor is introduced with the expression eq.~\ref{eq:butterworth}:

\begin{equation}
G(\omega) = \frac{1}{1 + (\omega/\omega_{cutoff})^{2s}},
\label{eq:butterworth}
\end{equation}

where $\omega$ is the frequency, $\omega_{cutoff}$ is the cutoff frequency, indicating the boundary frequency, from which the damping begins, and $s$ is the filter steepness parameter. The resulting expression is obtained with the introduction of penalizing factors $G(\omega) = G(k/N)$ into the series, representing derivatives:

\begin{equation}
u'(t_k) = \sum_{0 < k < \frac{N-1}{2}} G(k/N) \frac{2 \pi i}{T} k \left(\hat{u}_n exp(2 \pi i \frac{n k}{N}) - \hat{u}_{N - k} exp(-2 \pi i \frac{n k}{N})\right)
\label{eq:inv_discrete_Fourier_transform_diff_butterworth}
\end{equation}

The derivative of the higher orders can be calculated recursively from the lower order ones with the same filtering-based differentiation procedures, or, preferably, by the further multiplication with the integrating coefficient and IDFT. 

\subsection{Total variation regularization}

Variational principles provide an alternative method that incorporates inverse problem solution with the regularization of the variation of the gradient or its higher order analogues (e.g. Hessian). Rudin-Osher-Fatemi model \cite{rudin1992nonlinear} in its discrete formulation can be represented by the optimization problem of minimizing functional \ref{eq:TVR_func}.

\begin{equation}
 \vert D (\nabla \cdot u) \vert_{1} + \frac{\mu}{2} \vert K (\nabla \cdot u) - u \vert^{2}_{2} \longrightarrow \min_{u},
 \label{eq:TVR_func}
\end{equation}

where $\nabla \cdot u = (\frac{\partial u}{\partial t}, \frac{\partial u}{\partial x_1}, \; ...)$ is the gradient of the data field and $K$ and $D = (D_{t}, D_{x_1}, D_{x_2}, \; ...)$ represent discrete integration operators onf differentiation. Regularization of gradient variation is maintained with term $\vert D (\nabla \cdot u) \vert_{1} = \sum_{\Omega} \sqrt{\sum_{i, \;j} \frac{\partial^2 u)}{\partial x_i \partial x_j }}$.

\cite{chartrand2011numerical,chartrand2017numerical}

Although there are multiple approaches to the solution of the problem, we employ an approach, proposed in articles \cite{chartrand2011numerical,chartrand2017numerical}, that is designed for a function of one variable. While this approach can be generalized to the problems of higher dimensionality, the computational costs associated with the optimization limit the method's applicability to large datasets. To perform the functional optimization required in Eq.~\ref{eq:TVR_func}, the corresponding Euler-Lagrange equation has to be formed and solved.

\clearpage
\newpage

\section{ODE equation coefficients and Structural Hamming  Distances}
\label{app:ODE_res}

%ODE coeffs and Hamming
\begin{figure*}[ht!]
    \centering
    \includegraphics[width=.7\textwidth]{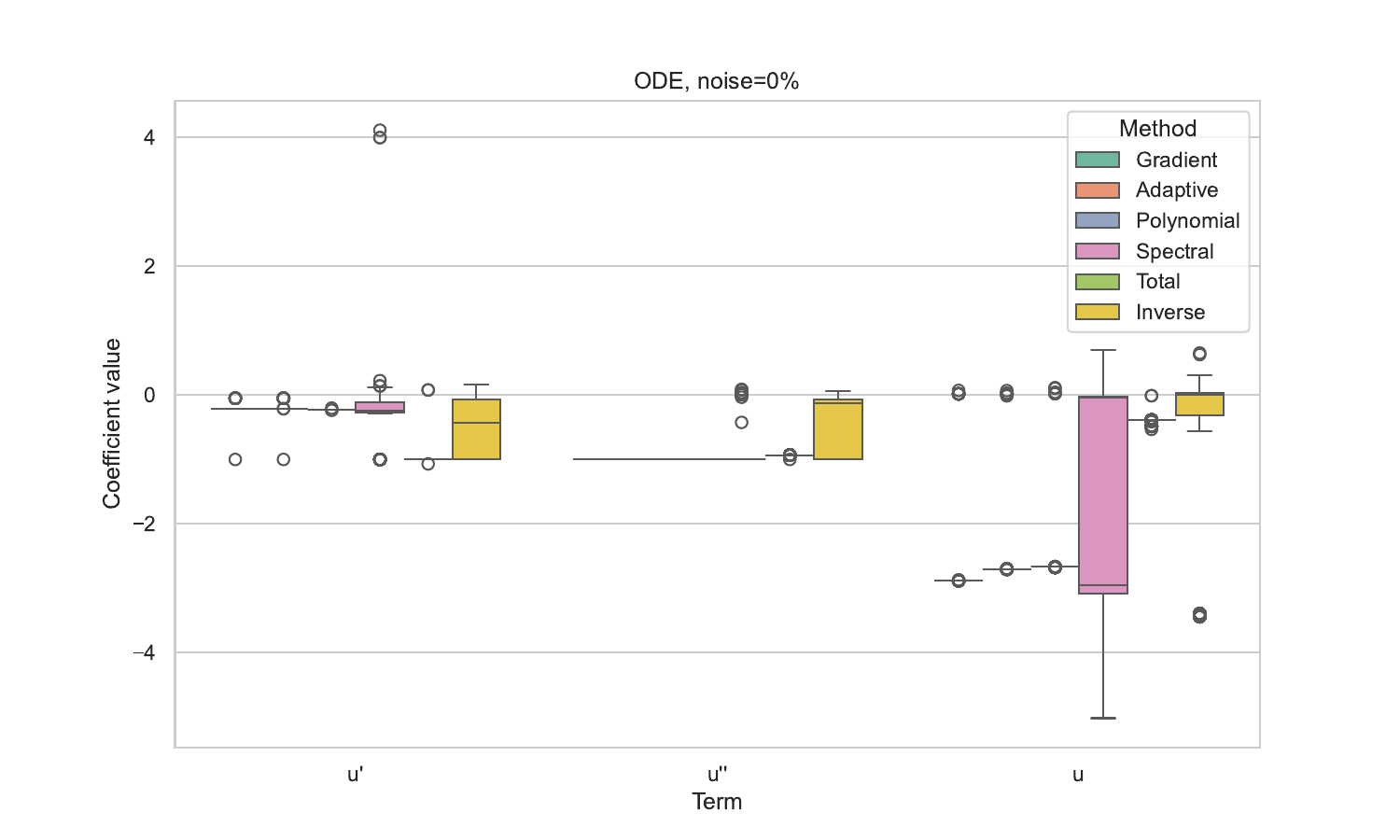}\
    \includegraphics[width=.7\textwidth]{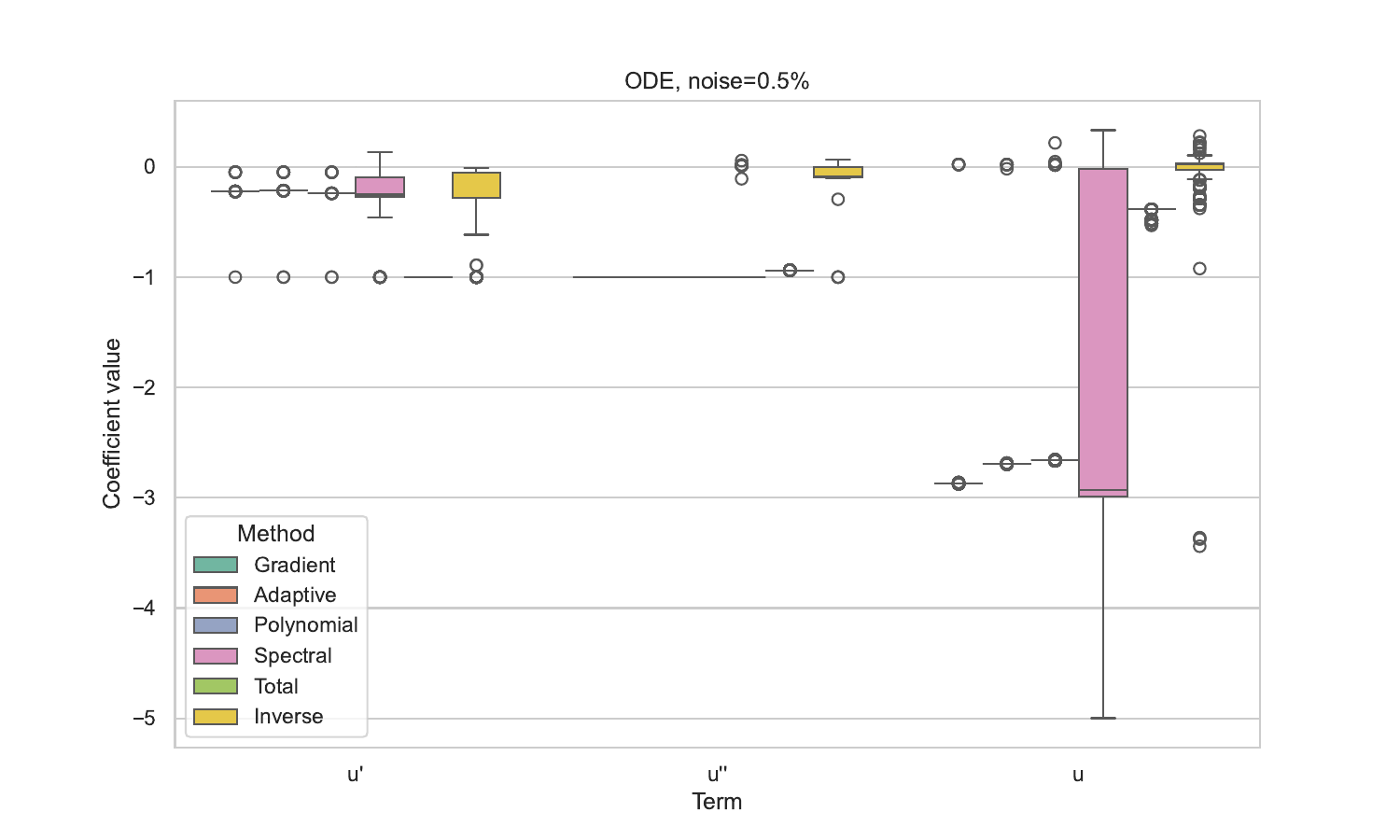}\
    \includegraphics[width=.7\textwidth]{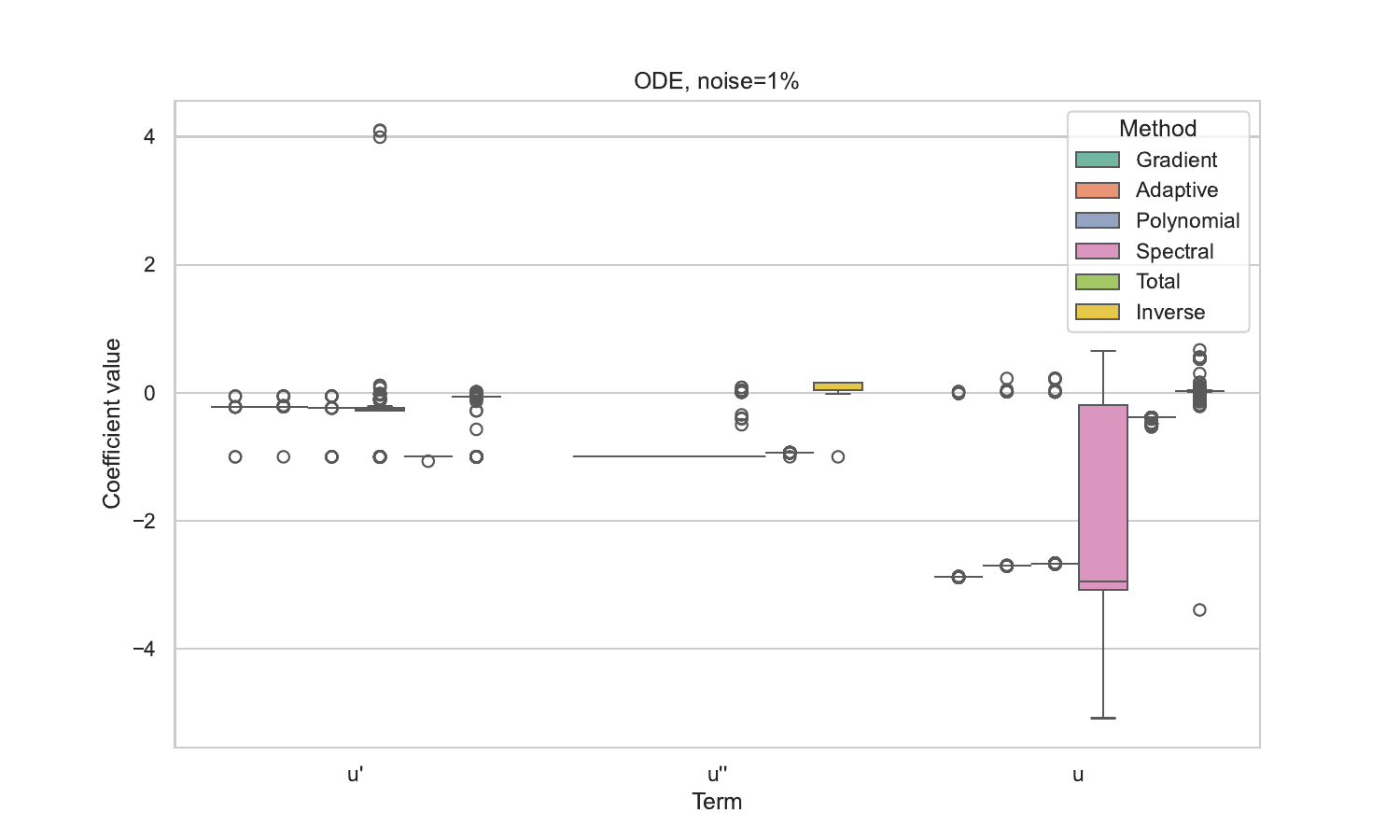}
    \caption{Distribution of coefficients values for different noise level}
    %label{fig:all}
\end{figure*}

\begin{figure*}[ht!]
    \centering
    \includegraphics[width=.7\textwidth]{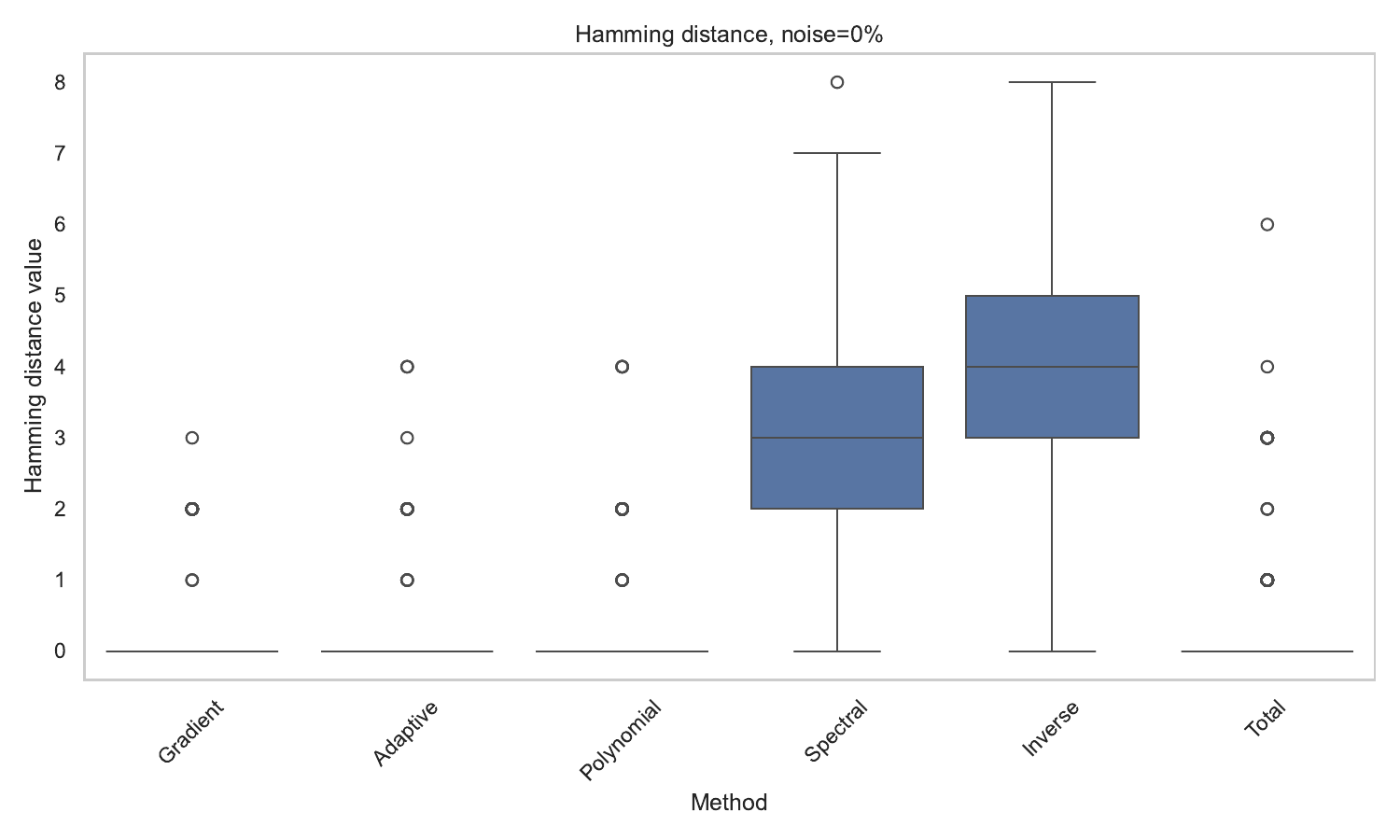}\
    \includegraphics[width=.7\textwidth]{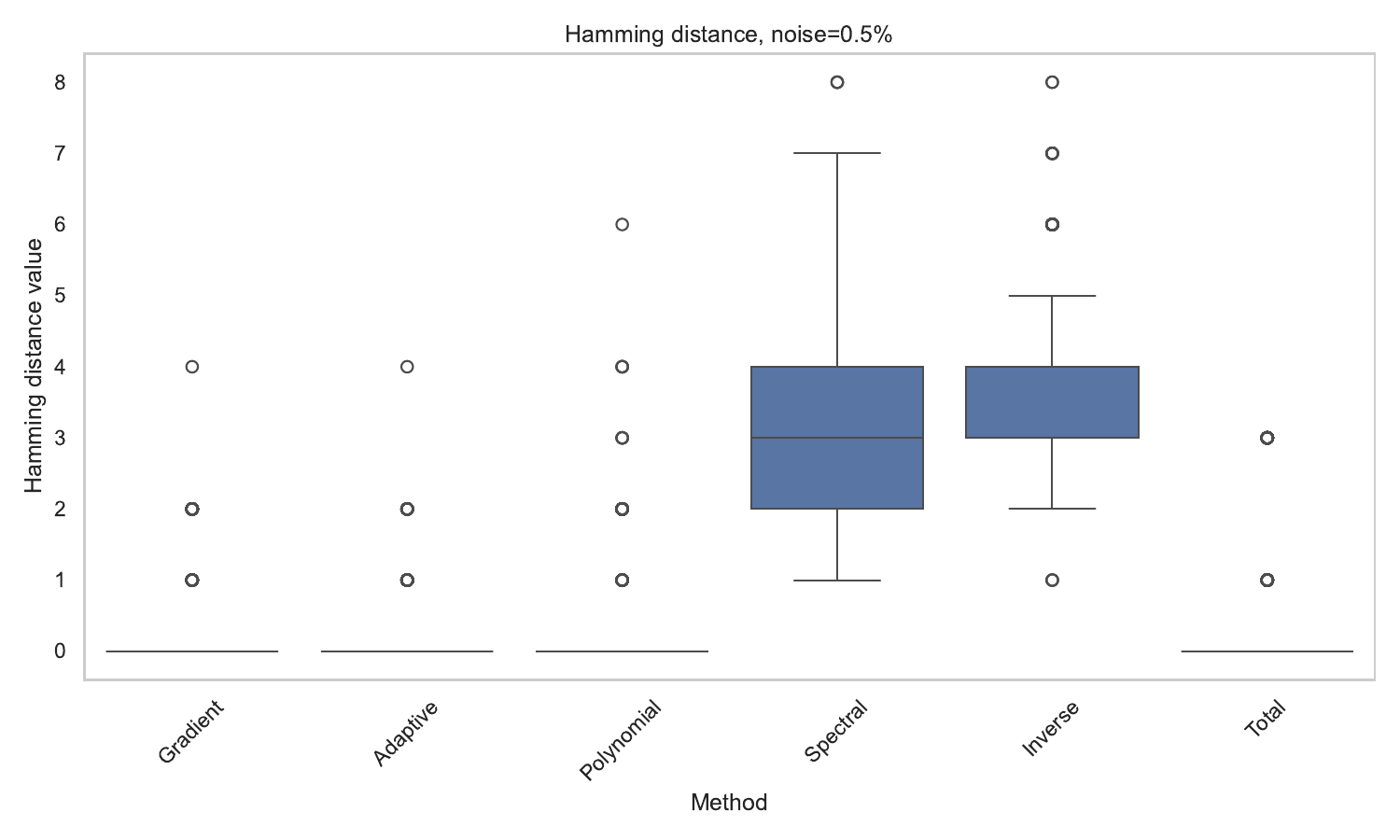}\
    \includegraphics[width=.7\textwidth]{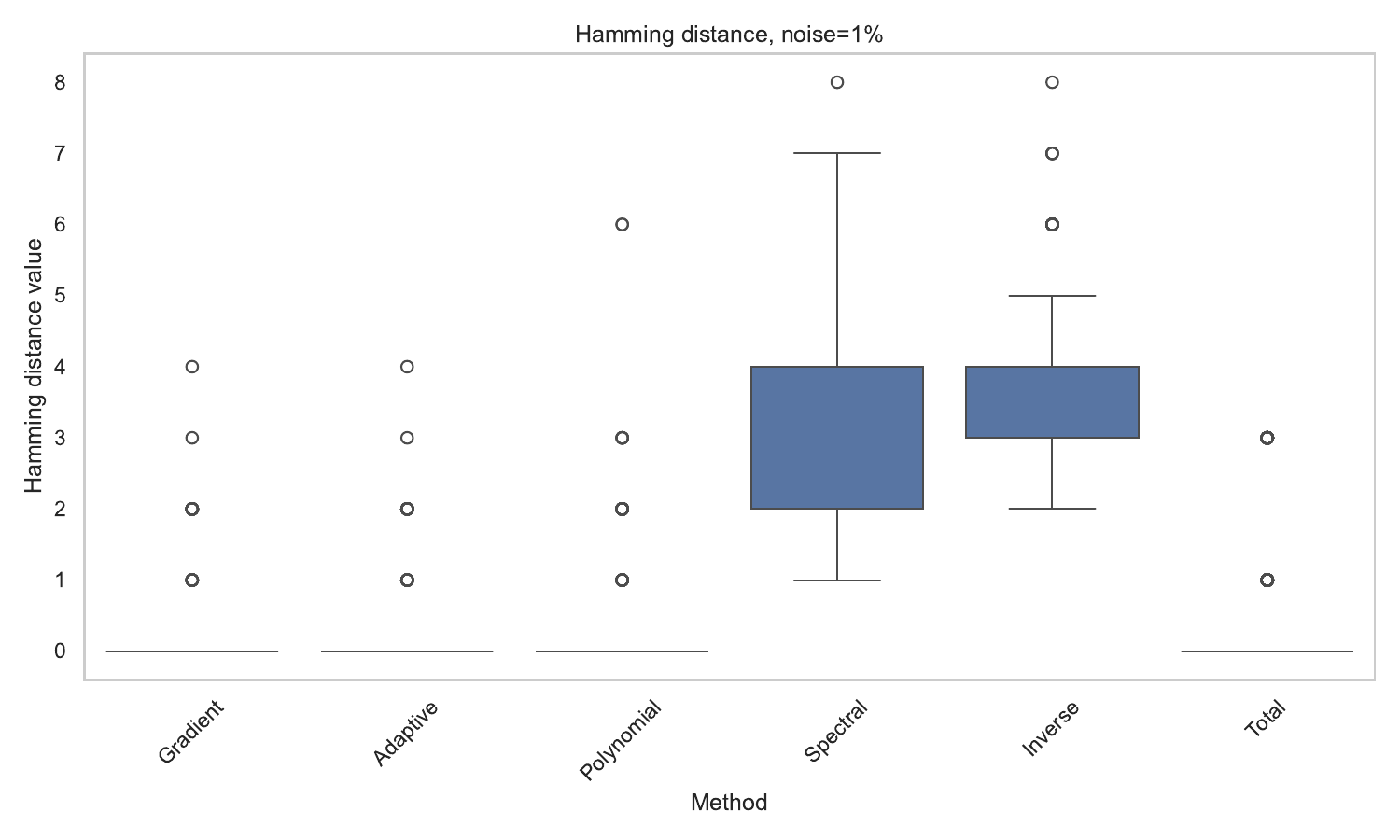}
    \caption{Distribution of coefficients values for different noise level}
    %label{fig:all}
\end{figure*}

\clearpage
\newpage
\section{KdV equation coefficients and Structural Hamming  Distances}
\label{app:KdV_res}

%KdV coeffs and Hamming
\begin{figure*}[ht!]
    \centering
    \includegraphics[width=.7\textwidth]{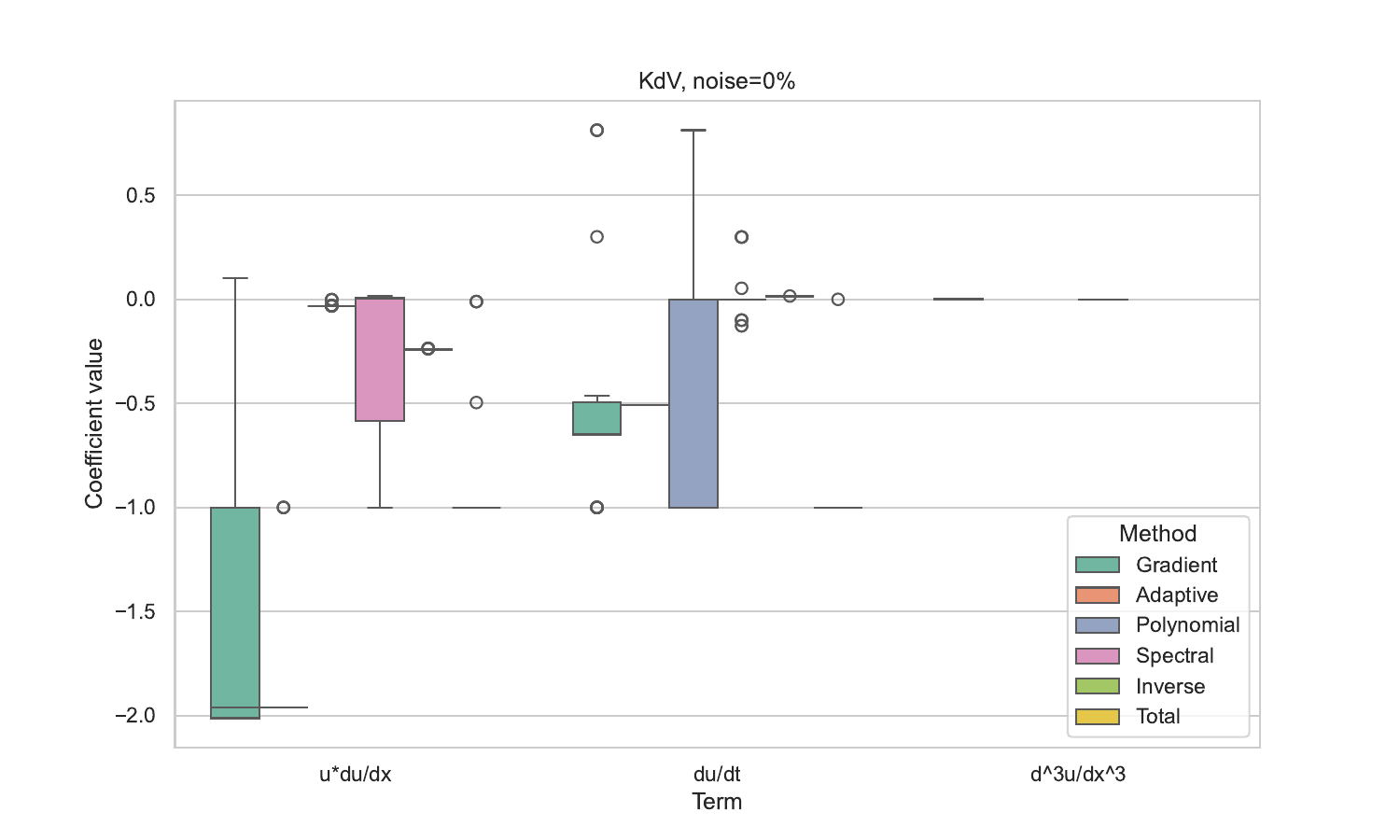}\
    \includegraphics[width=.7\textwidth]{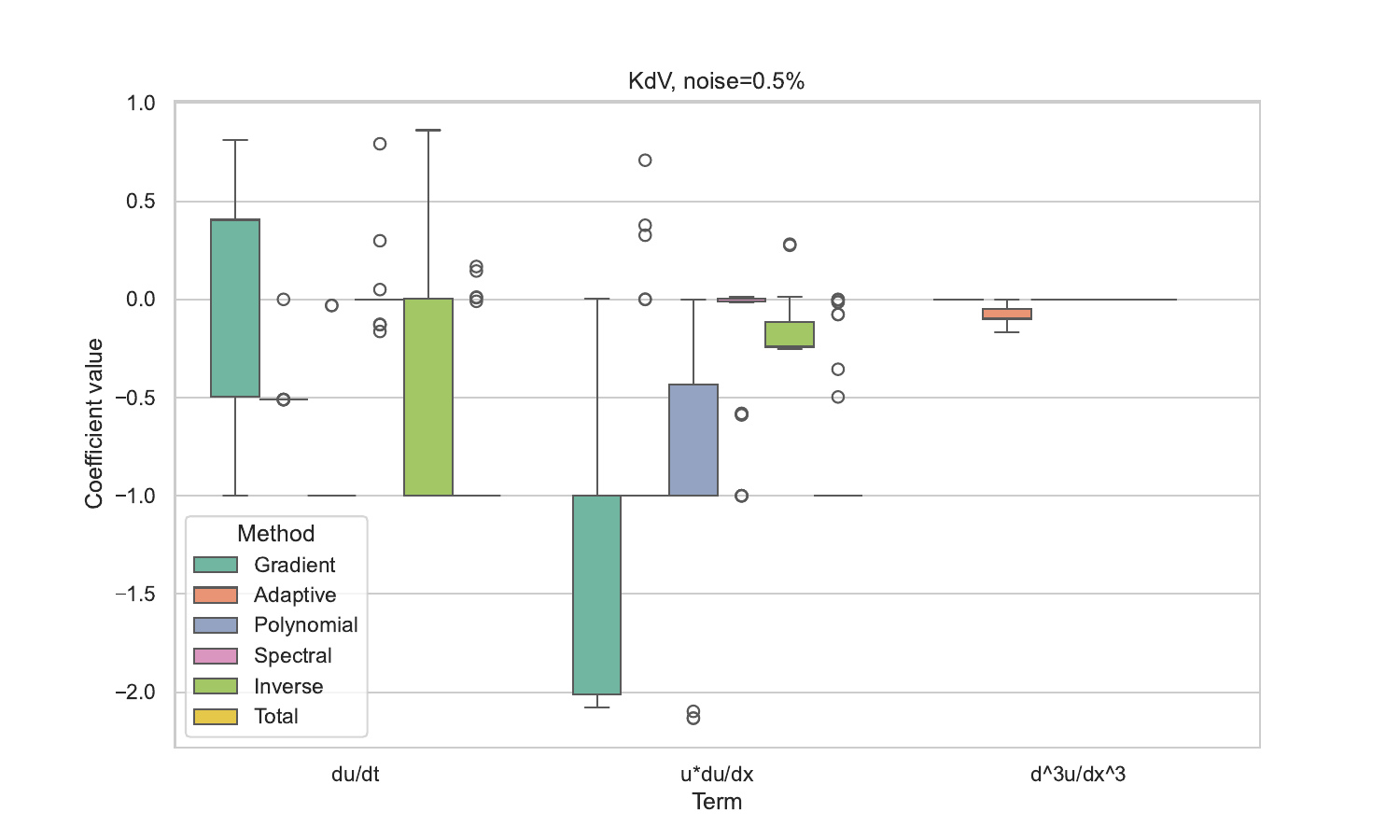}\
    \includegraphics[width=.7\textwidth]{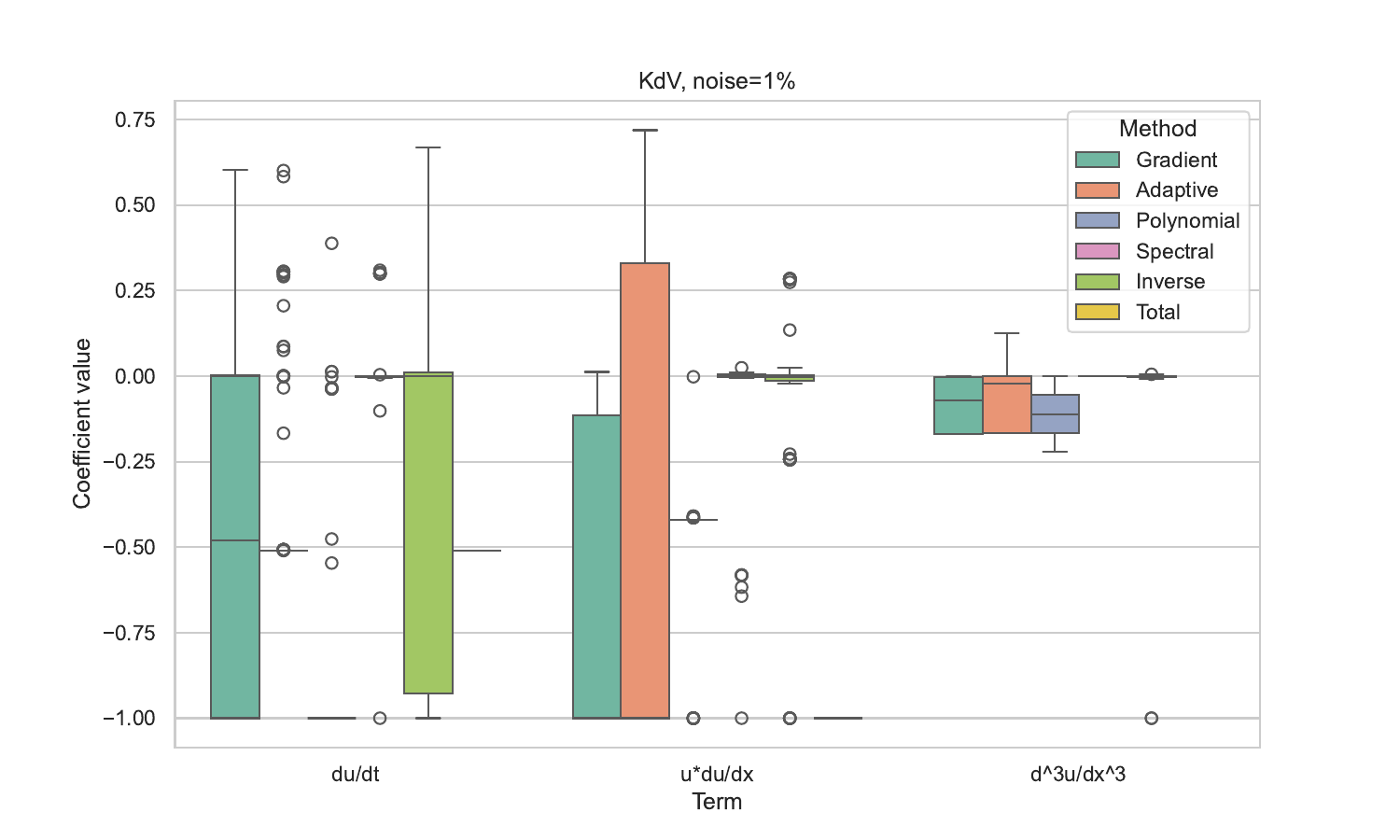}
    \caption{Distribution of coefficients values for different noise level}
    %label{fig:all}
\end{figure*}
\begin{figure*}[ht!]
    \centering
    \includegraphics[width=.7\textwidth]{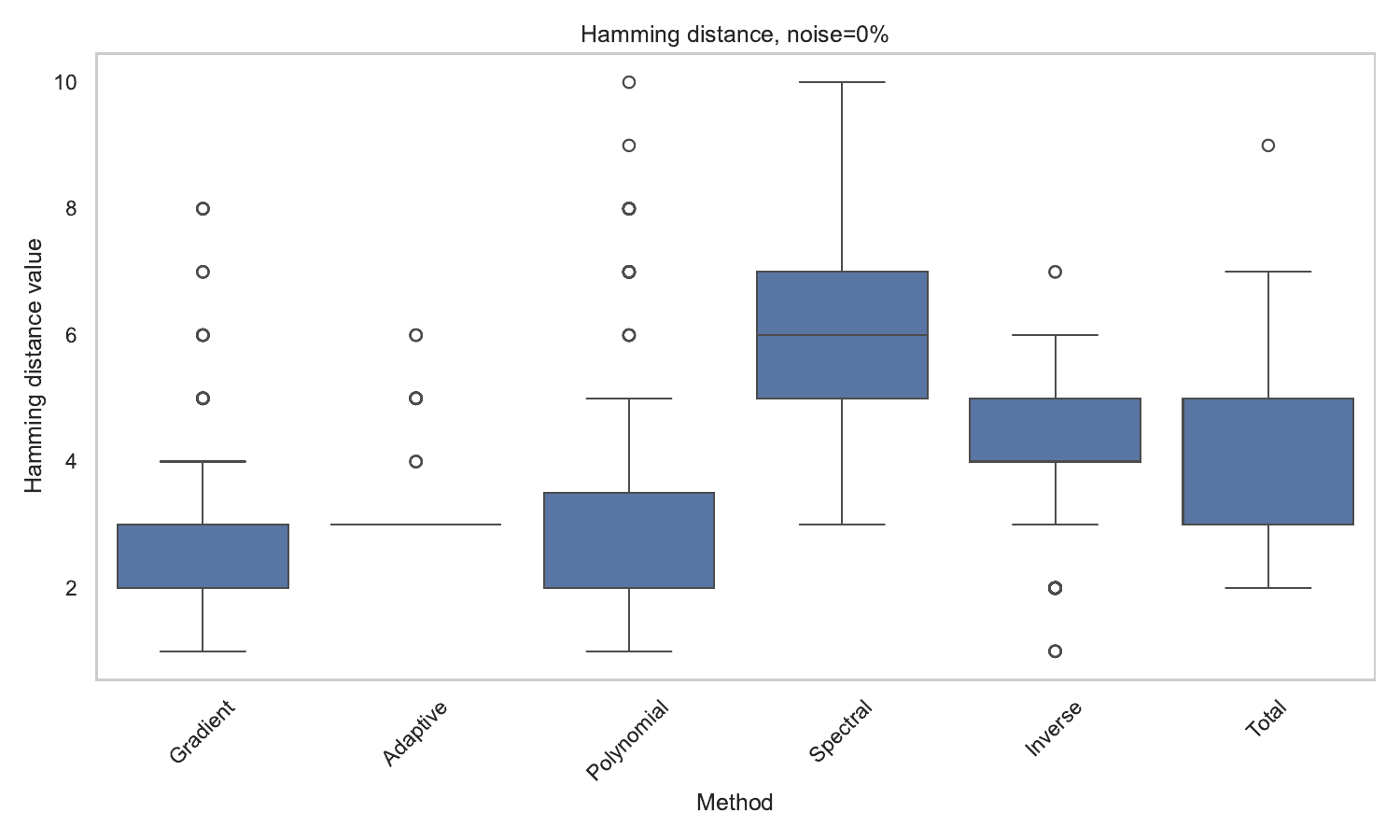}\
    \includegraphics[width=.7\textwidth]{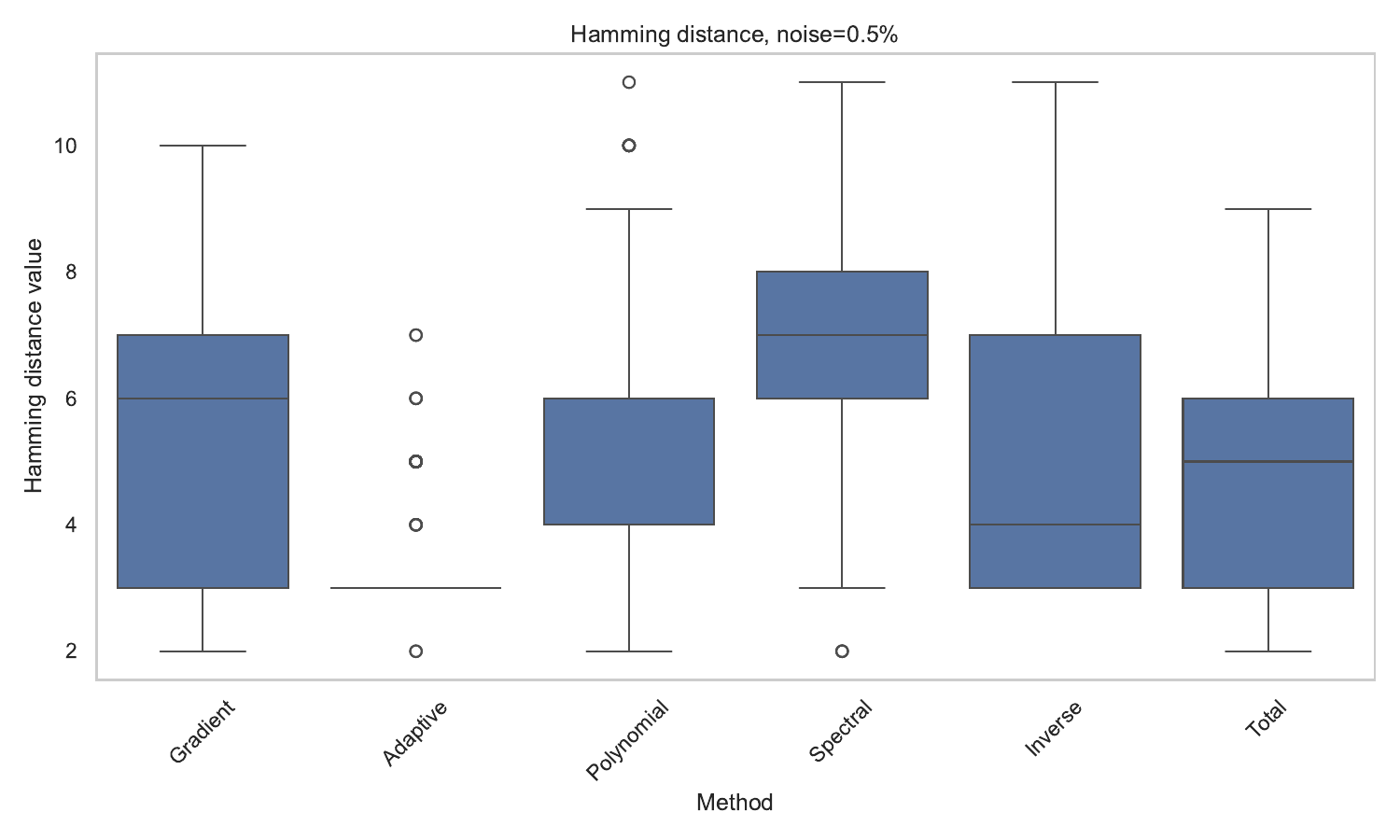}\
    \includegraphics[width=.7\textwidth]{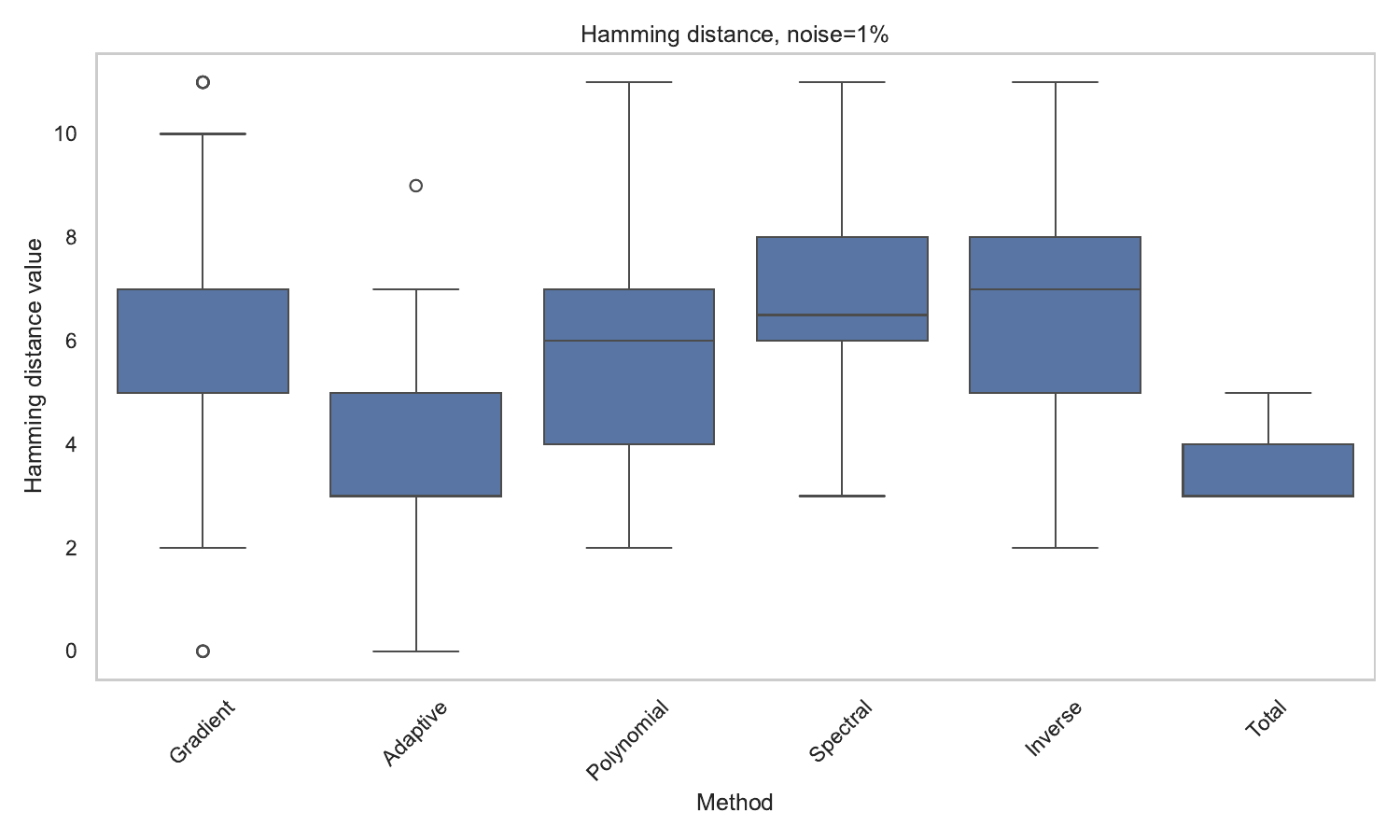}
    \caption{Distribution of coefficients values for different noise level}
    %label{fig:all}
\end{figure*}

\clearpage
\newpage

\section{Burgers equation coefficients and Structural Hamming  Distances}
\label{app:Burgers_res}

%Burgers coeffs and Hamming 
\begin{figure*}[ht!]
    \centering
    \includegraphics[width=.7\textwidth]{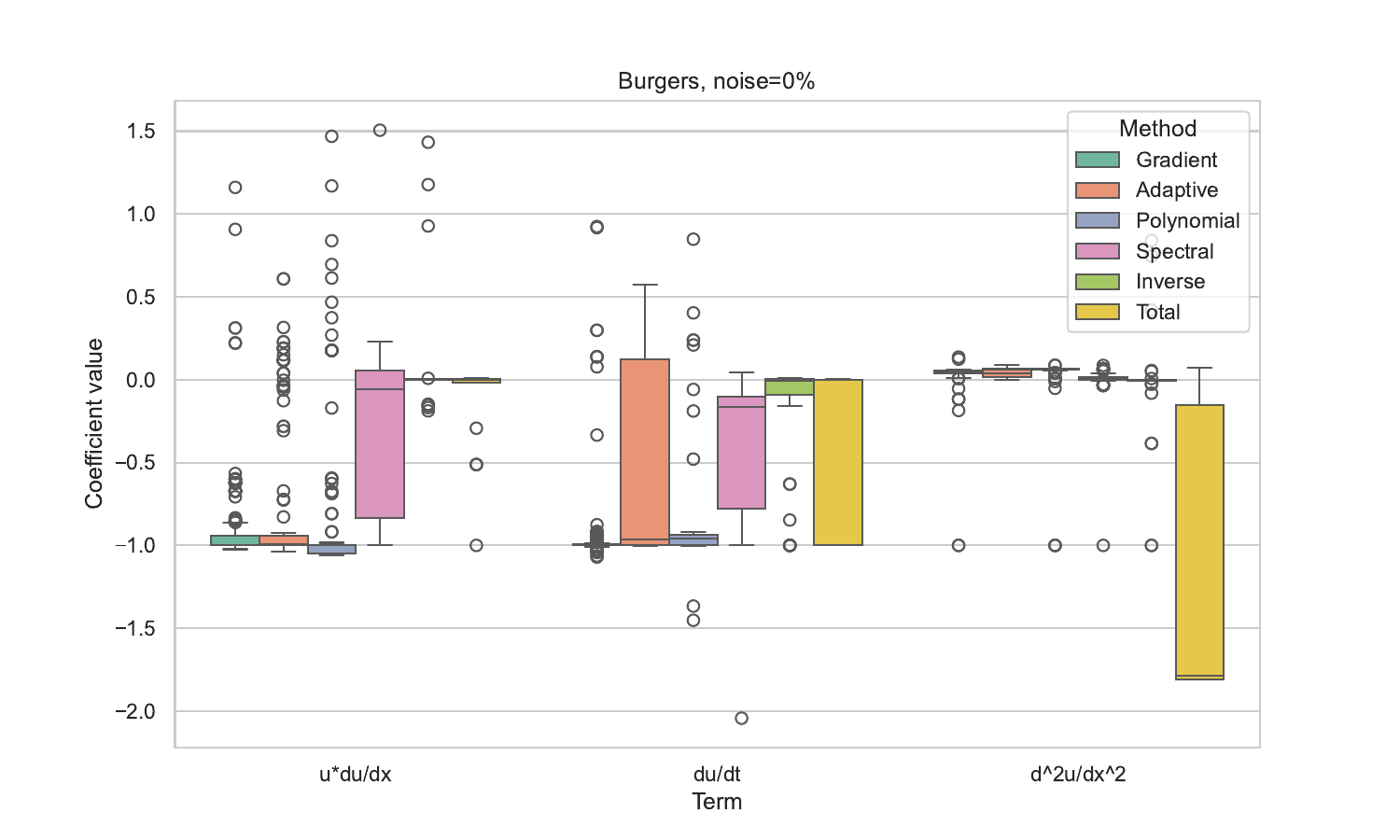}\
    \includegraphics[width=.7\textwidth]{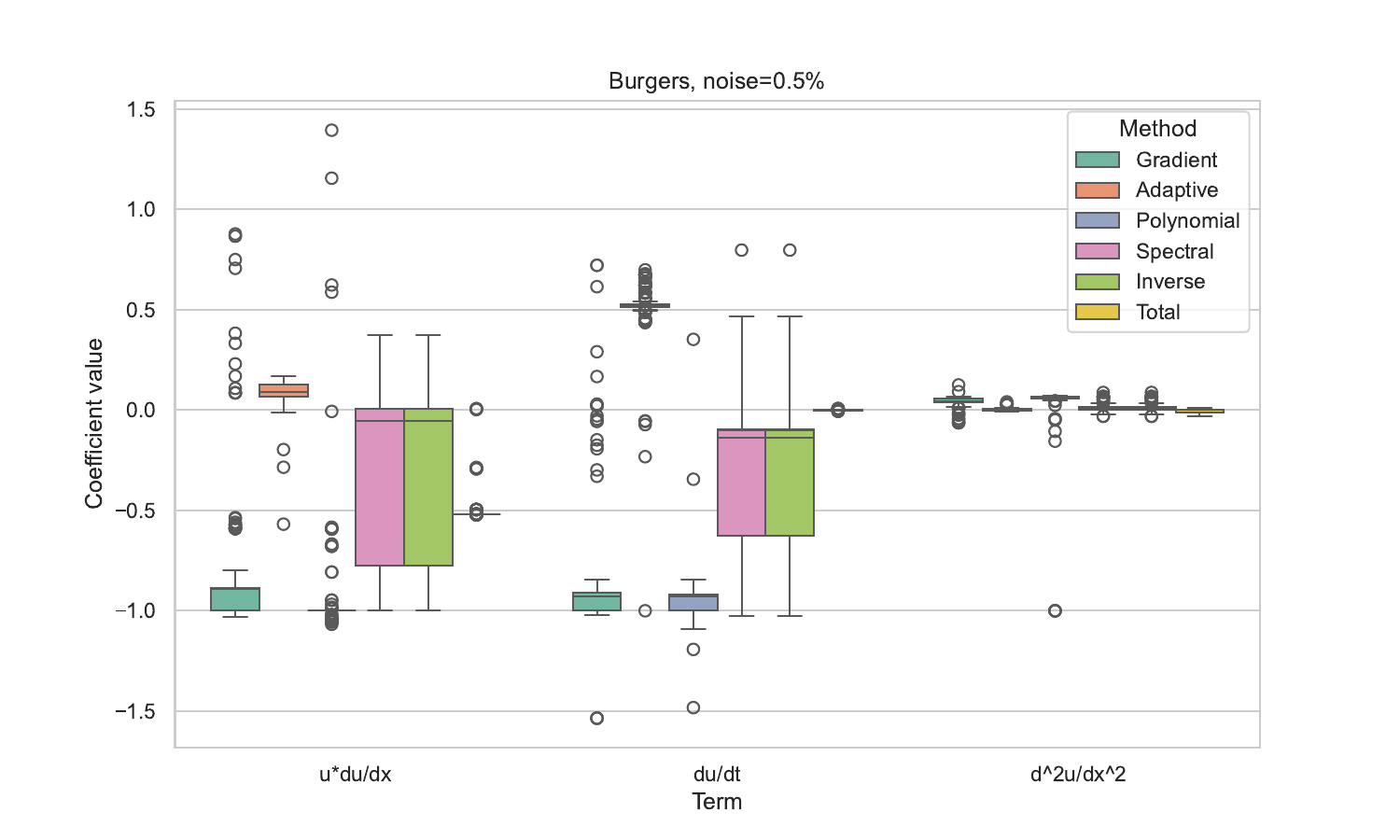}\
    \includegraphics[width=.7\textwidth]{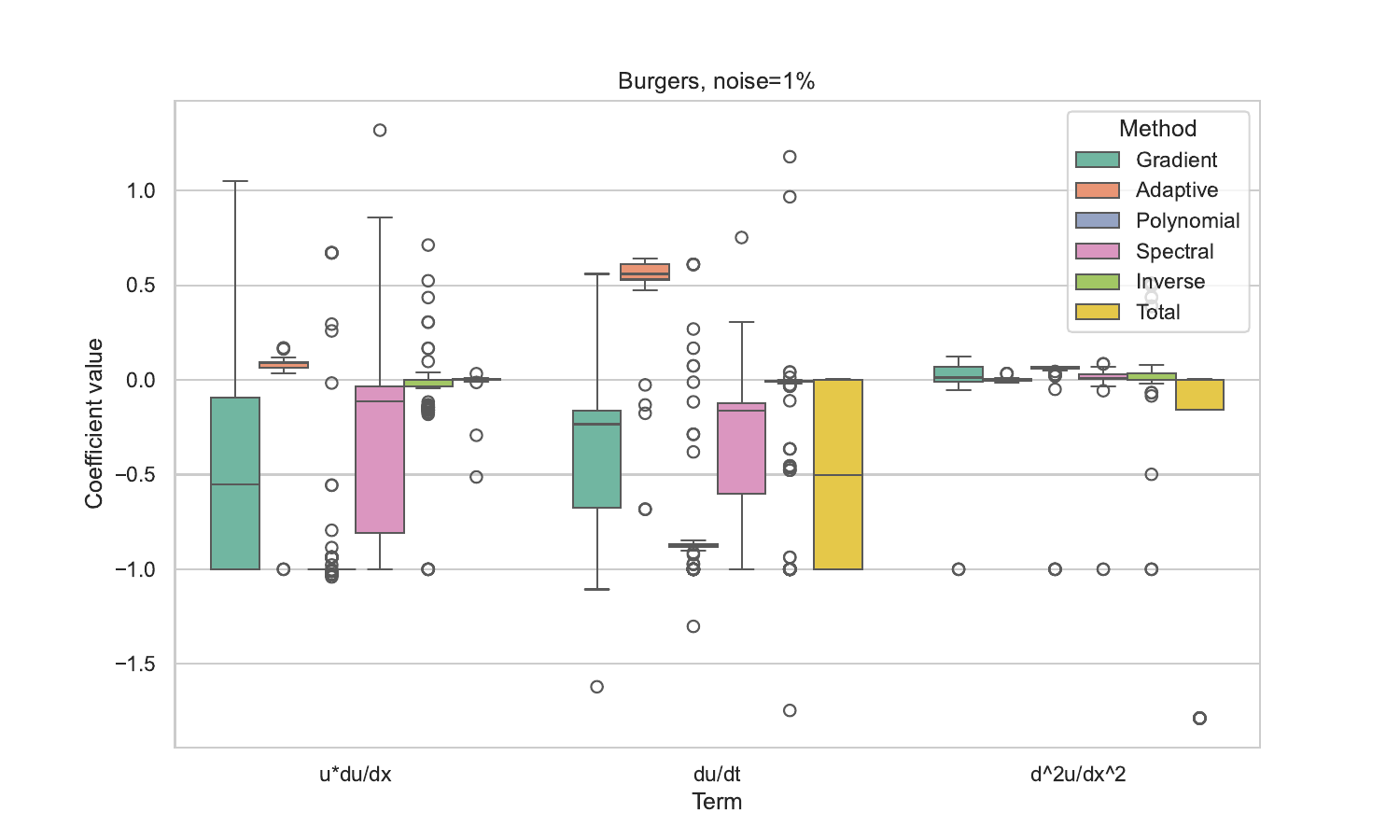}
    \caption{Distribution of coefficients values for different noise level}
    %label{fig:all}
\end{figure*}
\begin{figure*}[ht!]
    \centering
    \includegraphics[width=.7\textwidth]{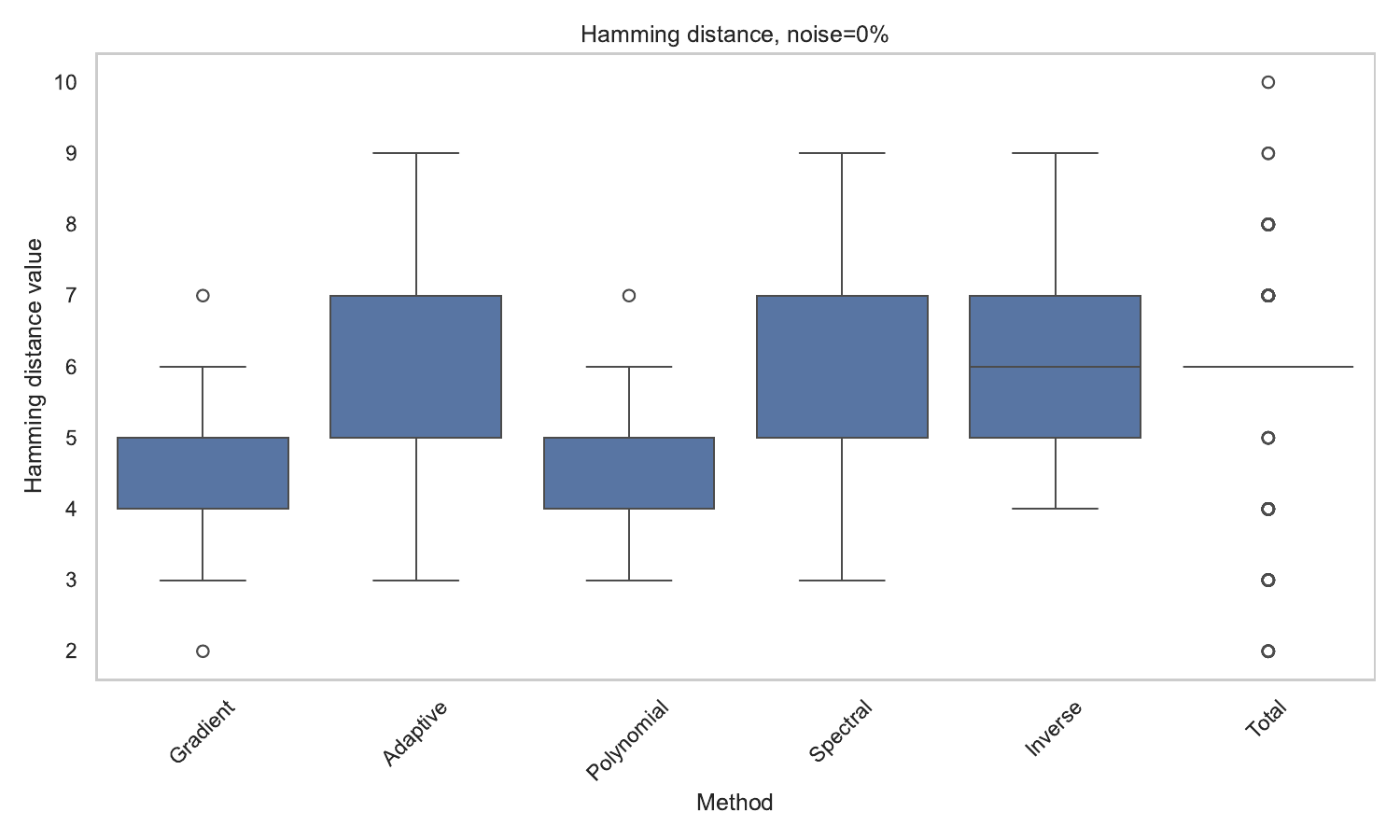}\
    \includegraphics[width=.7\textwidth]{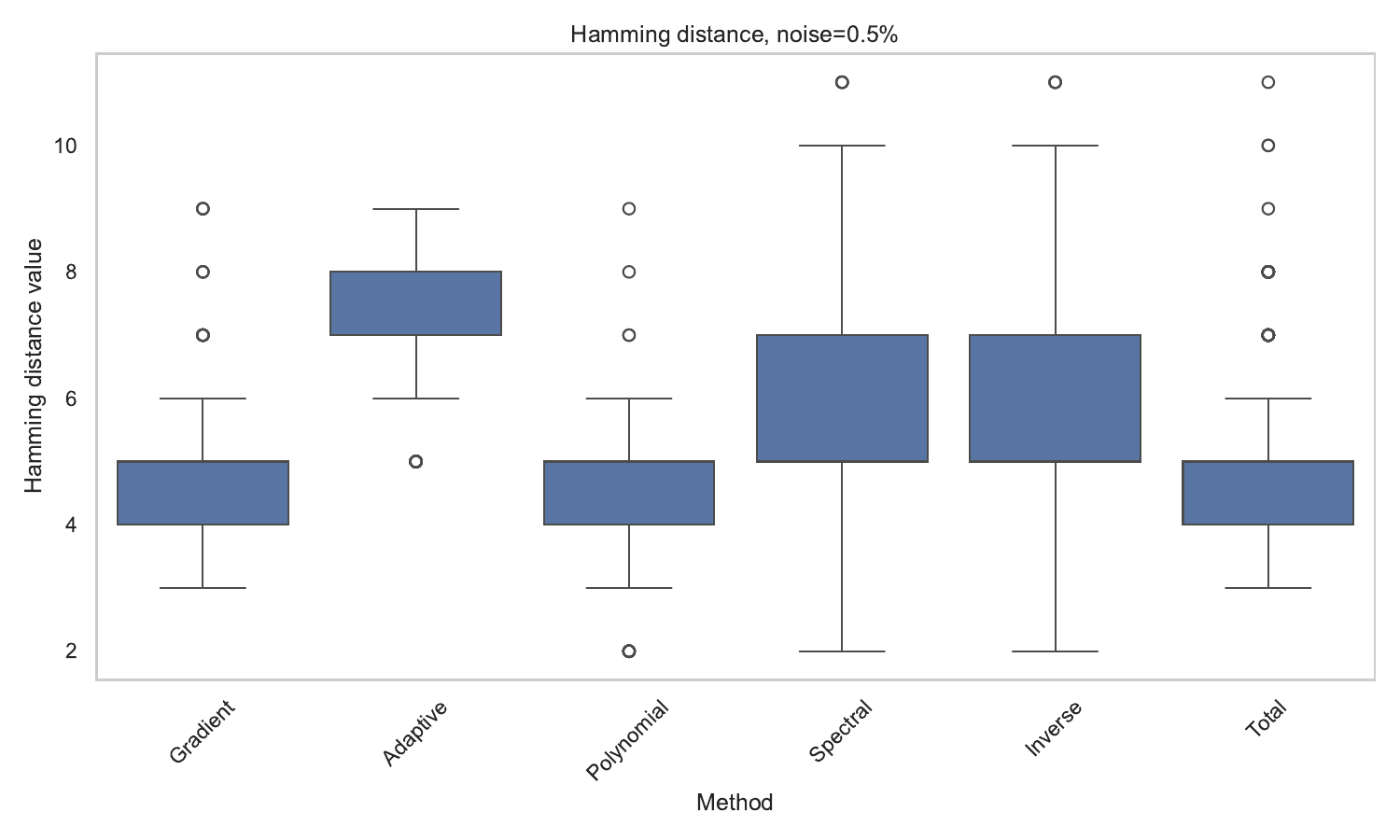}\
    \includegraphics[width=.7\textwidth]{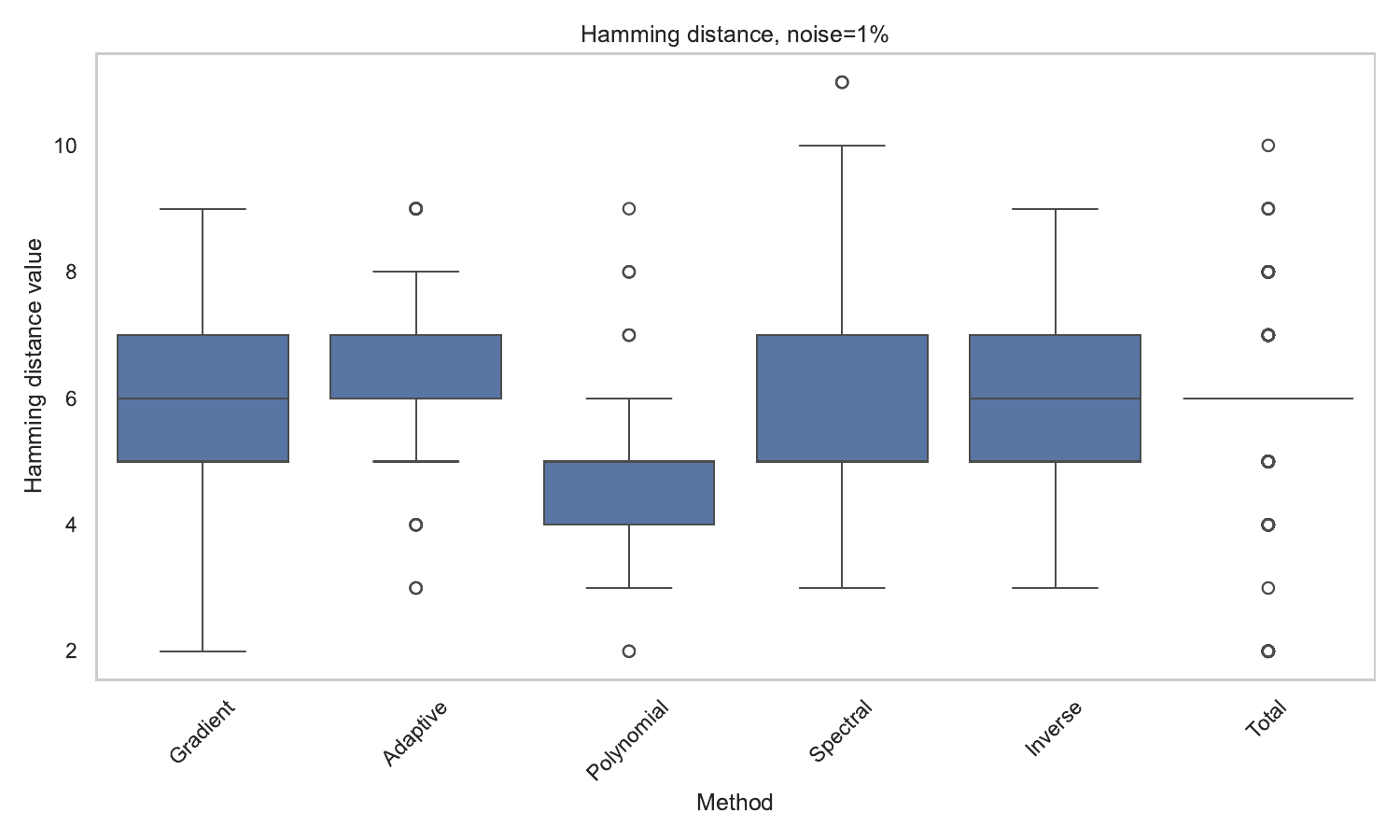}
    \caption{Distribution of coefficients values for different noise level}
    %label{fig:all}
\end{figure*}
\clearpage
\newpage
\section{Wave equation coefficients and Structural Hamming  Distances}
\label{app:Wave_res}

%Wave coeffs and Hamming
\begin{figure*}[ht!]
    \centering
    \includegraphics[width=.7\textwidth]{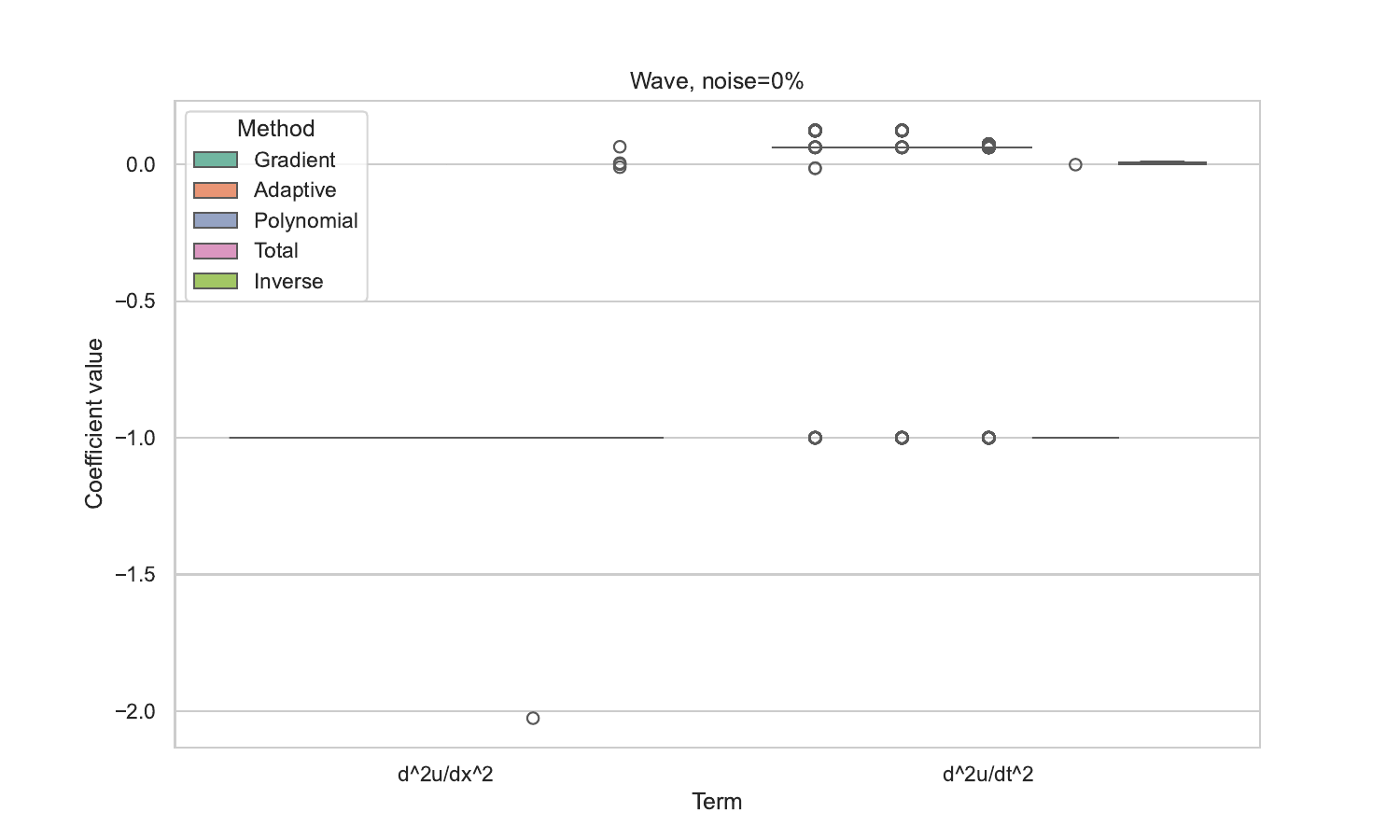}\
    \includegraphics[width=.7\textwidth]{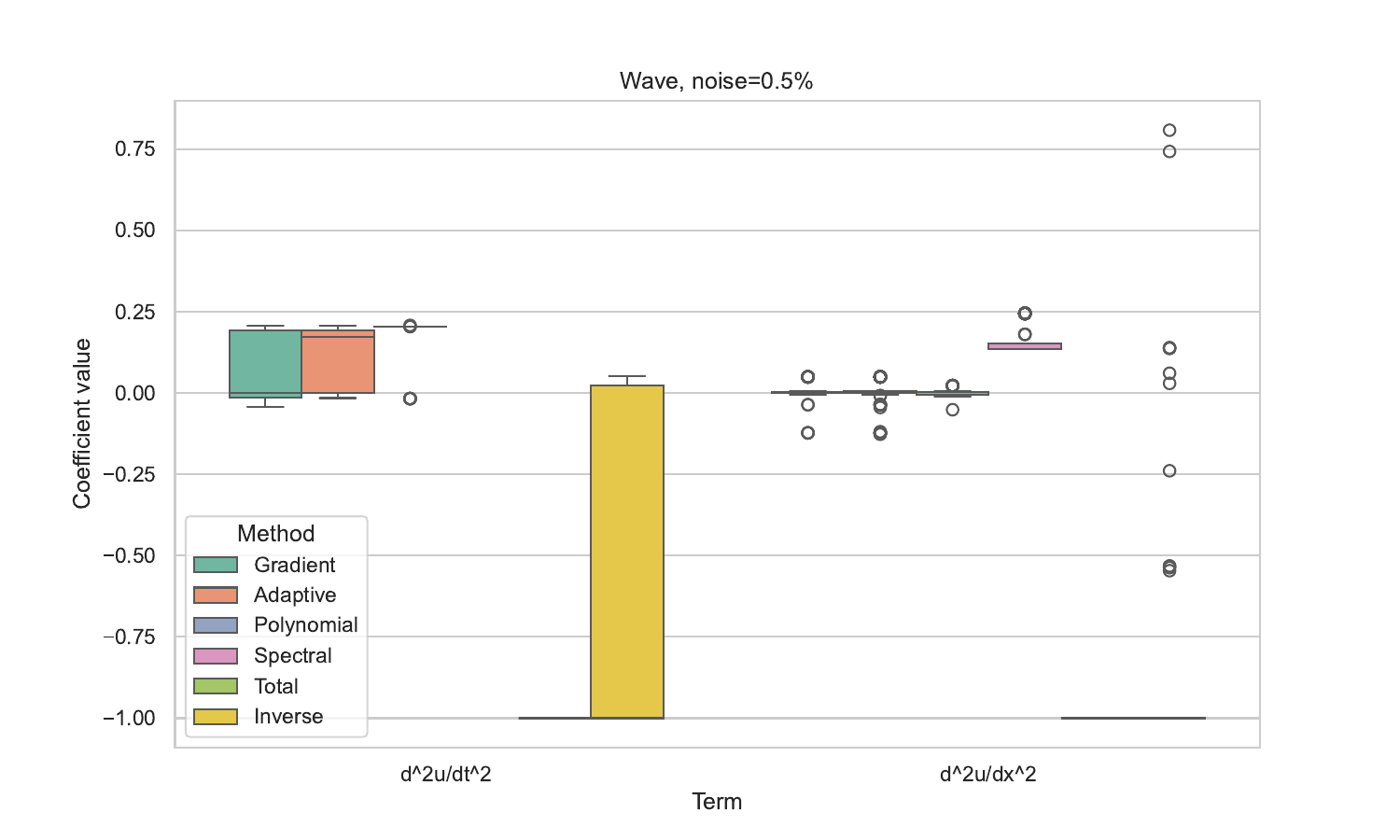}\
    \includegraphics[width=.7\textwidth]{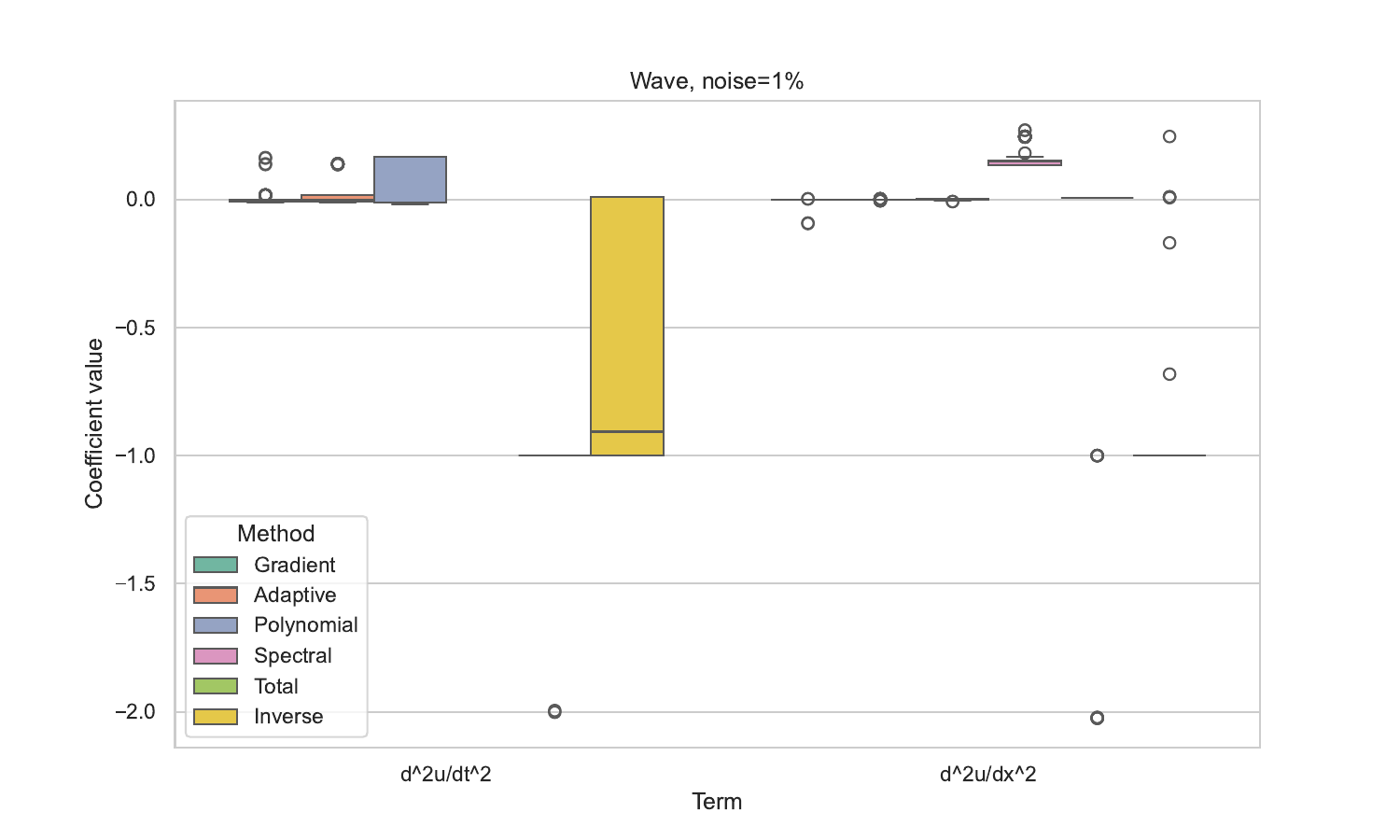}
    \caption{Distribution of coefficients values for different noise level}
    %label{fig:all}
\end{figure*}
\begin{figure*}[ht!]
    \centering
    \includegraphics[width=.7\textwidth]{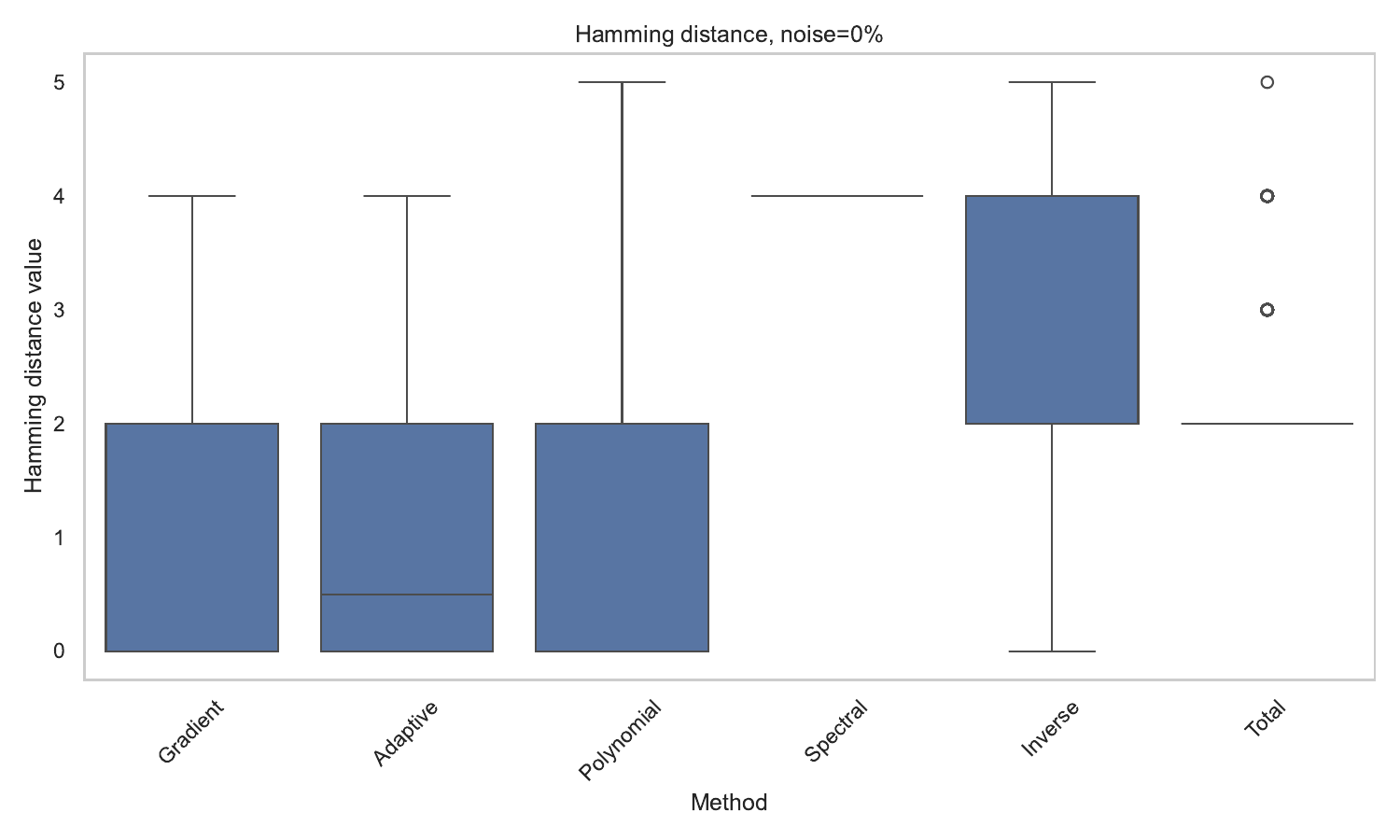}\
    \includegraphics[width=.7\textwidth]{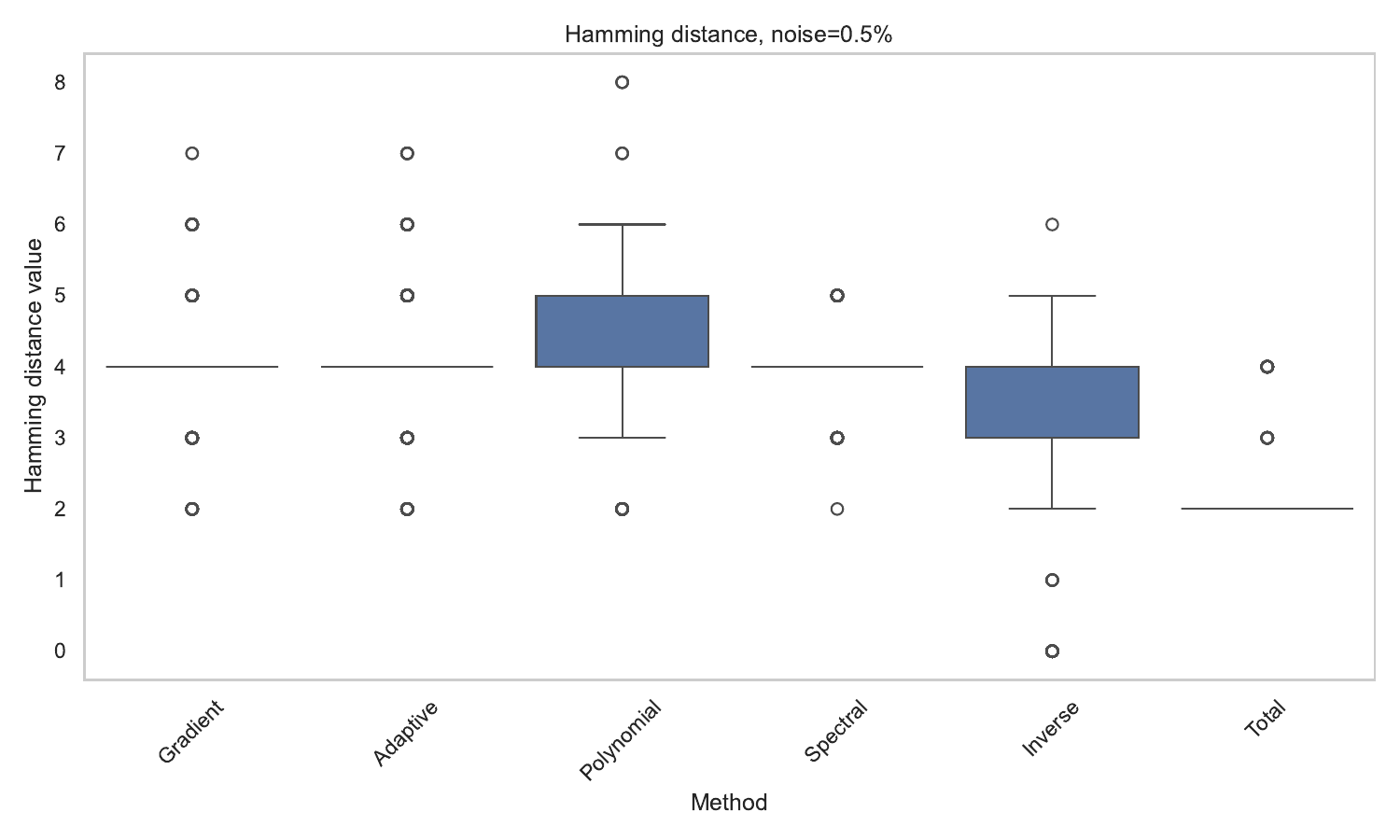}\
    \includegraphics[width=.7\textwidth]{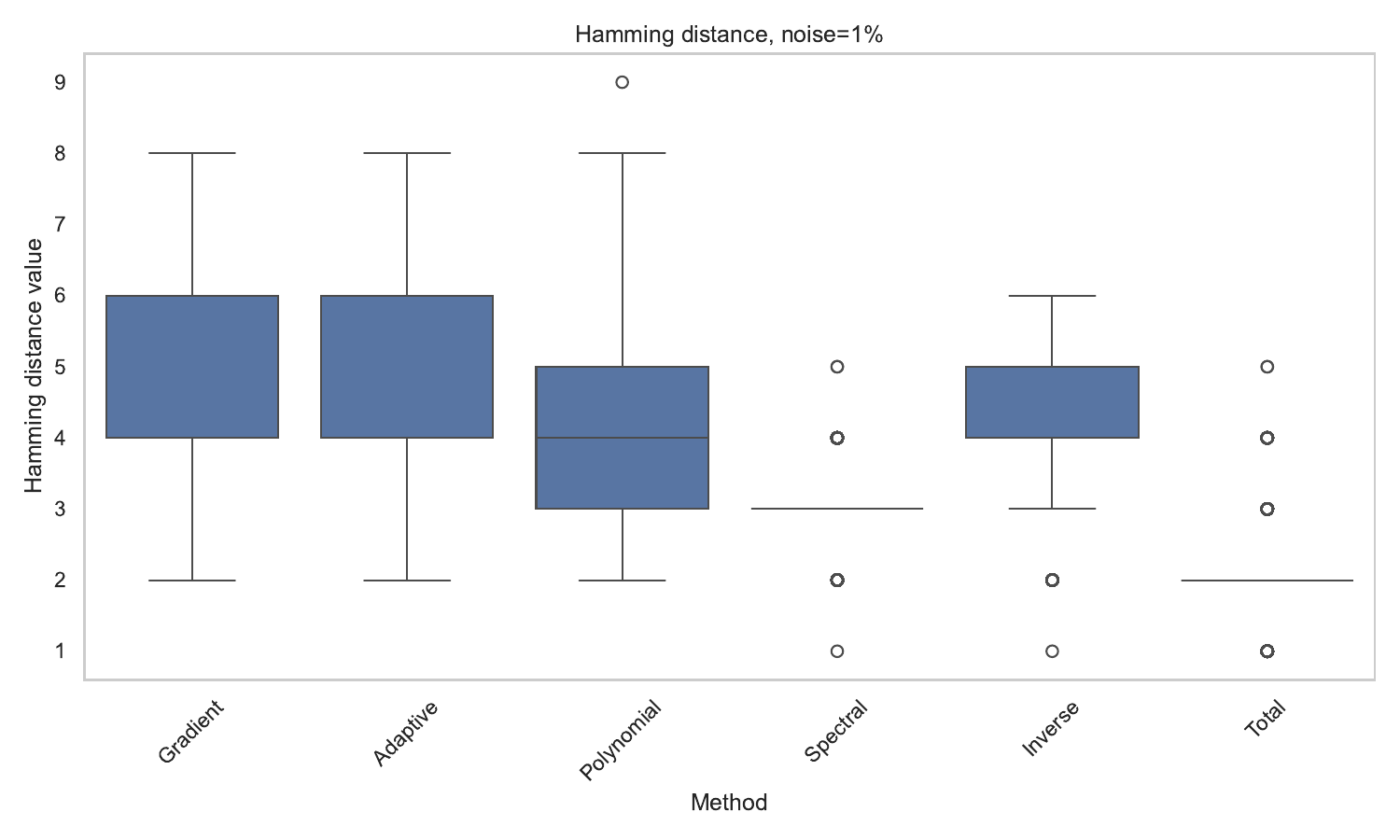}
    \caption{Distribution of coefficients values for different noise level}
    %label{fig:all}
\end{figure*}

\section{Laplace equation coefficients and Structural Hamming  Distances}
\label{app:Laplace_res}

%Laplace coeffs and Hamming
\begin{figure*}[ht!]
    \centering
    \includegraphics[width=.7\textwidth]{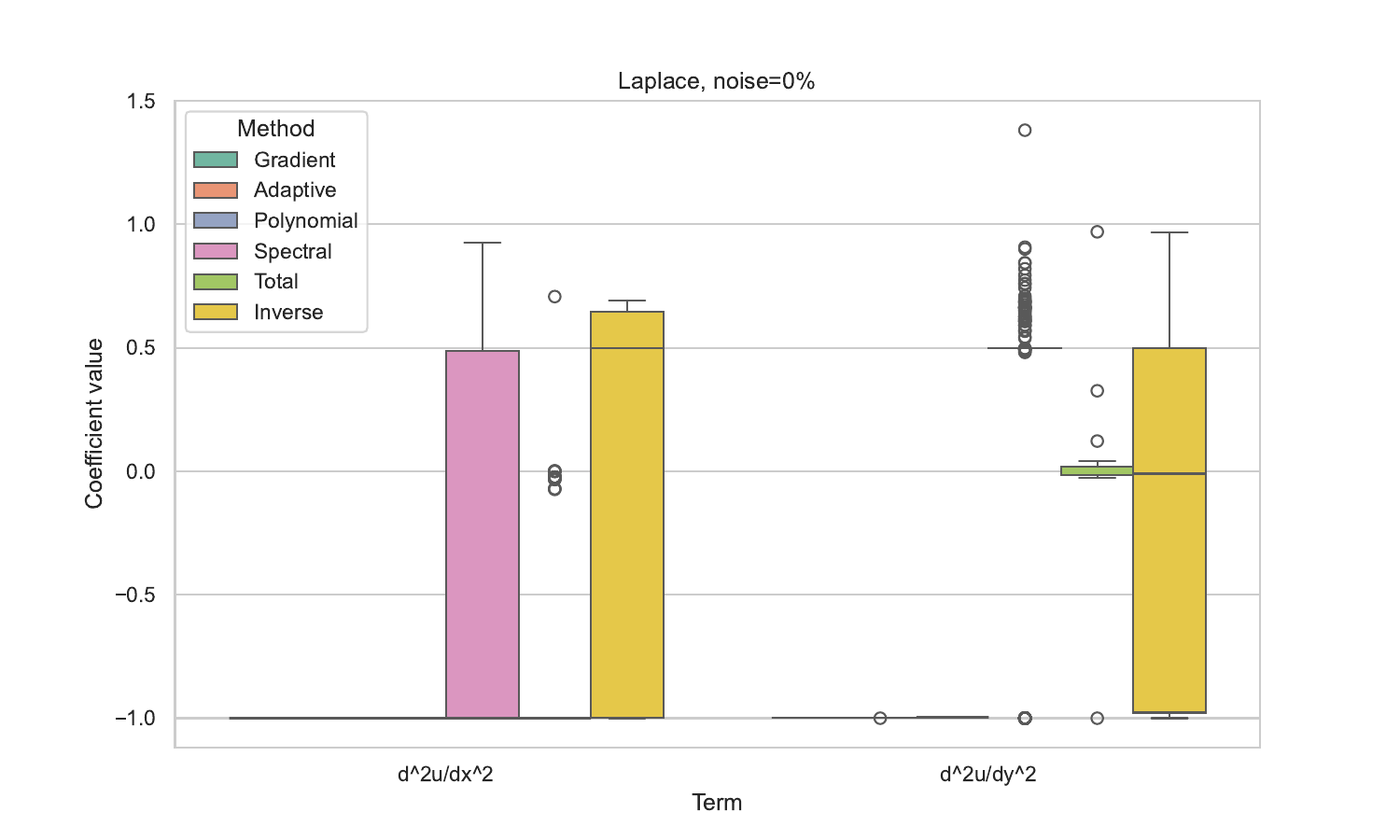}\
    \includegraphics[width=.7\textwidth]{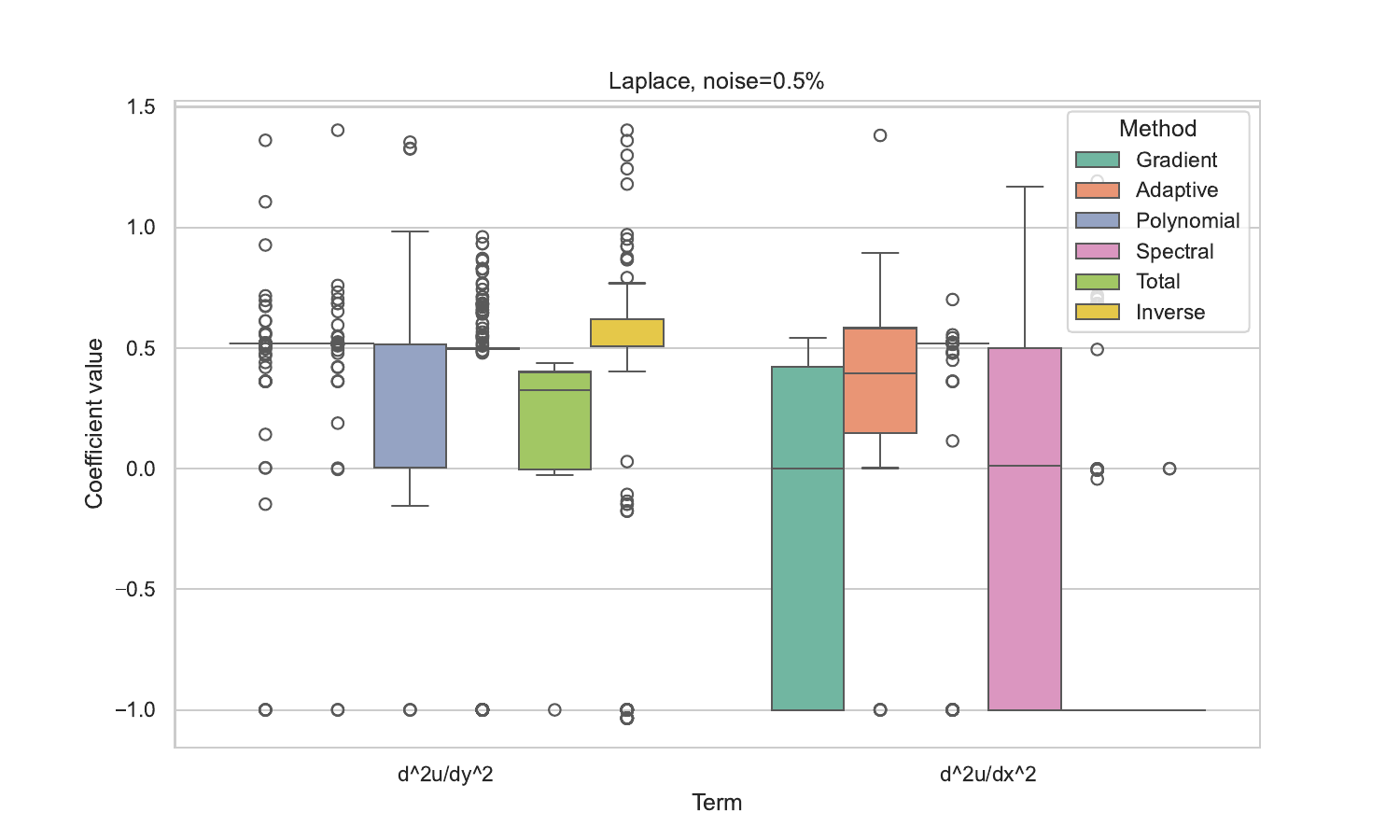}\
    \includegraphics[width=.7\textwidth]{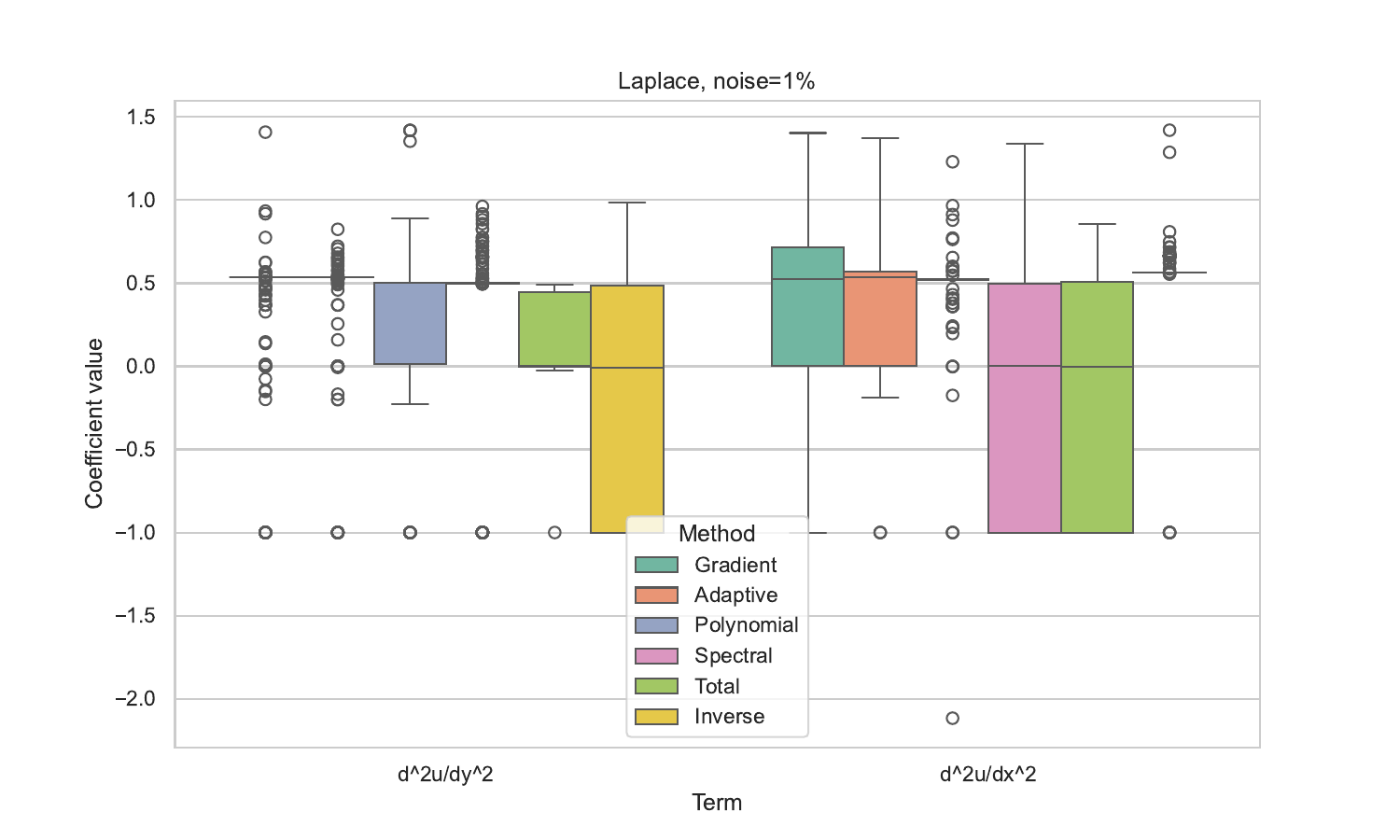}
    \caption{Distribution of coefficients values for different noise level}
    %label{fig:all}
\end{figure*}
\begin{figure*}[ht!]
    \centering
    \includegraphics[width=.7\textwidth]{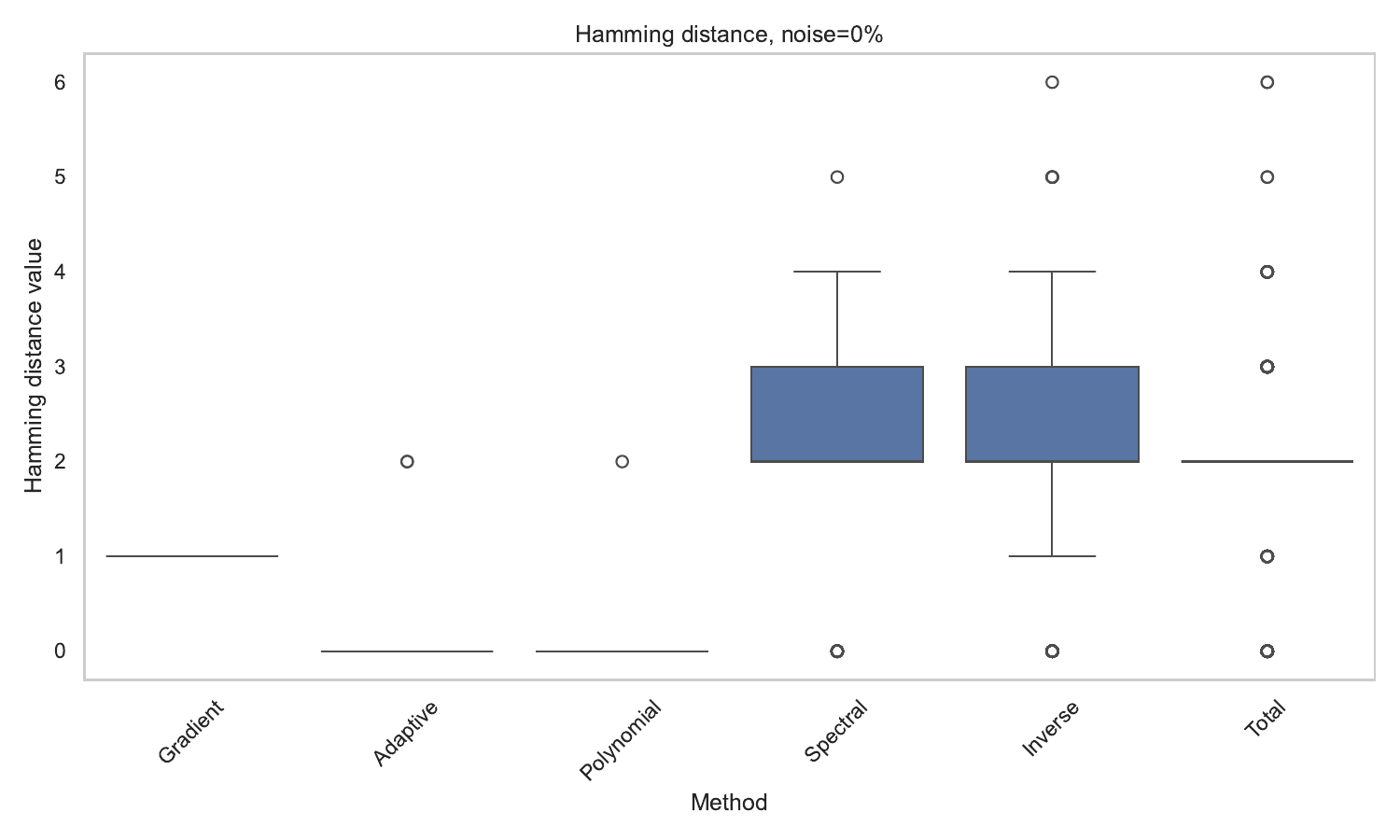}\
    \includegraphics[width=.7\textwidth]{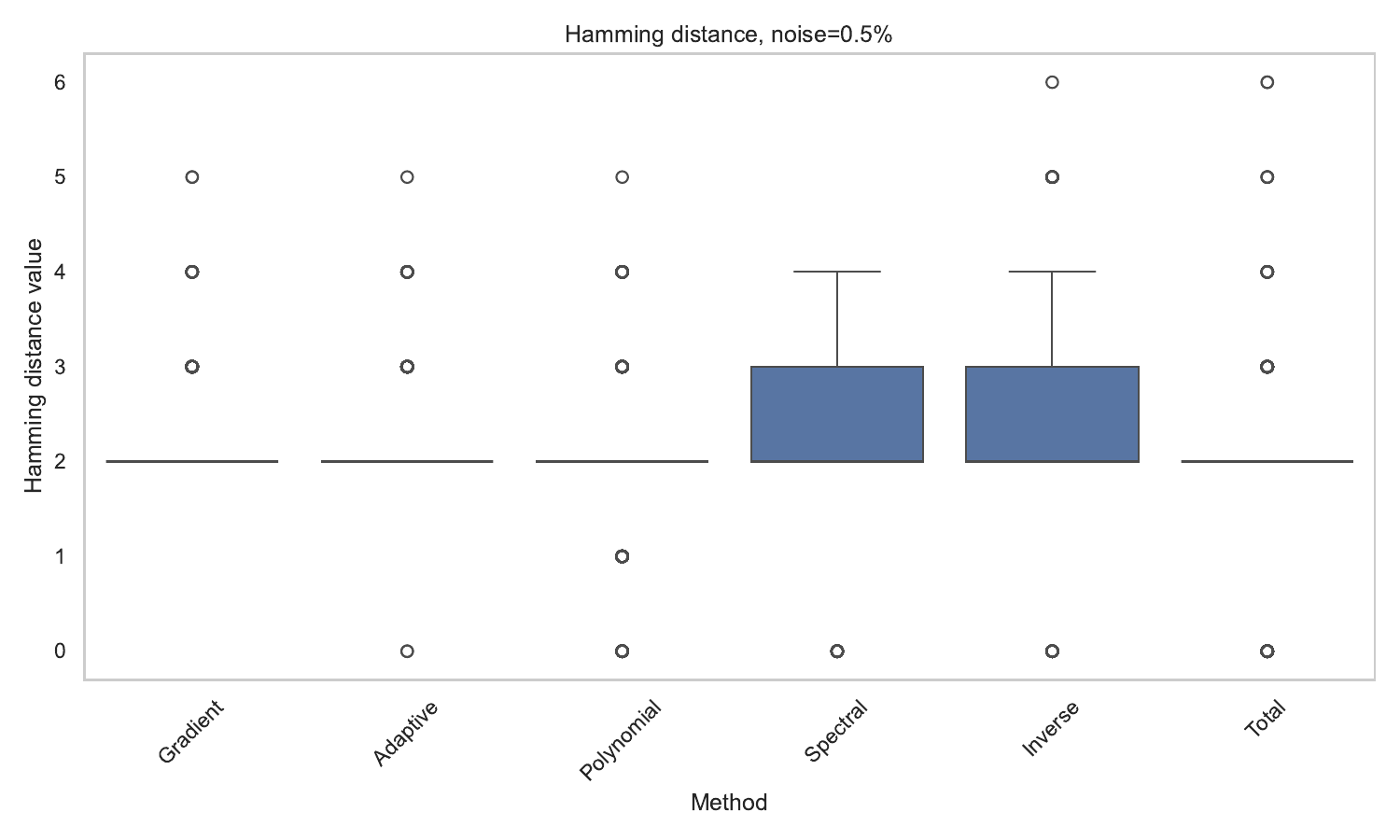}\
    \includegraphics[width=.7\textwidth]{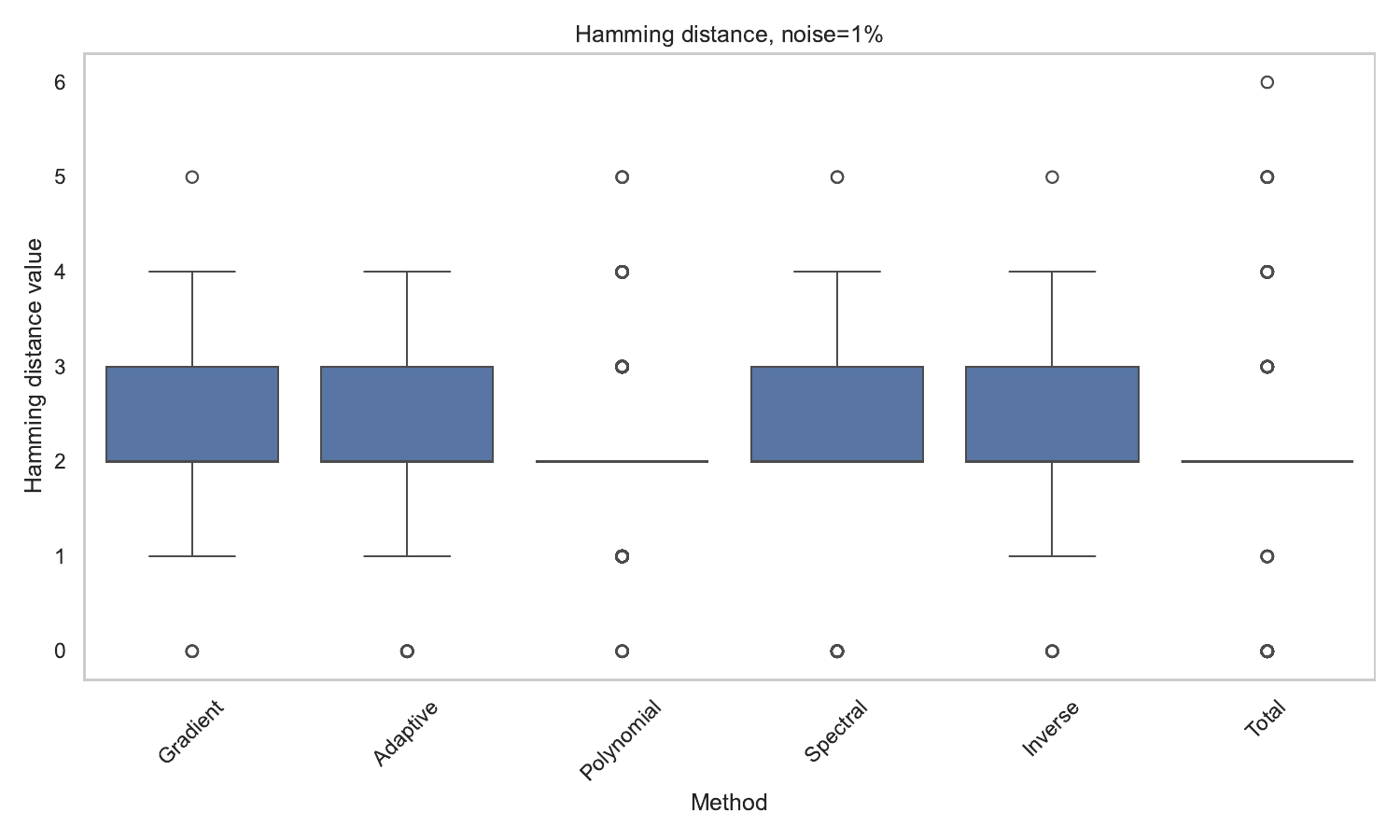}
    \caption{Distribution of coefficients values for different noise level}
    %label{fig:all}
\end{figure*}

\clearpage
\newpage

\section{Quasigeostrophic potential vorticity equation}
\label{app:pyqg}

\begin{figure*}[ht!]
    \includegraphics[width=1\textwidth]{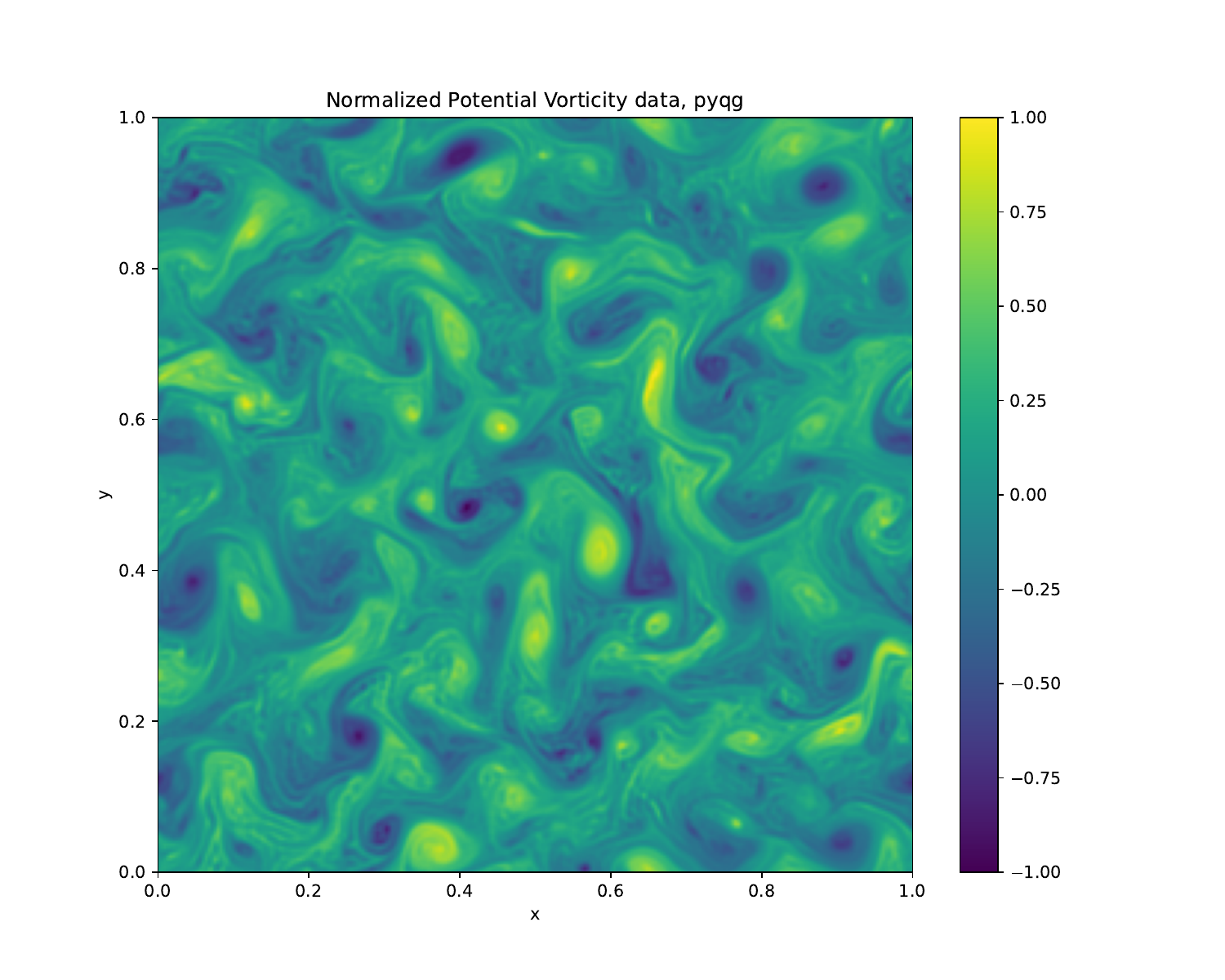}\hfill
    \caption{Normalized potential vorticity data, pyqg}
    %\label{fig:all}
\end{figure*}
\begin{figure*}[ht!]
    \centering
    \includegraphics[width=0.48\textwidth]{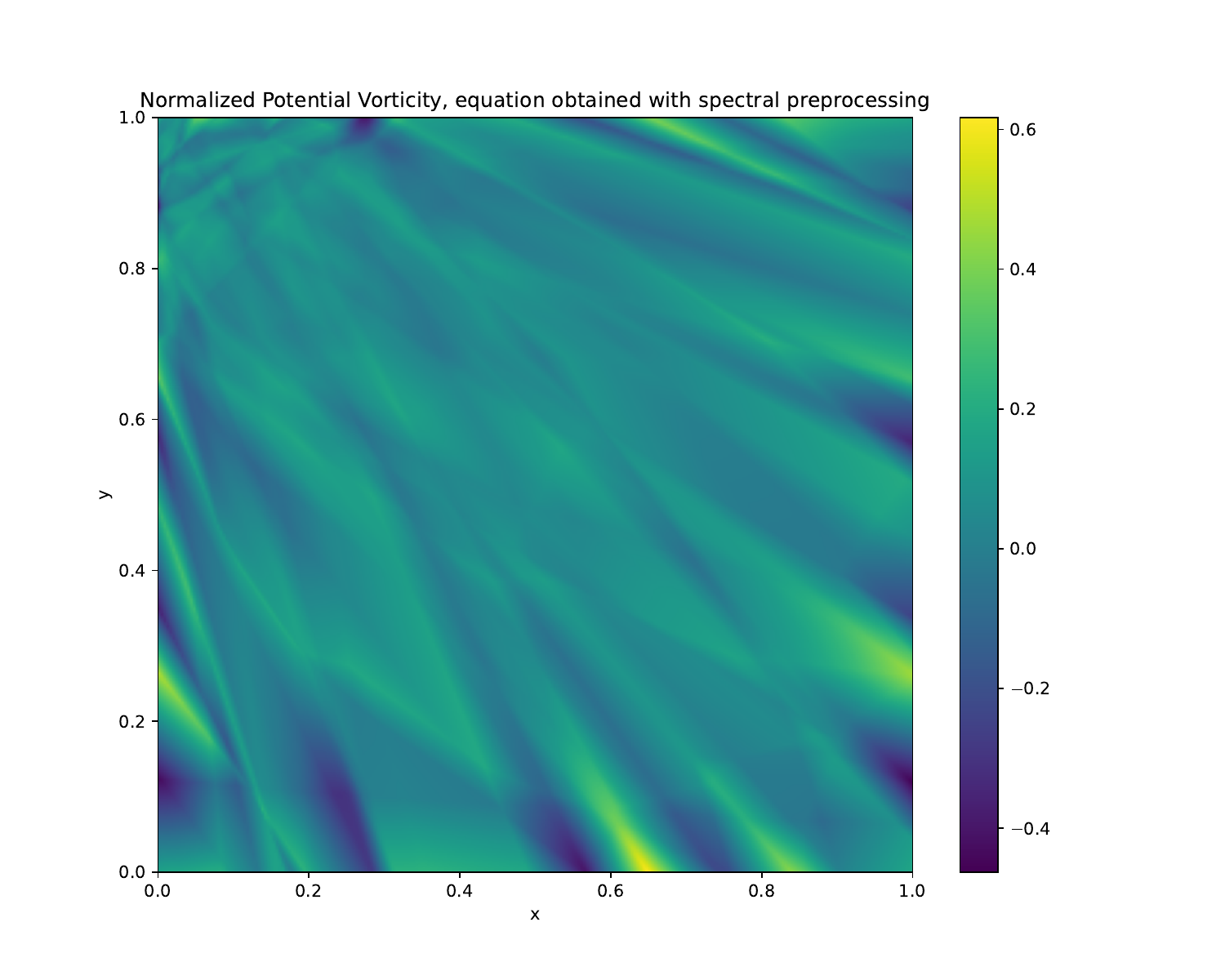}\hfill
    \includegraphics[width=0.48\textwidth]{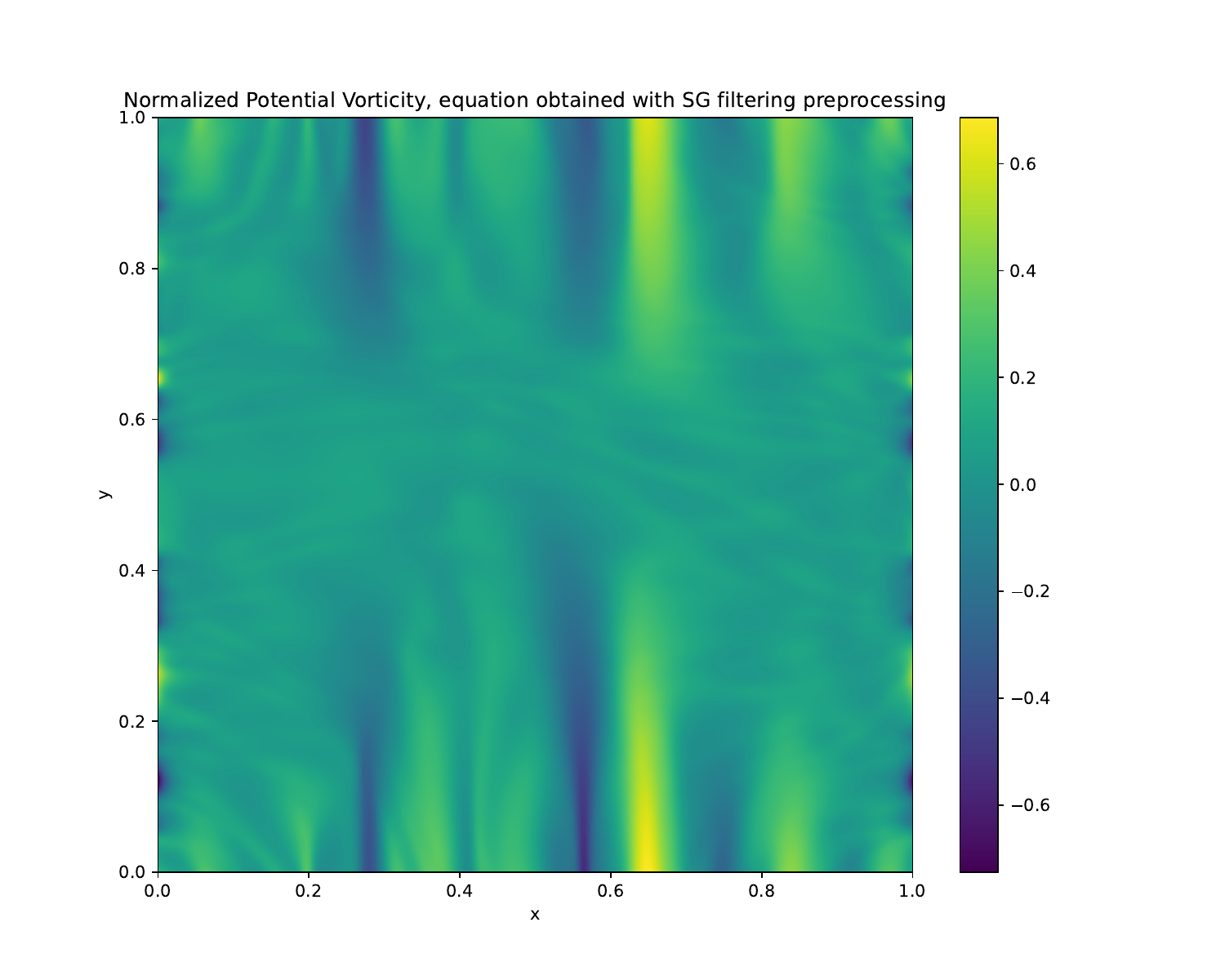}
    \caption{Normalized Potential Vorticity, equation obtained with spectral preprocessing (left) and SG filtering preprocessing (right)}
    \label{fig:all}
\end{figure*}
\begin{figure*}[ht!]
    \centering
    \includegraphics[width=0.48\textwidth]{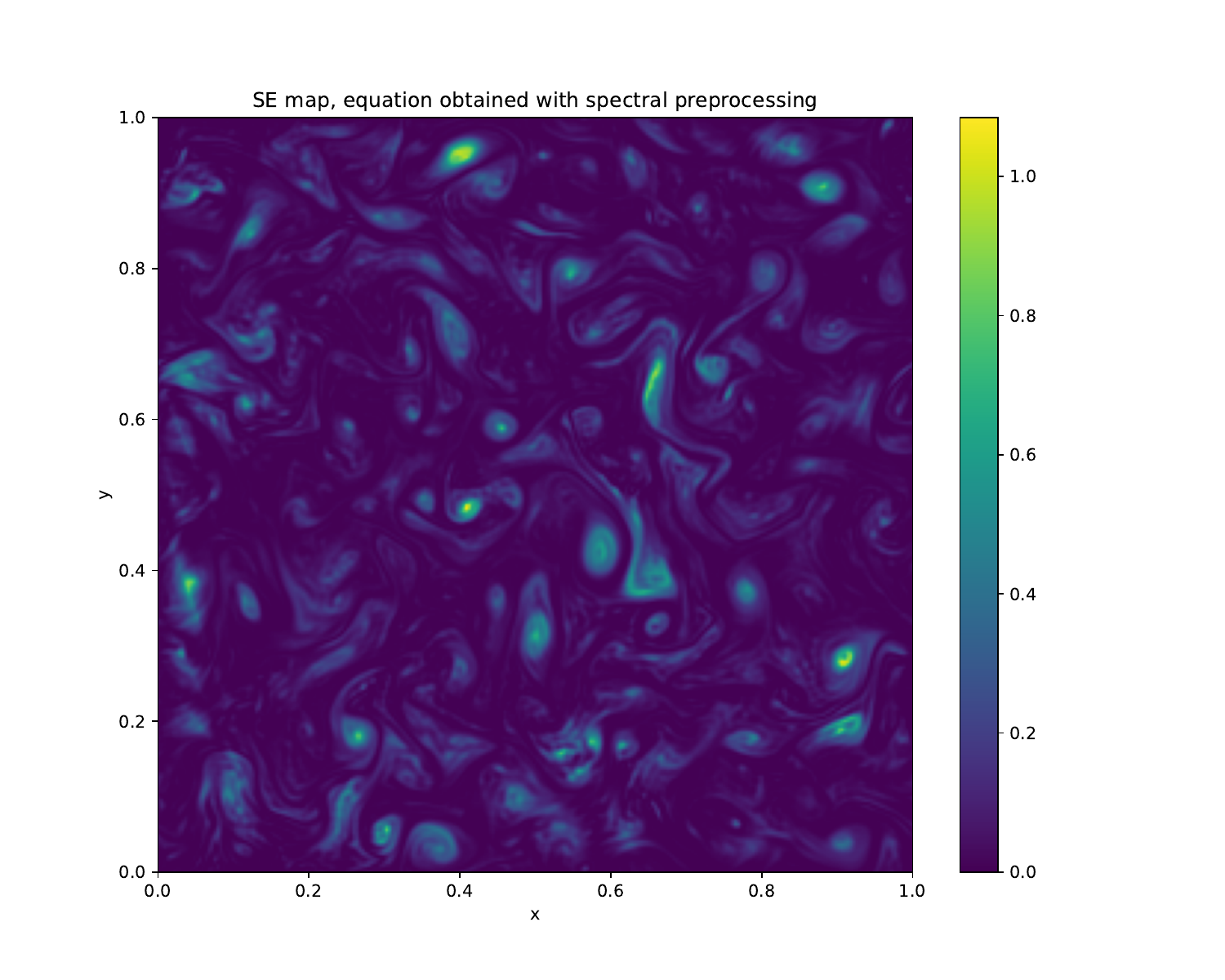}\hfill
    \includegraphics[width=0.48\textwidth]{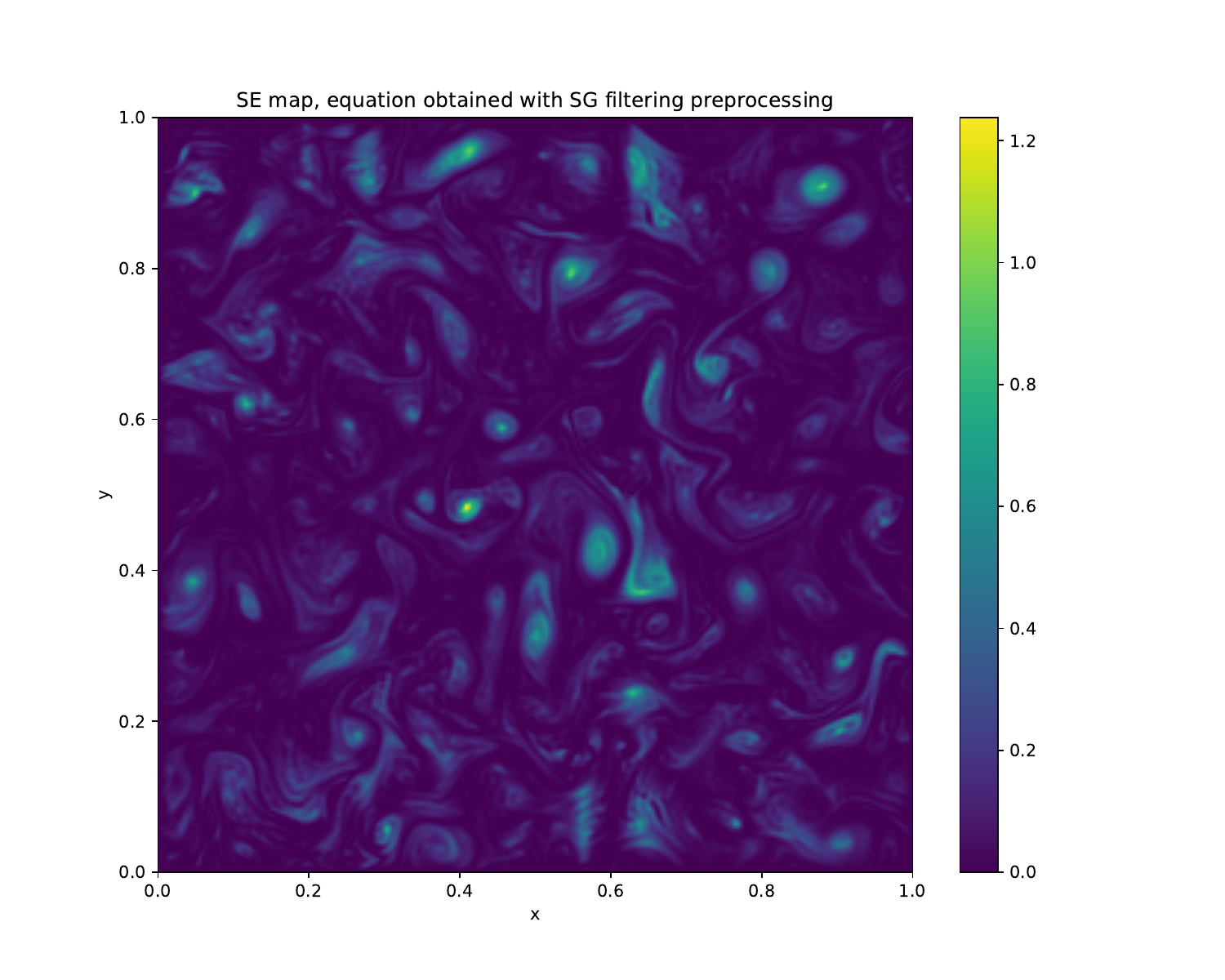}
    \caption{SE map, equation obtained with spectral preprocessing (left, MSE = 0.057) and SG filtering preprocessing(right, MSE = 0.065)}
    \label{fig:all}
\end{figure*}

\clearpage
\newpage

\section{Structural Hamming  Distances}
\label{app:SHD}

\begin{table}[ht!]
\caption{SHD for equations calculated with EPDE, noise = 0\%}
%\label{tab:my-table}
\centering
\begin{tabular}{|c|c|c|c|c|c|}
\hline
Methods/Equations & Burgers    & KdV        & Laplace    & ode        & Wave       \\ \hline
Gradient          & 4 ± 0.091  & 3 ± 0.1302 & 1          & 0 + 0.0491 & 1 ± 0.1206 \\ \hline
Adaptive          & 6 ± 0.1607  & 3 ± 0.0525 & 0 + 0.0159 & 0 + 0.0636 & 1 ± 0.1189 \\ \hline
Polynomial        & 4 ± 0.272  & 3 ± 0.1824 & 0 + 0.0112 & 0 + 0.0654 & 1 ± 0.119  \\ \hline
Spectral          & 6 ± 0.1993 & 6 ± 0.1698 & 3 ± 0.0999 & 3 ± 0.1855 & 4          \\ \hline
Inverse           & 6 ± 0.3554 & 4 ± 0.1267 & 2 ± 0.1432 & 4 ± 0.2077 & 3 ± 0.1314 \\ \hline
Total\_var             & 6 ± 0.1299 & 4 ± 0.1332 & 2 ± 0.1169 & 0 ± 0.0749 & 2 ± 0.0726 \\ \hline
\end{tabular}
\end{table}

\begin{table}[ht!]
\caption{SHD for equations calculated with EPDE, noise = 0.5\%}
%\label{tab:my-table}
\centering
\begin{tabular}{|c|c|c|c|c|c|}
\hline
Methods/Equations & Burgers    & KdV        & Laplace    & ode        & Wave       \\ \hline
Gradient          & 5 ± 0.1412 & 5 ± 0.2209 & 2 ± 0.0629 & 0 + 0.0535 & 4 ± 0.0979 \\ \hline
Adaptive          & 7 ± 0.107 & 3 ± 0.078 & 2 ± 0.0637 & 0 + 0.0458 & 4 ± 0.0671 \\ \hline
Polynomial        & 4 ± 0.1334 & 5 ± 0.1935 & 2 ± 0.1062 & 0 + 0.0734 & 4 ± 0.1237 \\ \hline
Spectral          & 6 ± 0.2107 & 7 ± 0.2087 & 3 ± 0.0992 & 3 ± 0.1971 & 4 ± 0.0631 \\ \hline
Inverse           & 6 ± 0.2107 & 5 ± 0.2632 & 3 ± 0.124  & 4 ± 0.1531 & 3 ± 0.1305 \\ \hline
Total\_var             & 5 ± 0.1413 & 5 ± 0.1394 & 2 ± 0.1084 & 0 + 0.0783 & 2 ± 0.0571 \\ \hline
\end{tabular}
\end{table}

\begin{table}[ht!]
\caption{SHD for equations calculated with EPDE, noise = 1\%}
%\label{tab:my-table}
\centering
\begin{tabular}{|c|c|c|c|c|c|}
\hline
Methods/Equations & Burgers    & KdV        & Laplace    & ode        & Wave       \\ \hline
Gradient          & 6 ± 0.2134 & 6 ± 0.2824 & 2 ± 0.0843 & 0 + 0.0424 & 5 ± 0.1186 \\ \hline
Adaptive          & 7 ± 0.1292  & 4 ± 0.1484 & 2 ± 0.0815 & 0 + 0.0433 & 5 ± 0.0833 \\ \hline
Polynomial        & 4 ± 0.1248 & 6 ± 0.247  & 2 ± 0.102  & 0 + 0.0627 & 4 ± 0.1252 \\ \hline
Spectral          & 6 ± 0.2096 & 7 ± 0.1827 & 3 ± 0.0974 & 3 ± 0.1904 & 3 ± 0.0687 \\ \hline
Inverse           & 6 ± 0.2057 & 7 ± 0.2509 & 2 ± 0.0883 & 4 ± 0.1231 & 4 ± 0.1004 \\ \hline
Total\_var             & 6 ± 0.1283 & 5 ± 0.133  & 2 ± 0.1064 & 0 ± 0.0792 & 2 ± 0.0847 \\ \hline
\end{tabular}
\end{table}

\begin{table}[ht!]
\caption{SHD for equations calculated with SINDy, noise = 0\%}
%\label{tab:my-table}
\centering
\begin{tabular}{|c|c|c|c|c|c|}
\hline
Methods/Equations & Burgers & KdV & Laplace & ode & Wave \\ \hline
Gradient          & 0       & 1   & 0       & 0   & 0    \\ \hline
Adaptive          & 0       & 3   & 0       & 0   & 1    \\ \hline
Polynomial        & 1       & 3   & 0       & 0   & 1    \\ \hline
Spectral          & 1       & 2   & 1       & 1   & 1    \\ \hline
Inverse           & 1       & 1   & 1       & 0   & 1    \\ \hline
Total\_var             & 5       & 4   & 2       & 0   & 2    \\ \hline
\end{tabular}
\end{table}

\begin{table}[ht!]
\caption{SHD for equations calculated with SINDy, noise = 0.5\%}
%\label{tab:my-table}
\centering
\begin{tabular}{|c|c|c|c|c|c|}
\hline
Methods/Equations & Burgers & KdV & Laplace & ode & Wave \\ \hline
Gradient          & 0       & 2   & 2       & 0   & 2    \\ \hline
Adaptive          & 0       & 2   & 2       & 0   & 2    \\ \hline
Polynomial        & 1       & 2   & 2       & 0   & 2    \\ \hline
Spectral          & 1       & 2   & 2       & 1   & 1    \\ \hline
Inverse           & 3       & 3   & 4       & 0   & 1    \\ \hline
Total\_var             & 5       & 4   & 2       & 0   & 2    \\ \hline
\end{tabular}
\end{table}

\begin{table}[ht!]
\caption{SHD for equations calculated with SINDy, noise = 1\%}
%\label{tab:my-table}
\centering
\begin{tabular}{|c|c|c|c|c|c|}
\hline
Methods/Equations & Burgers & KdV & Laplace & ode & Wave \\ \hline
Gradient          & 2       & 2   & 3       & 0   & 2    \\ \hline
Adaptive          & 2       & 2   & 1       & 0   & 2    \\ \hline
Polynomial        & 0       & 5   & 4       & 0   & 2    \\ \hline
Spectral          & 1       & 2   & 2       & 1   & 1    \\ \hline
Inverse           & 1       & 4   & 4       & 0   & 2    \\ \hline
Total\_var             & 3       & 4   & 2       & 0   & 2    \\ \hline
\end{tabular}
\end{table}

\clearpage
\newpage

\section{Differentiation errors}
\label{app:diff_error}

%KdV diff errors
\begin{figure*}[ht!]
    \begin{multicols}{3}
          \includegraphics[width=.3\textwidth]{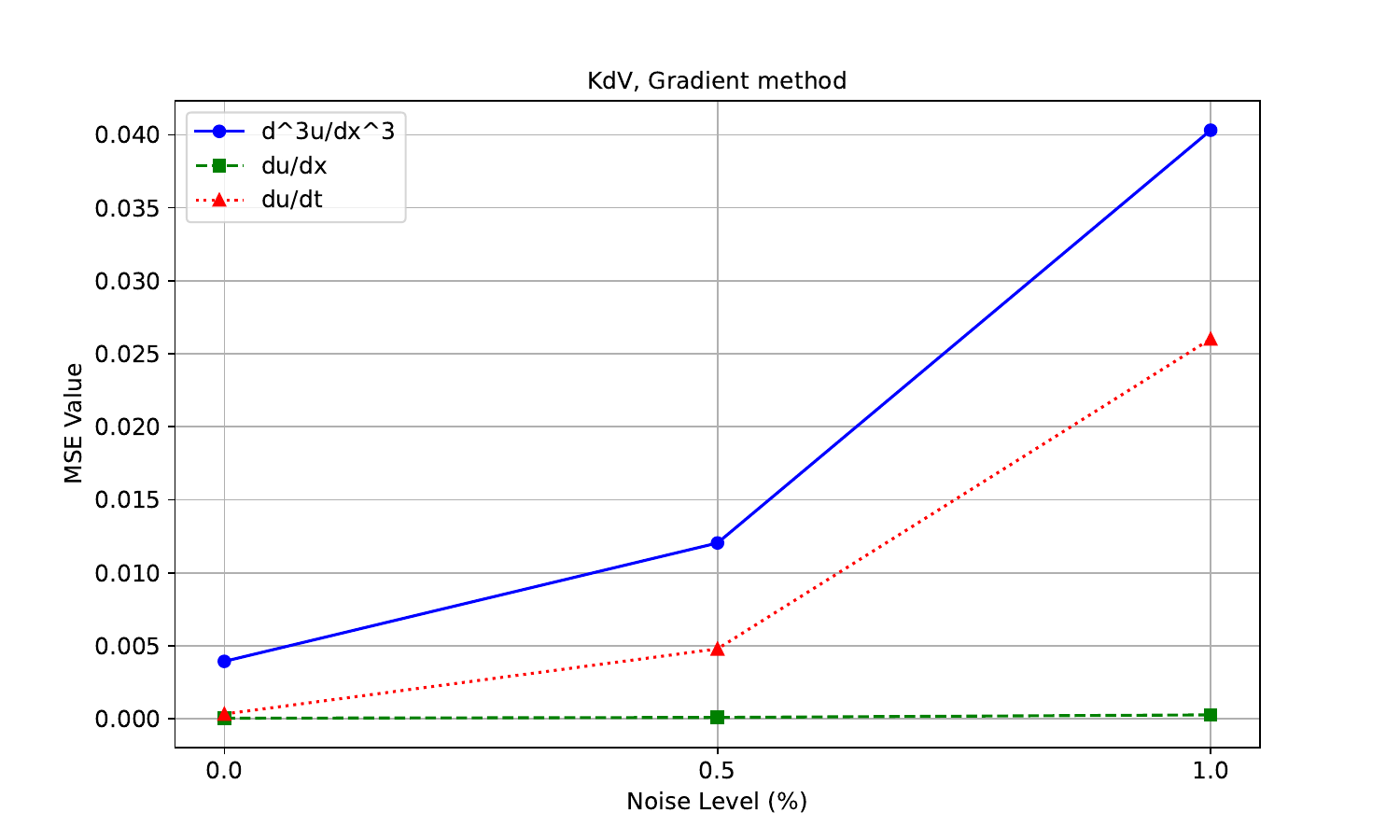}\hfill
          \includegraphics[width=.3\textwidth]{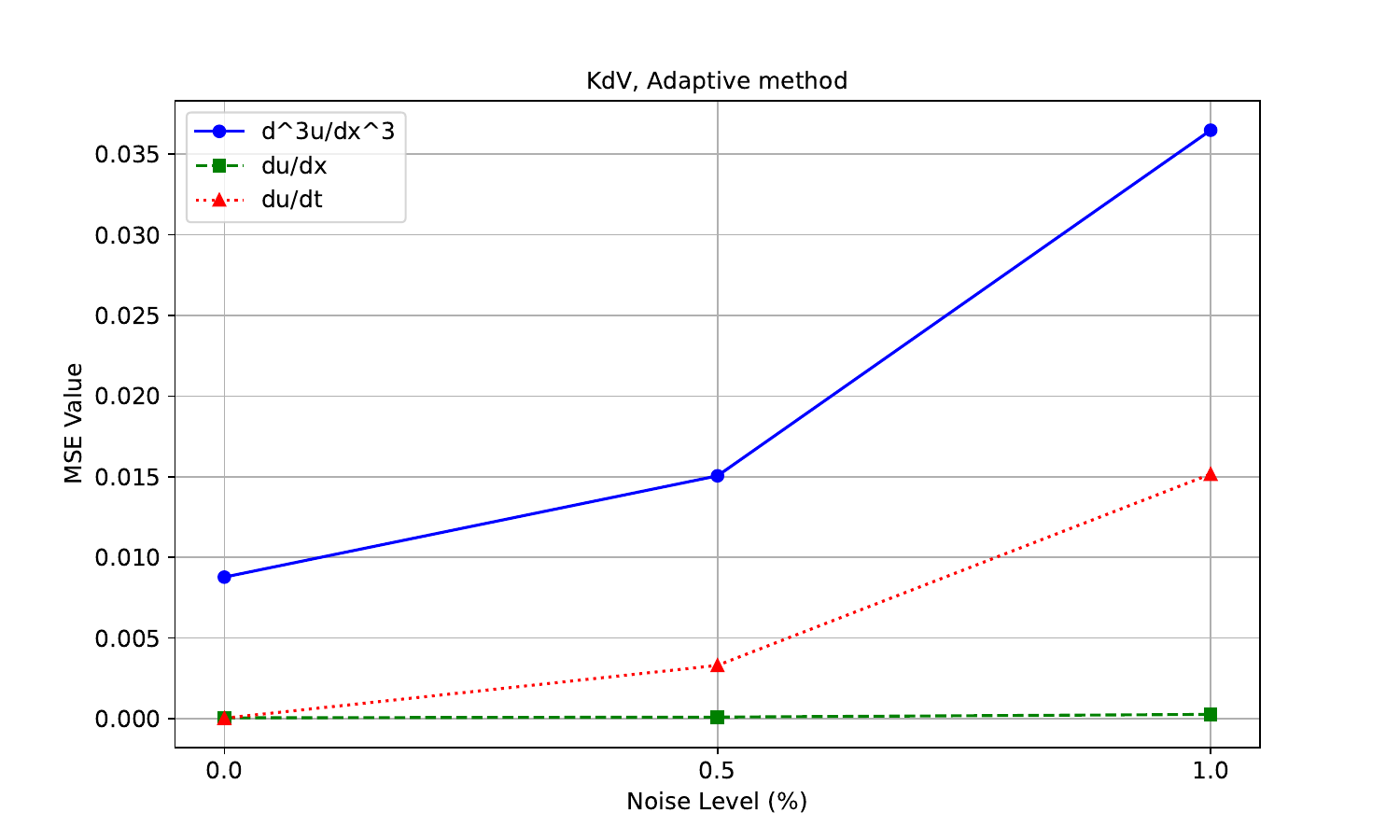}\hfill
          \includegraphics[width=.3\textwidth]{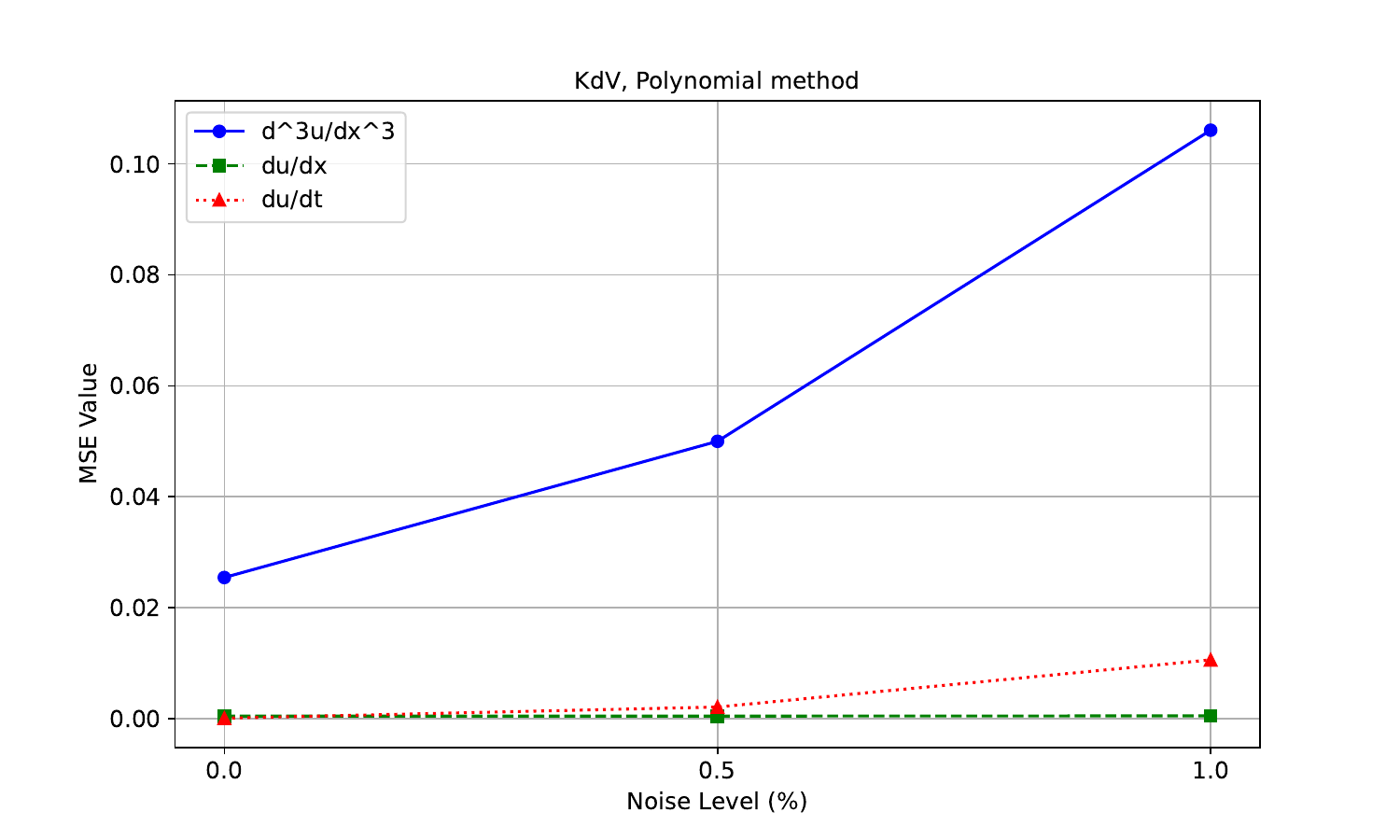}\hfill
          \includegraphics[width=.3\textwidth]{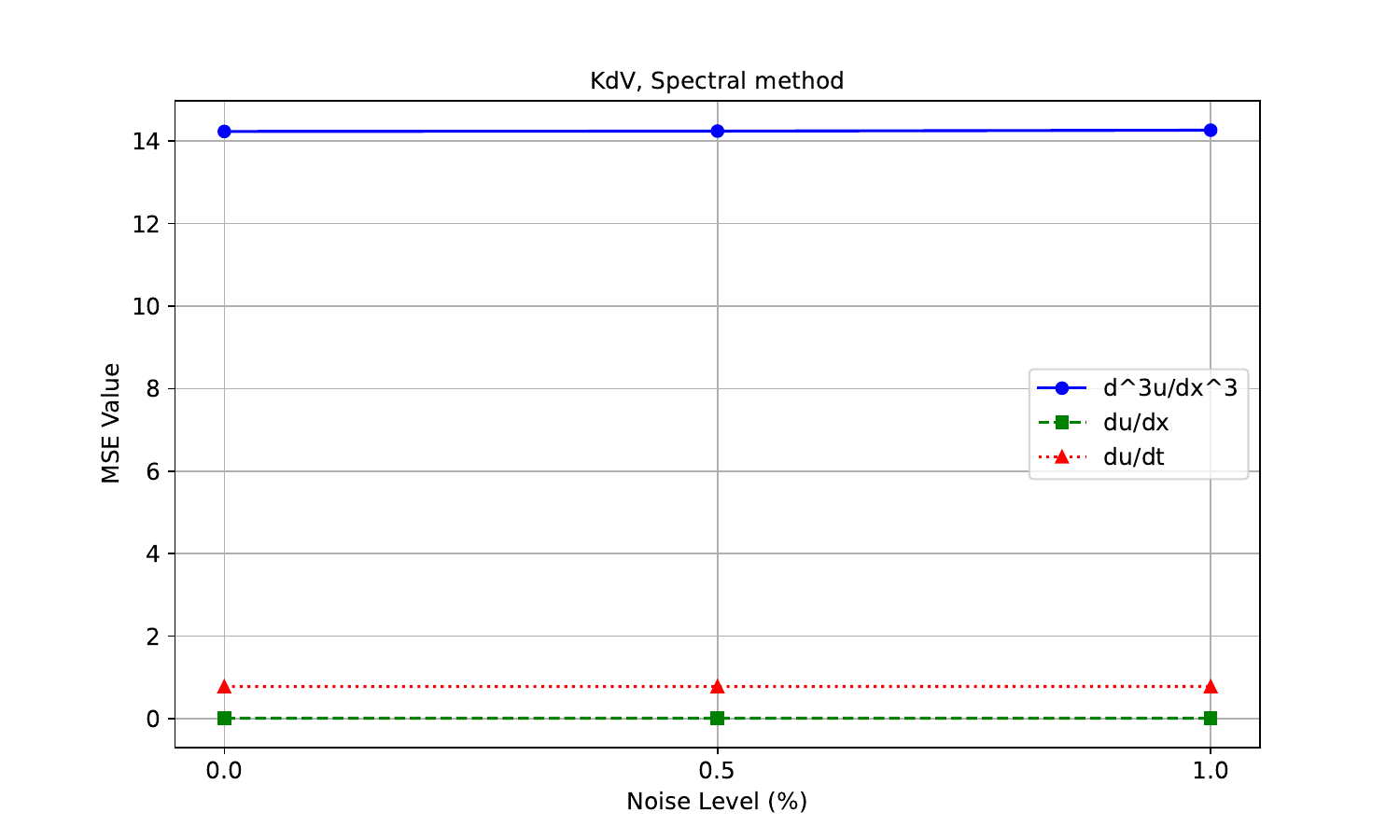}\hfill
          \includegraphics[width=.3\textwidth]{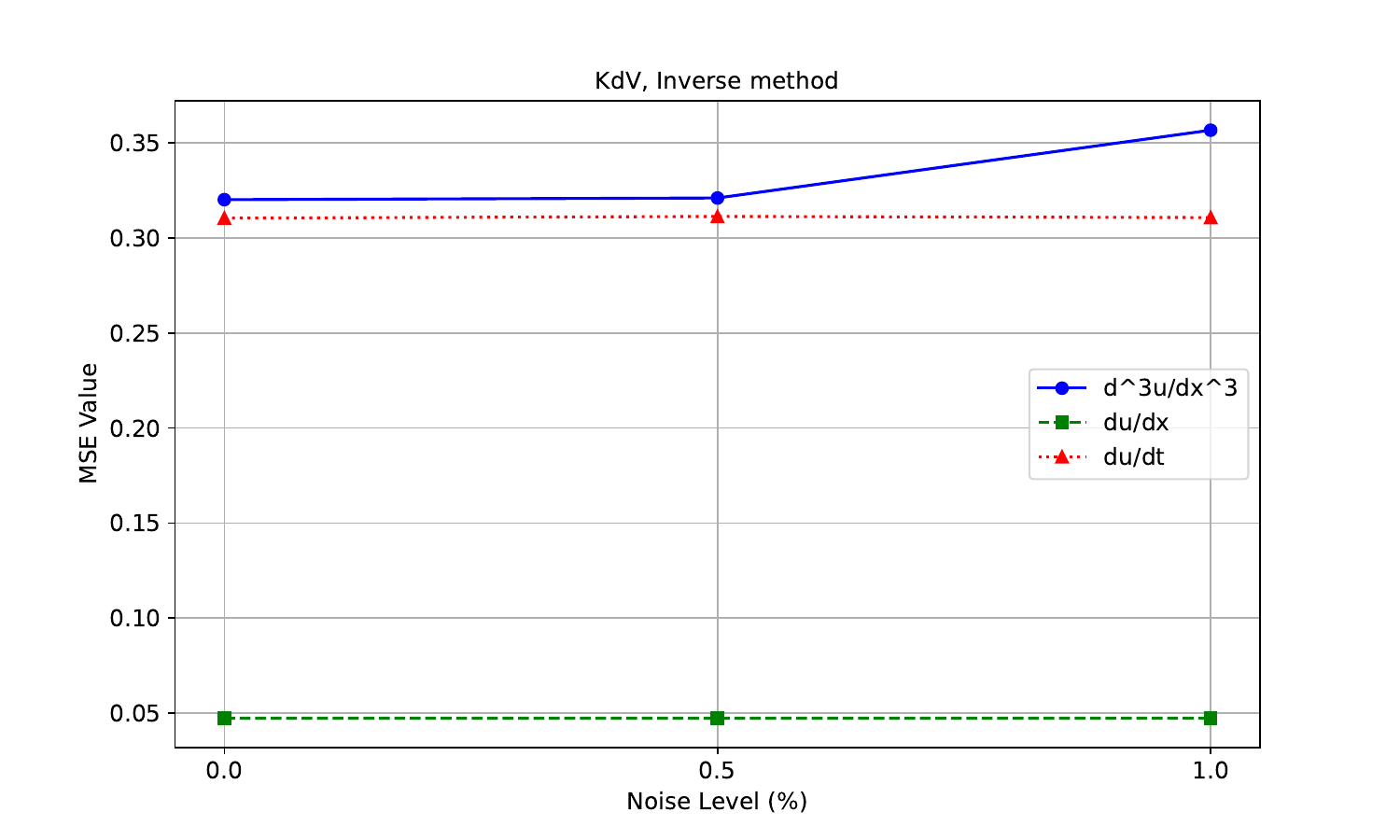}\hfill
          \includegraphics[width=.3\textwidth]{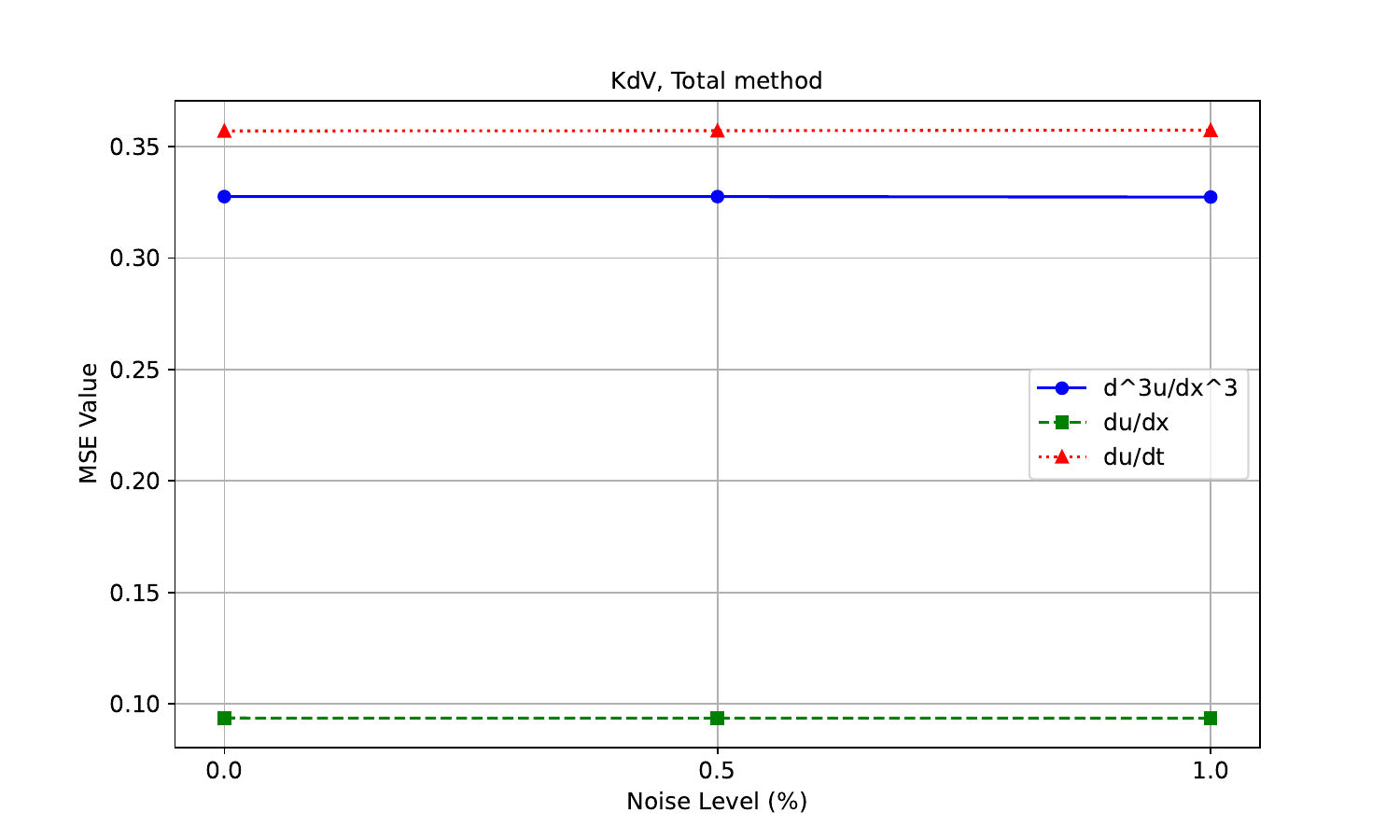}
    \end{multicols}
    \caption{Differentiation errors (MSE) for KdV equation with different noise level}
    %\label{fig:all}
\end{figure*}

%Burgers diff errors
\begin{figure*}[ht!]
    \begin{multicols}{3}
          \includegraphics[width=.3\textwidth]{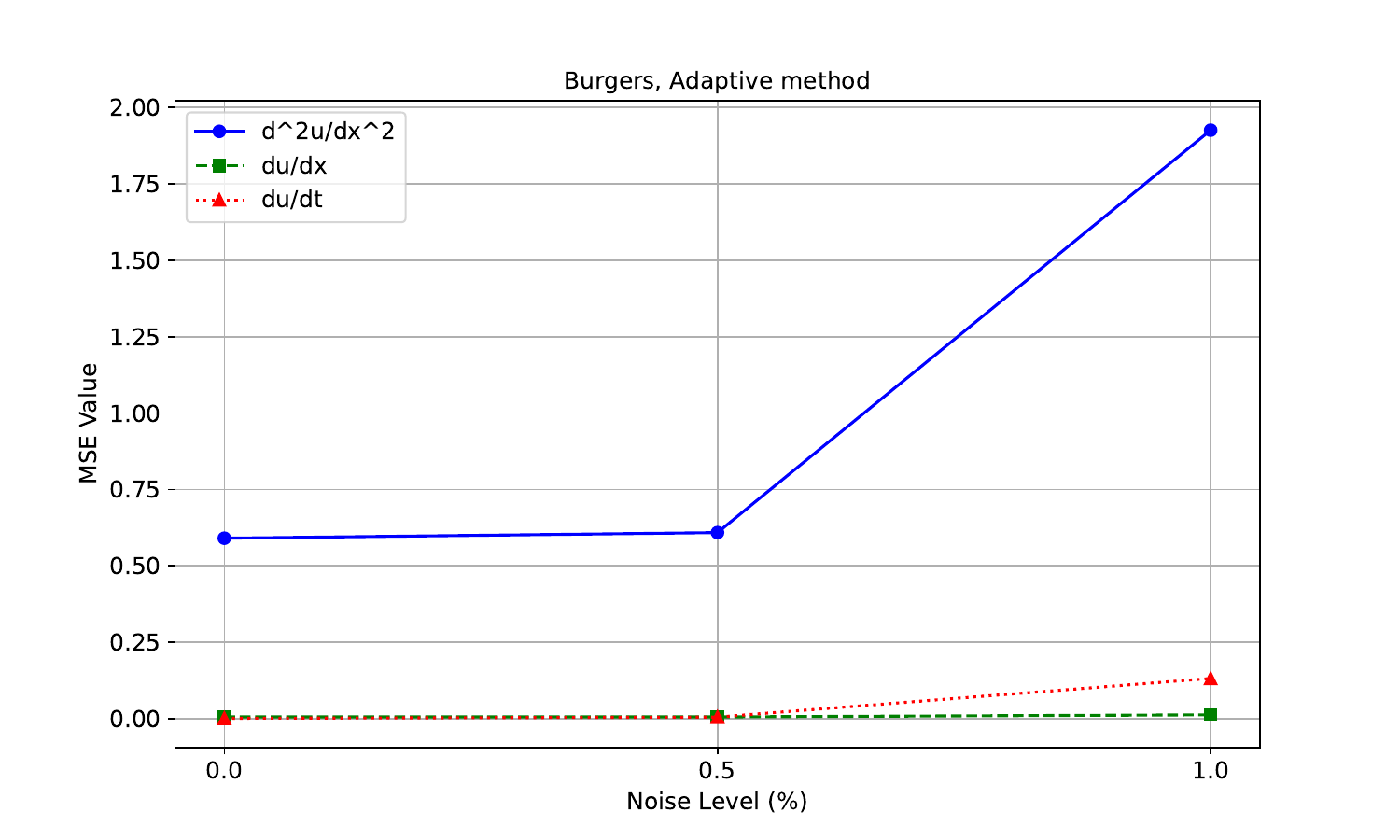}\hfill
          \includegraphics[width=.3\textwidth]{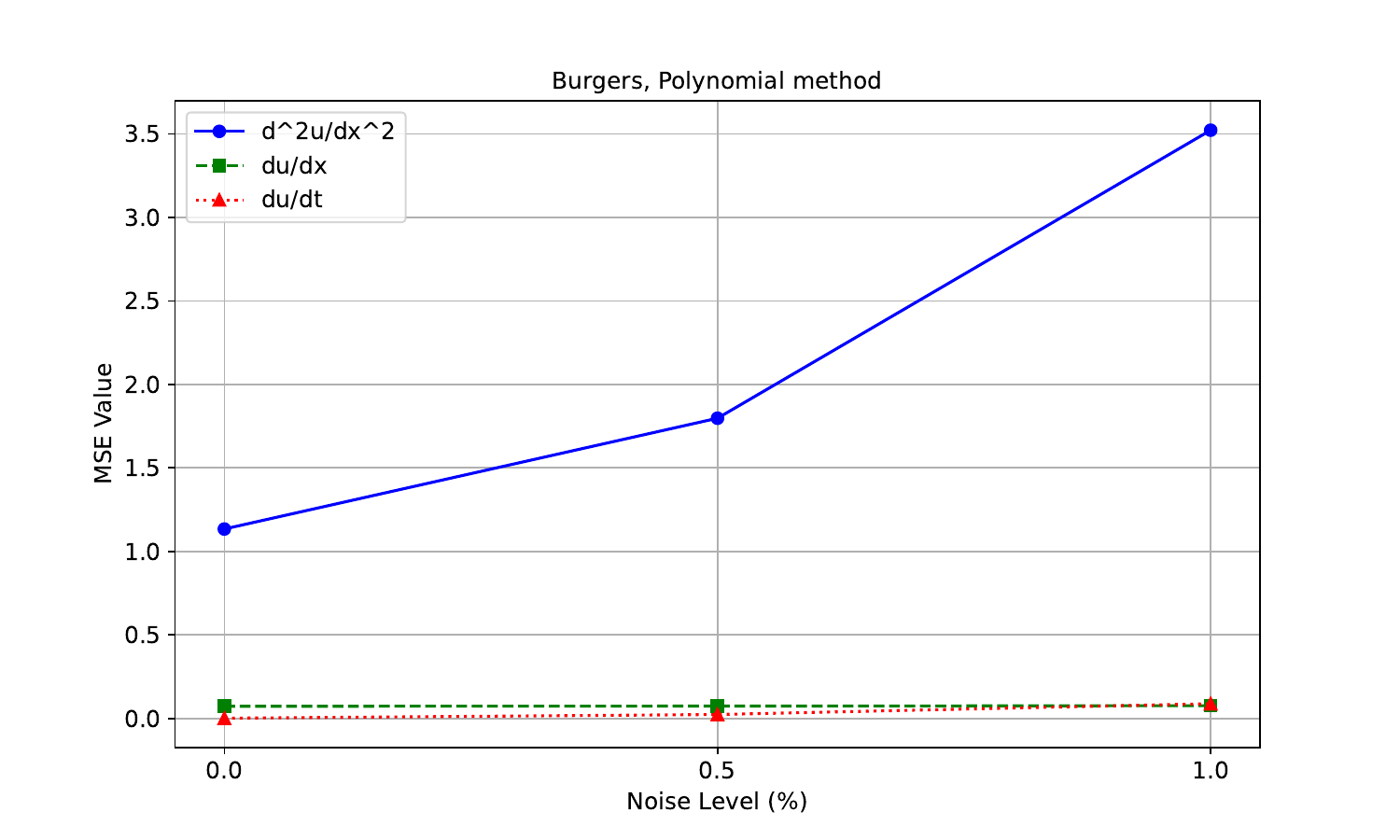}\hfill
          \includegraphics[width=.3\textwidth]{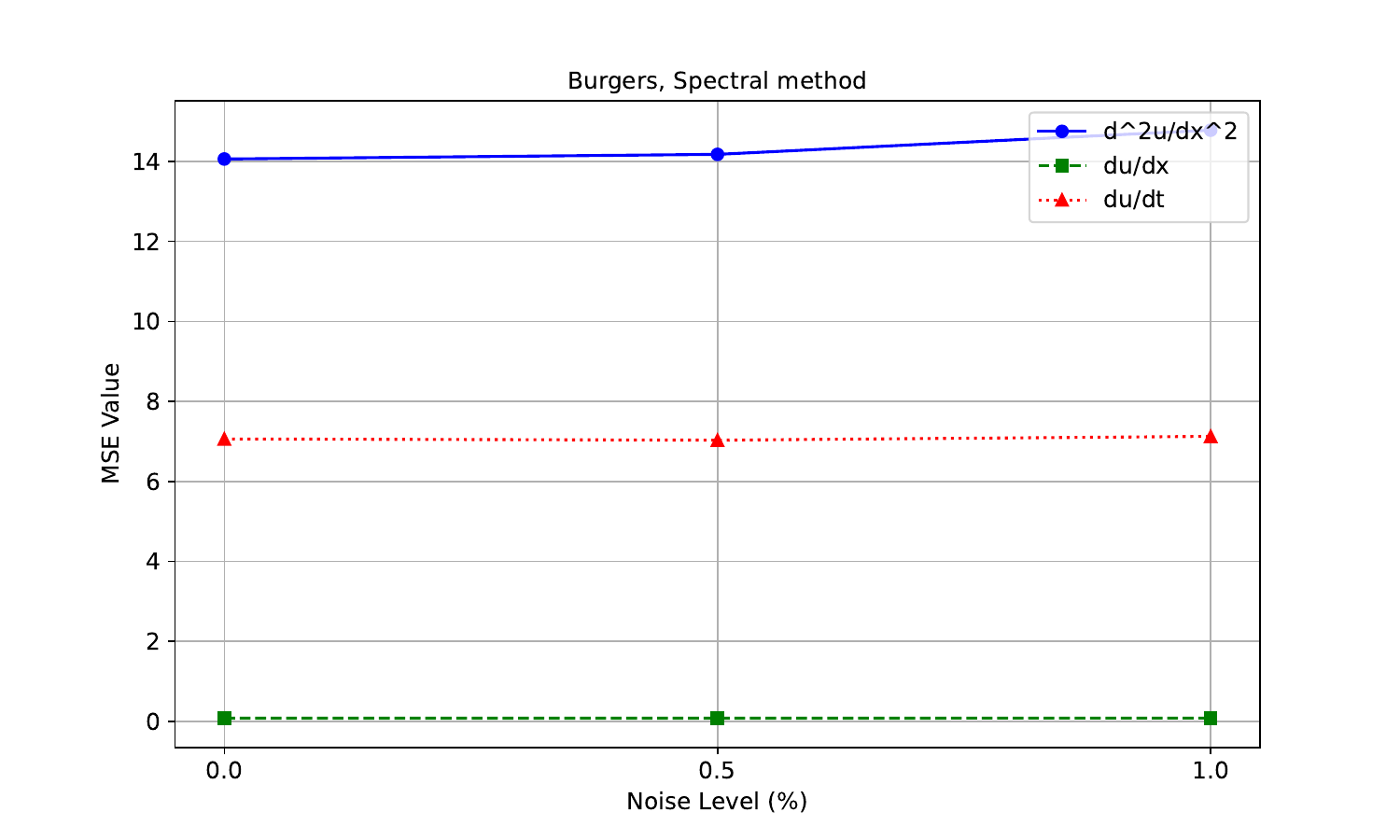}\hfill
          \includegraphics[width=.3\textwidth]{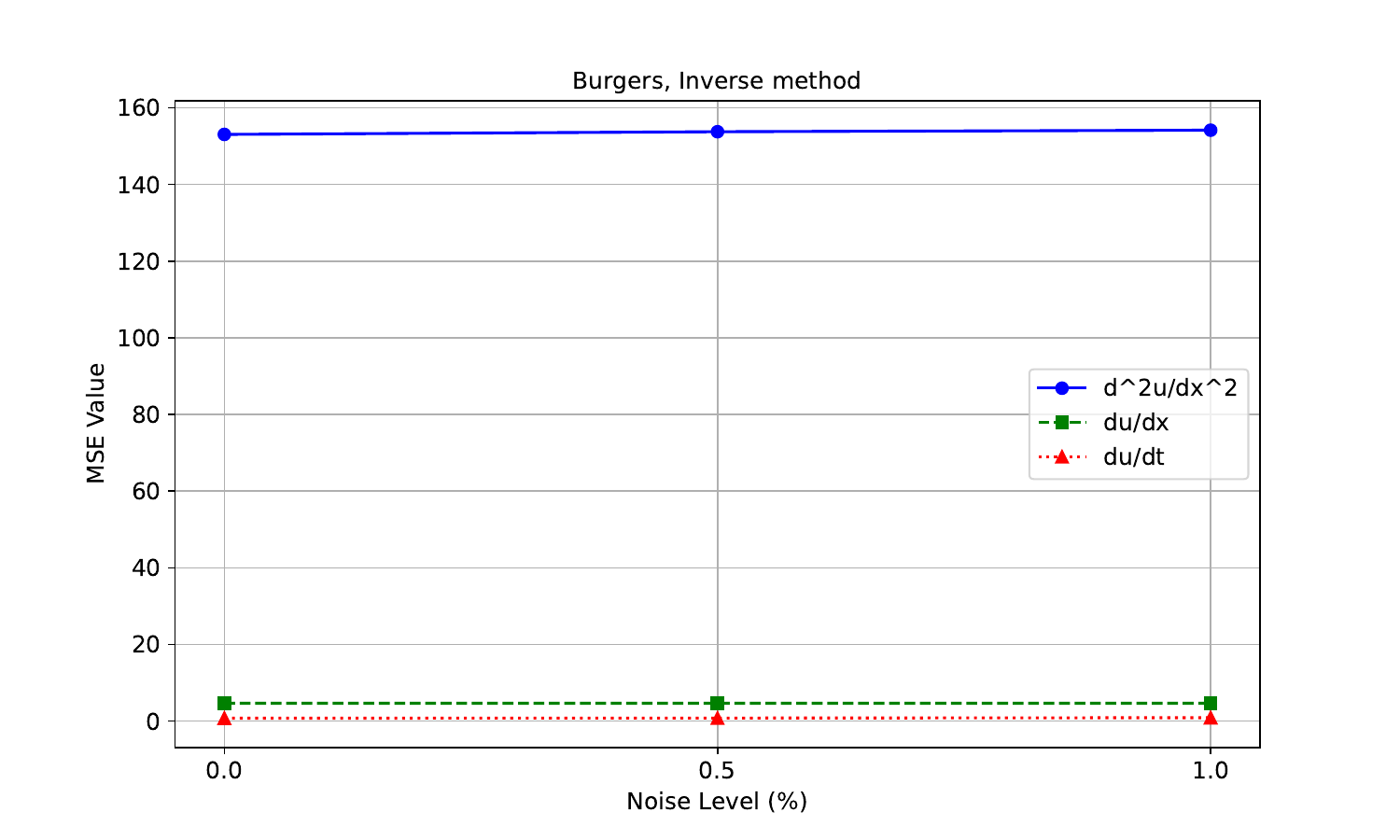}\hfill
          \includegraphics[width=.3\textwidth]{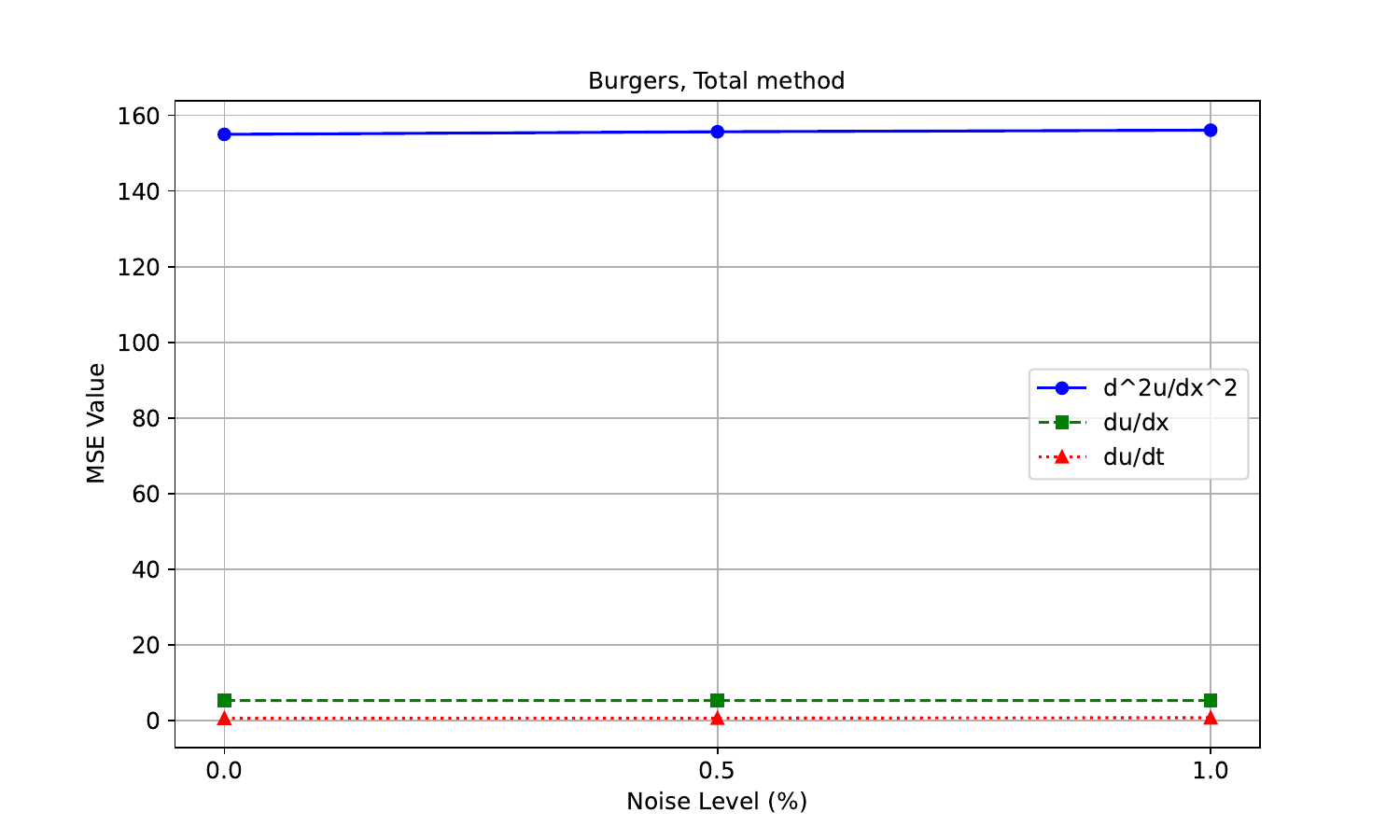}
    \end{multicols}
    \caption{Differentiation errors (MSE) for Burgers equation with different noise level}
    %\label{fig:all}
\end{figure*}

%Laplace diff errors
\begin{figure*}[ht!]
    \begin{multicols}{3}
          \includegraphics[width=.3\textwidth]{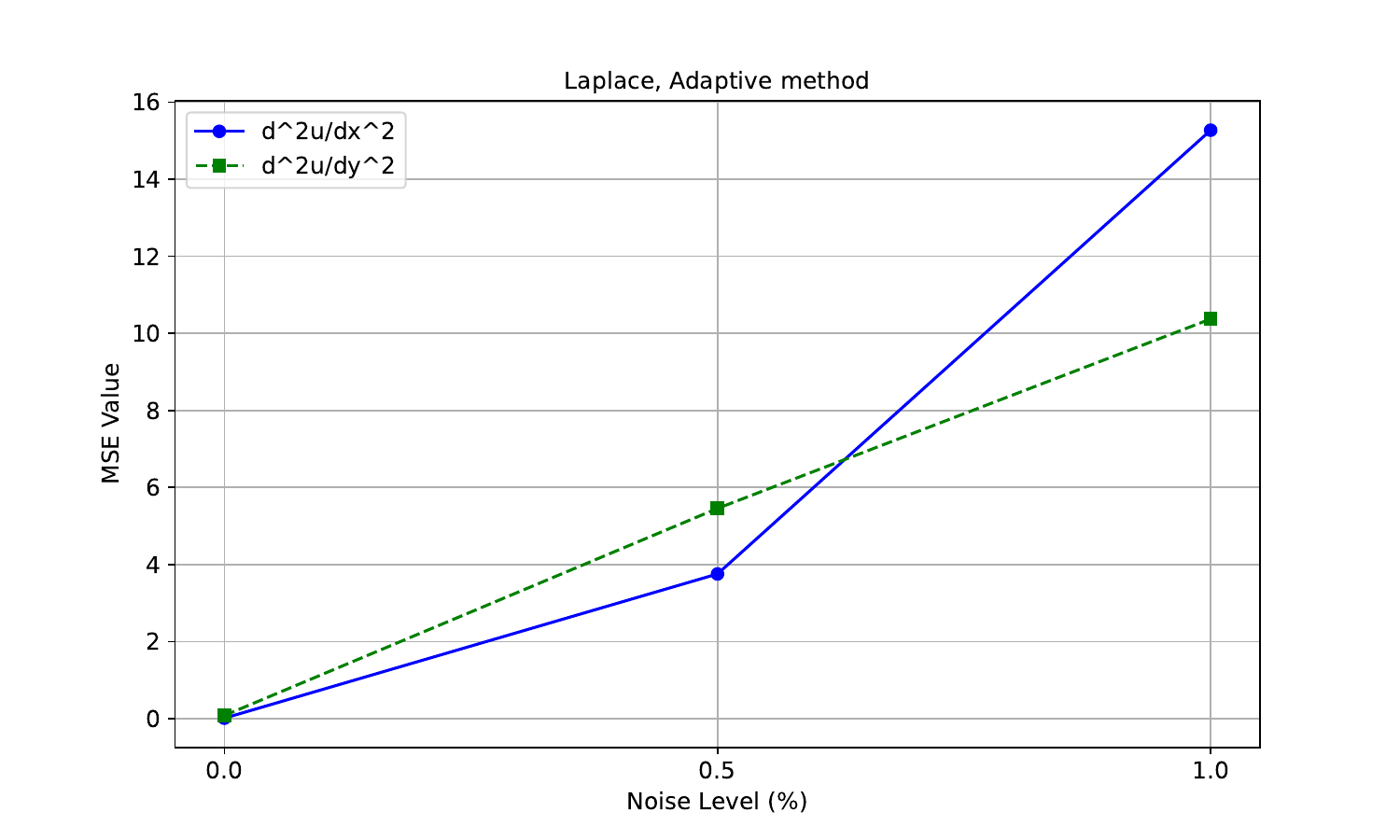}\hfill
          \includegraphics[width=.3\textwidth]{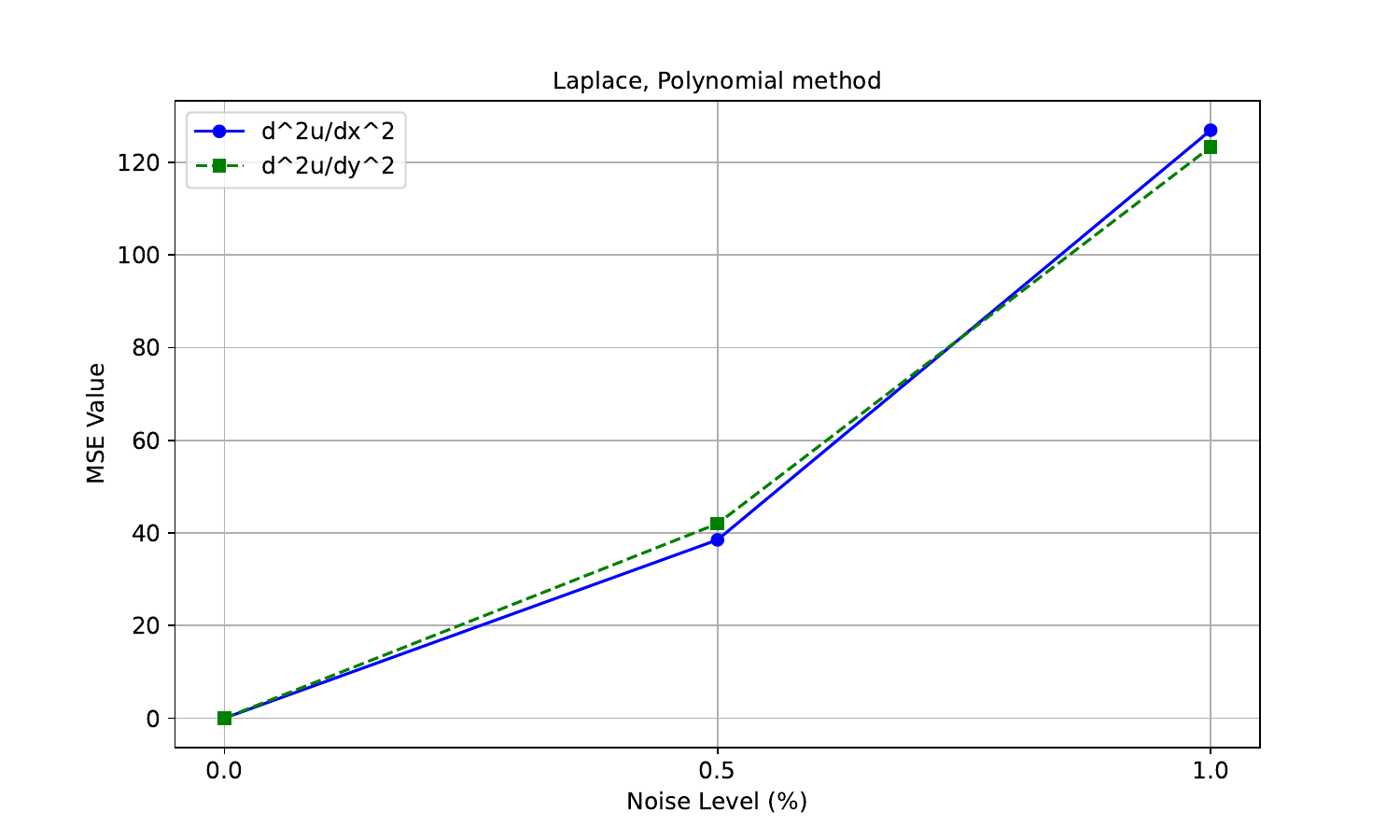}\hfill
          \includegraphics[width=.3\textwidth]{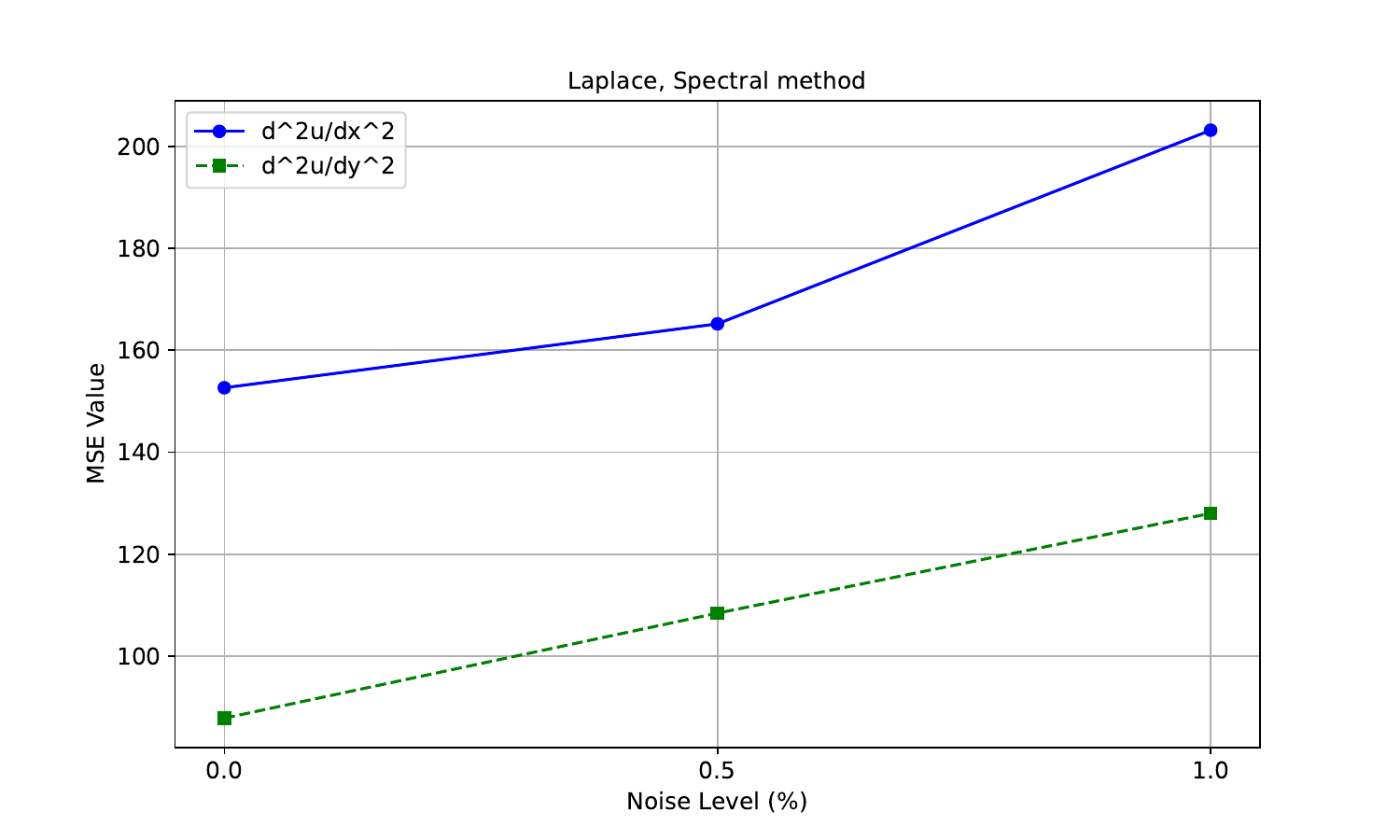}\hfill
          \includegraphics[width=.3\textwidth]{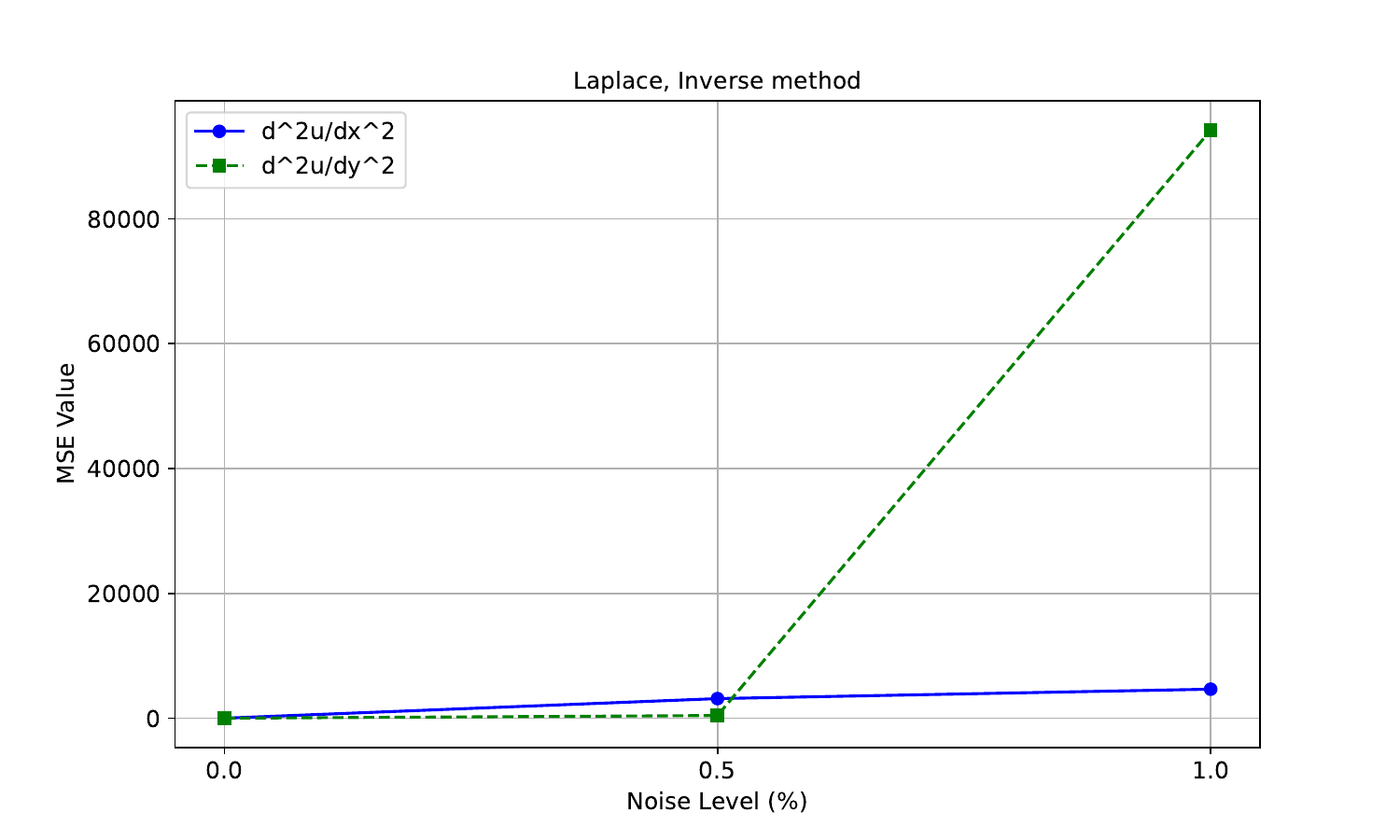}\hfill
          \includegraphics[width=.3\textwidth]{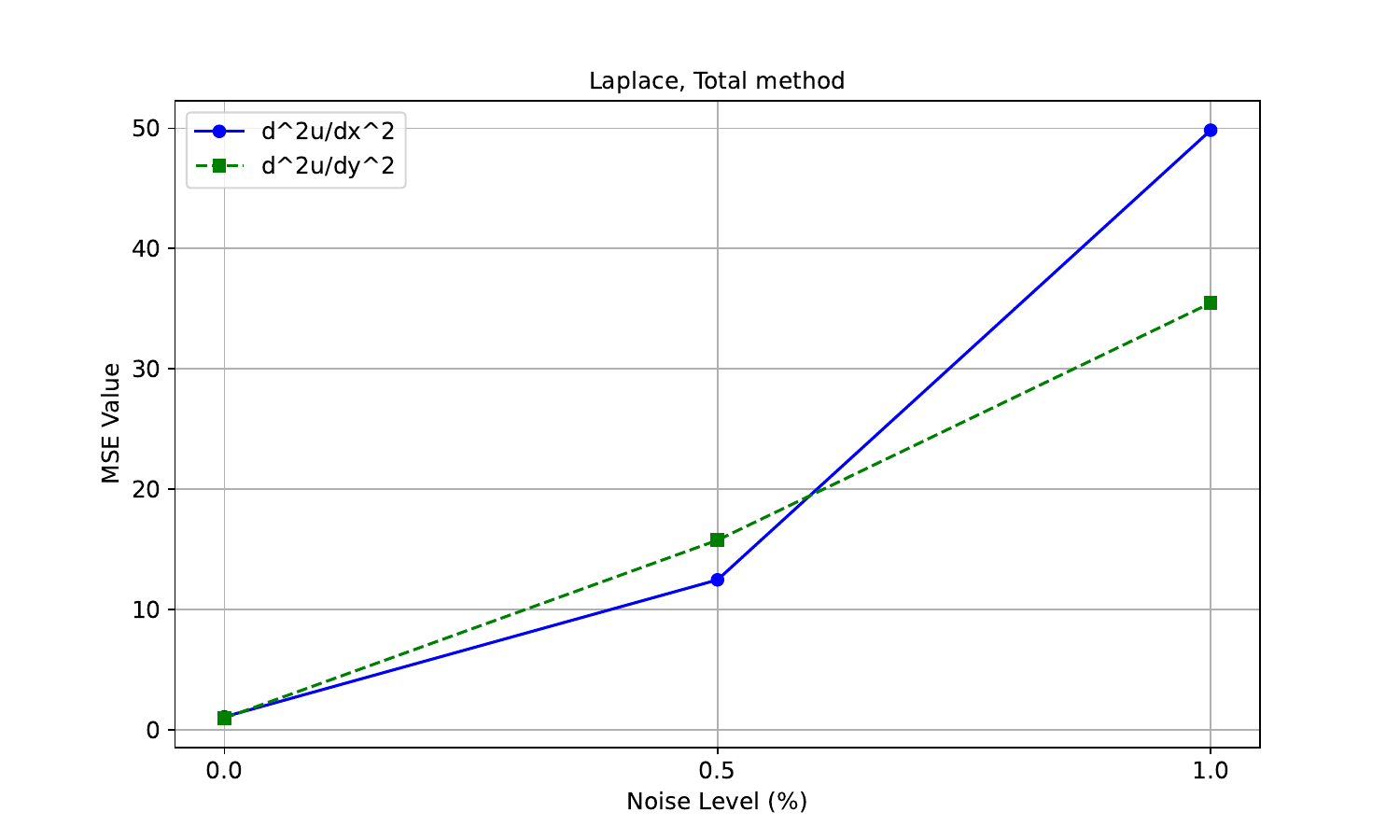}
    \end{multicols}
    \caption{Differentiation errors (MSE) for Laplace equation with different noise level}
    %\label{fig:all}
\end{figure*}

%ODE diff errors
\begin{figure*}[ht!]
    \begin{multicols}{3}
          \includegraphics[width=.3\textwidth]{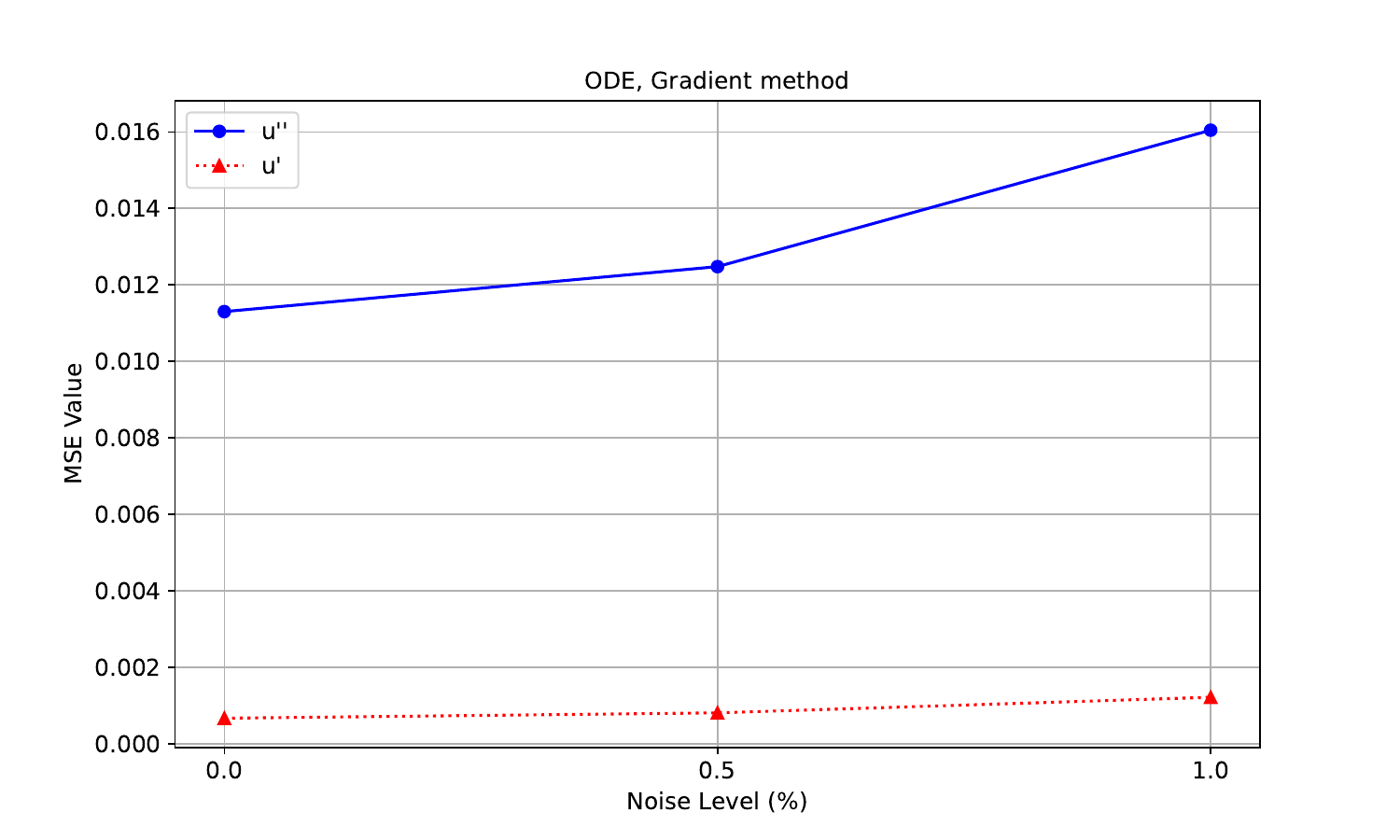}\hfill
          \includegraphics[width=.3\textwidth]{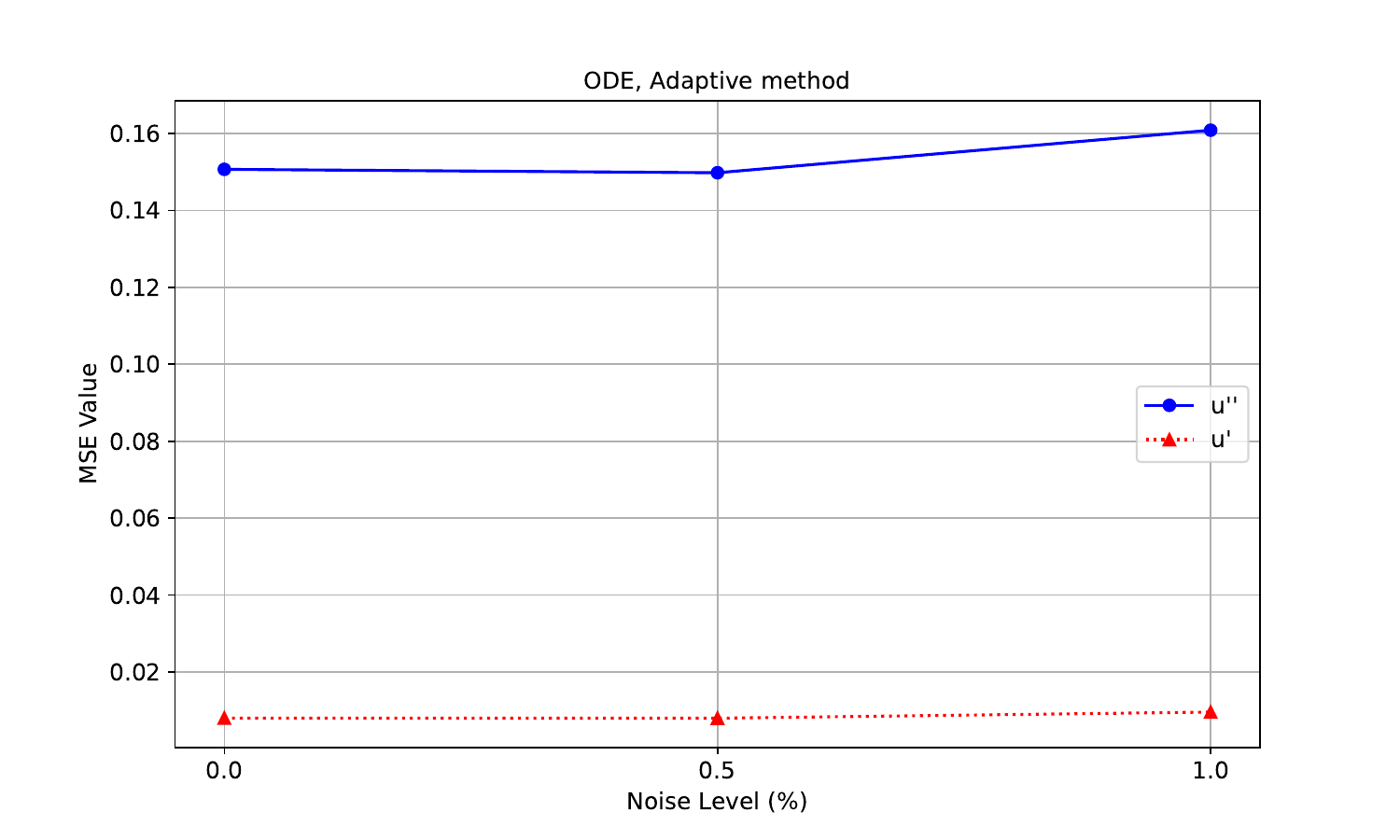}\hfill
          \includegraphics[width=.3\textwidth]{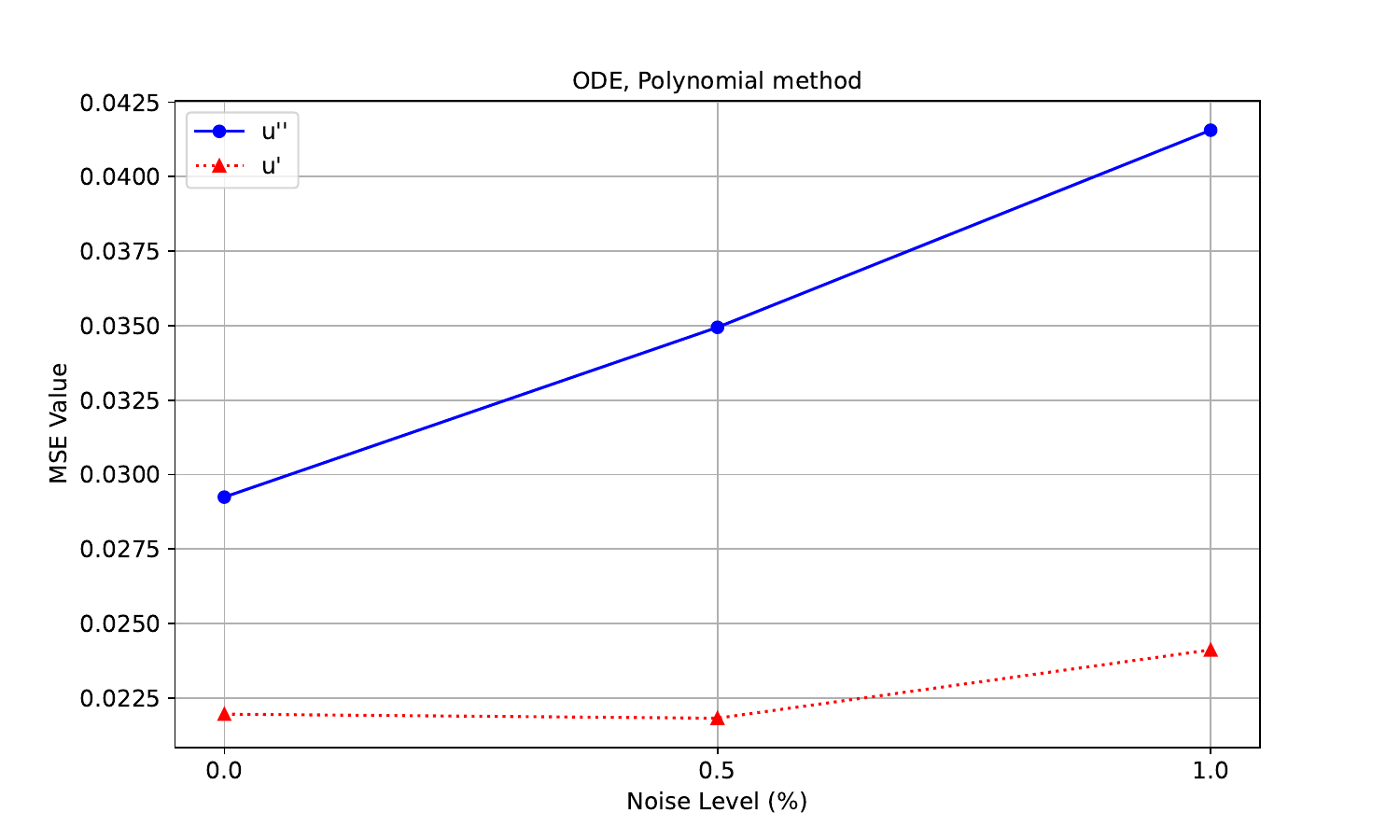}\hfill
          \includegraphics[width=.3\textwidth]{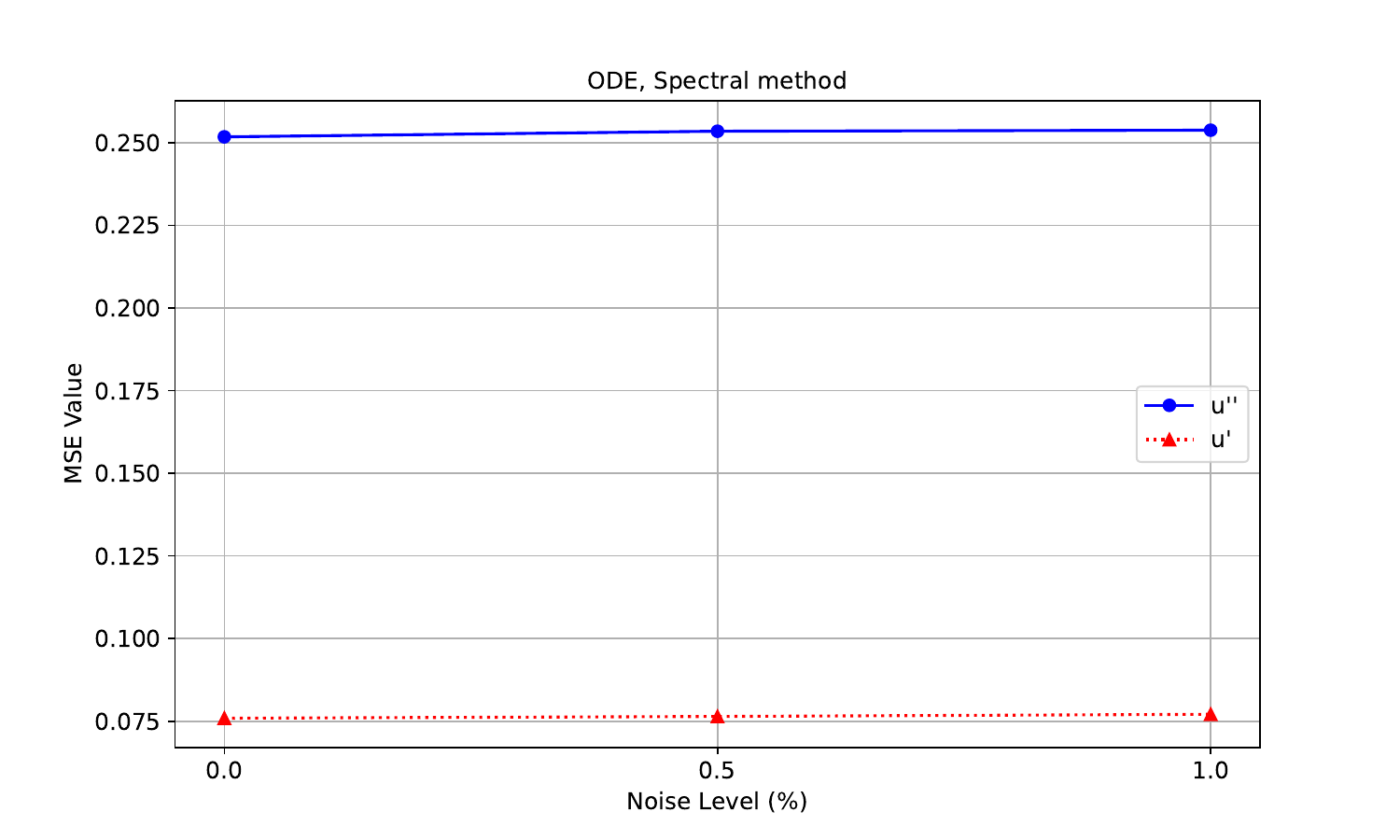}\hfill
          \includegraphics[width=.3\textwidth]{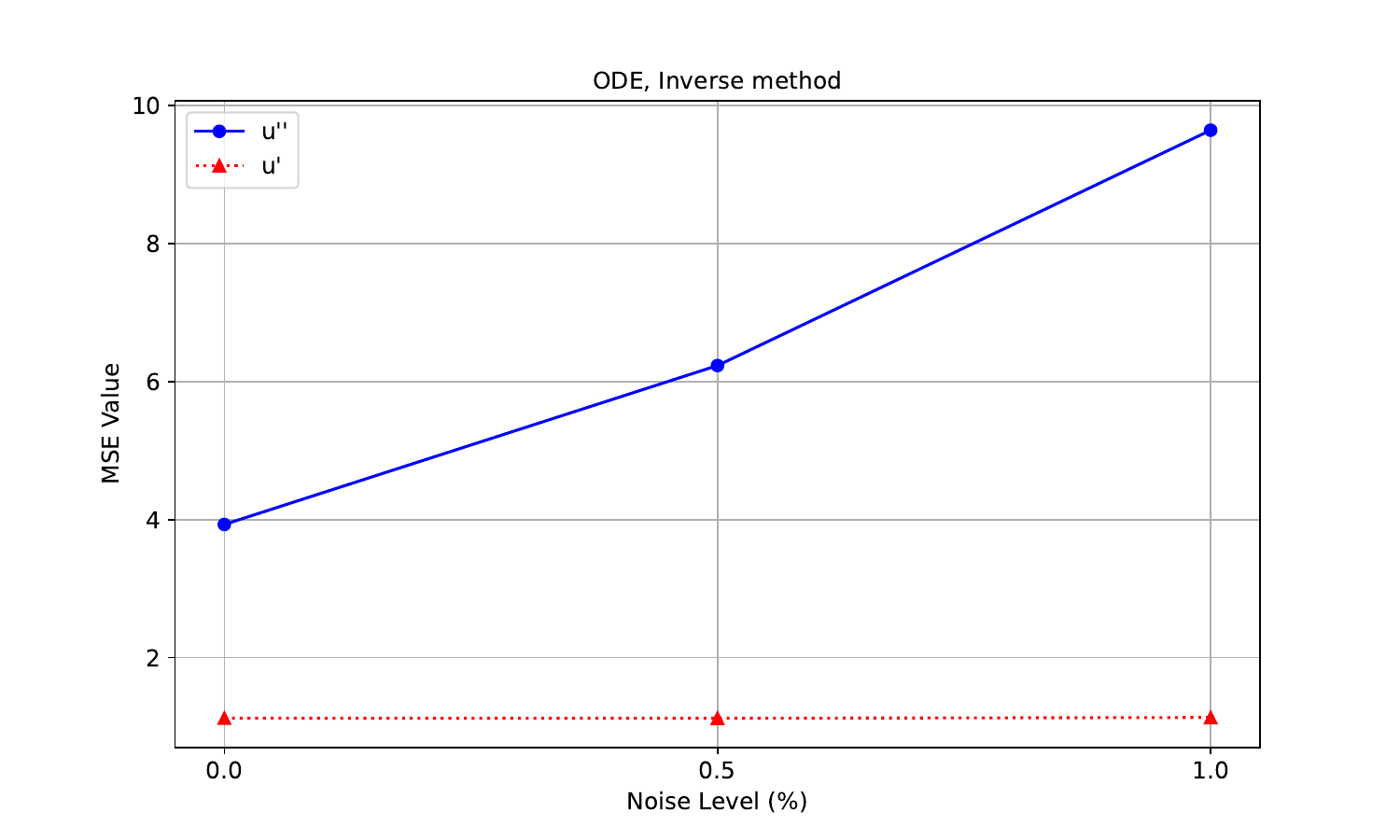}\hfill
          \includegraphics[width=.3\textwidth]{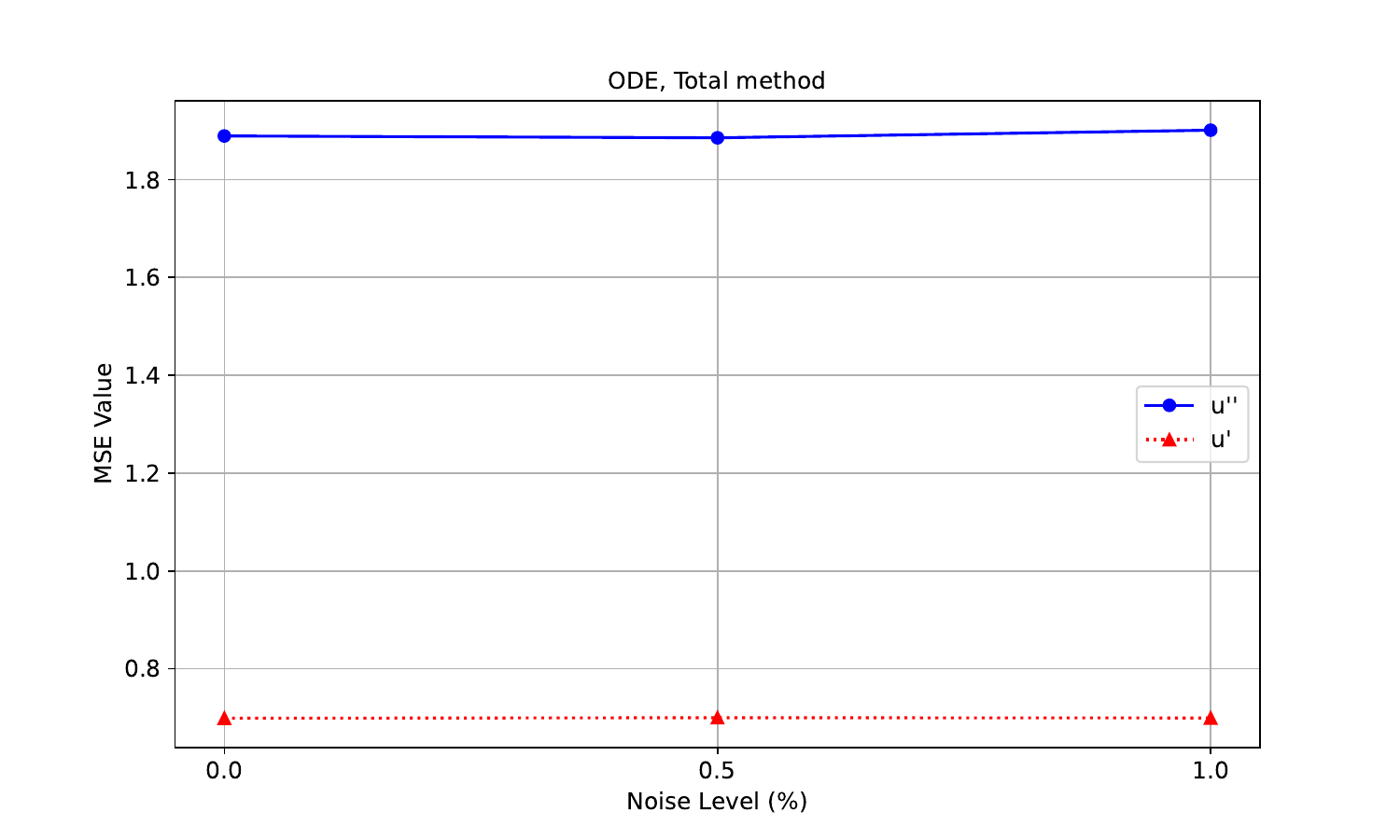}
    \end{multicols}
    \caption{Differentiation errors (MSE) for ODE equation with different noise level}
    %\label{fig:all}
\end{figure*}

%Wave diff errors
\begin{figure*}[ht!]
    \begin{multicols}{3}
          \includegraphics[width=.3\textwidth]{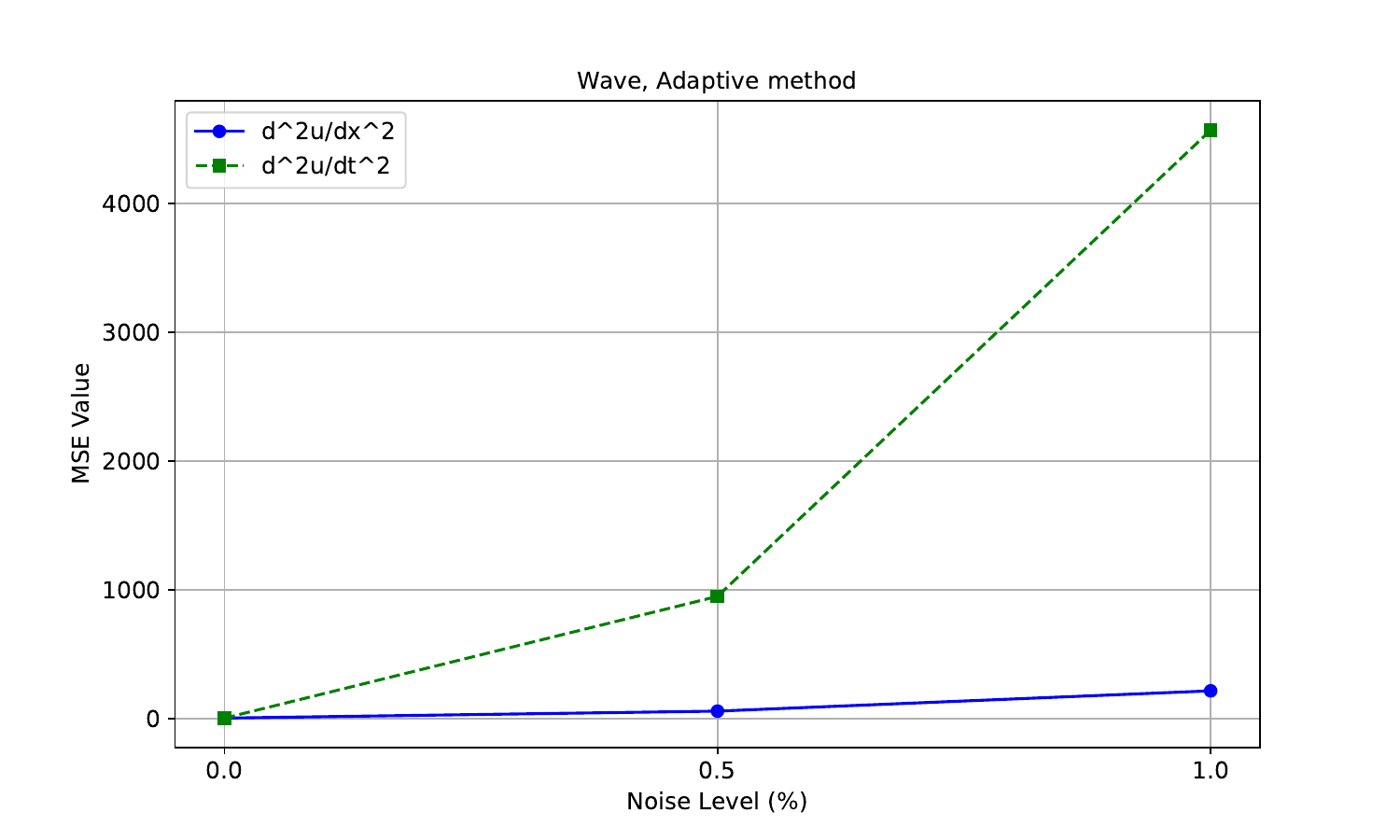}\hfill
          \includegraphics[width=.3\textwidth]{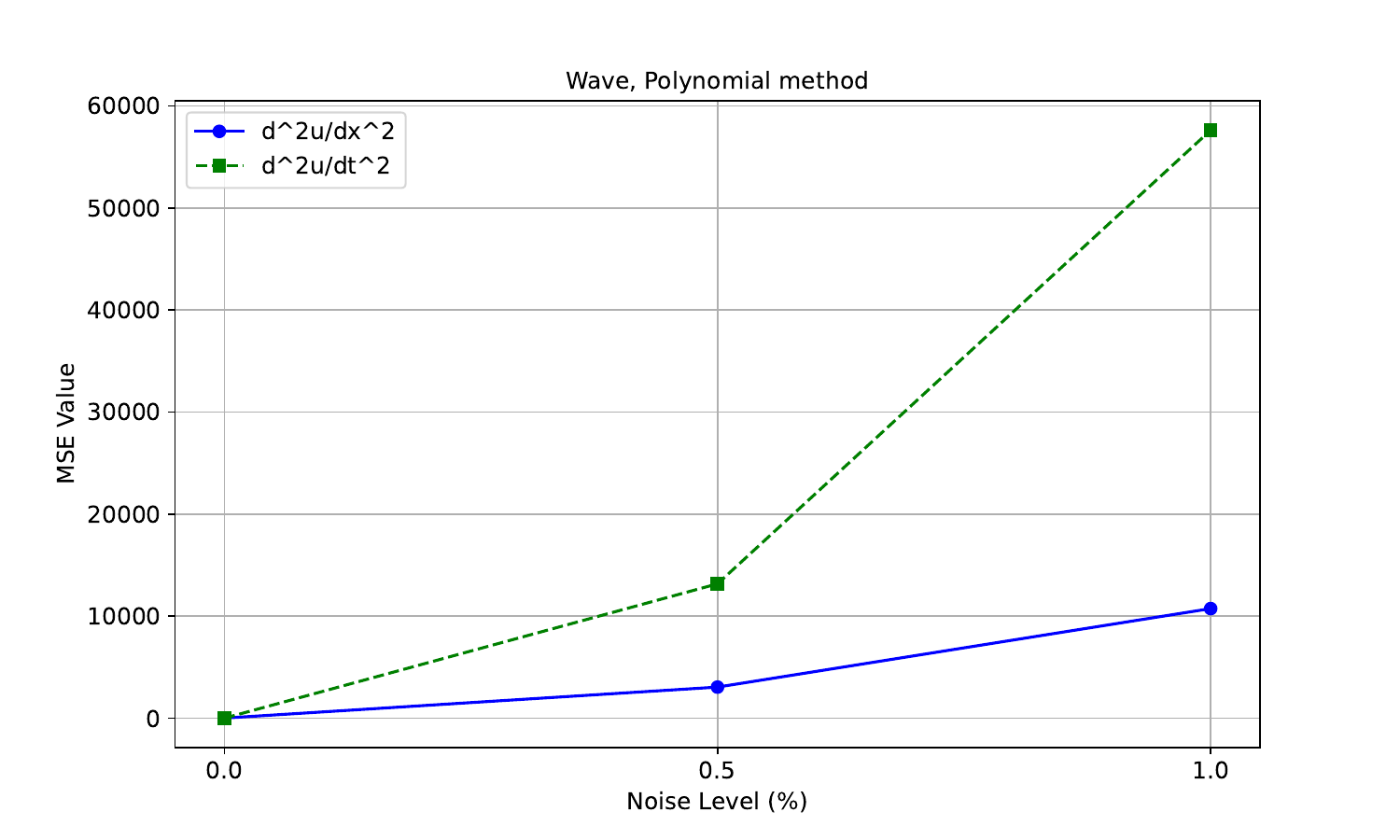}\hfill
          \includegraphics[width=.3\textwidth]{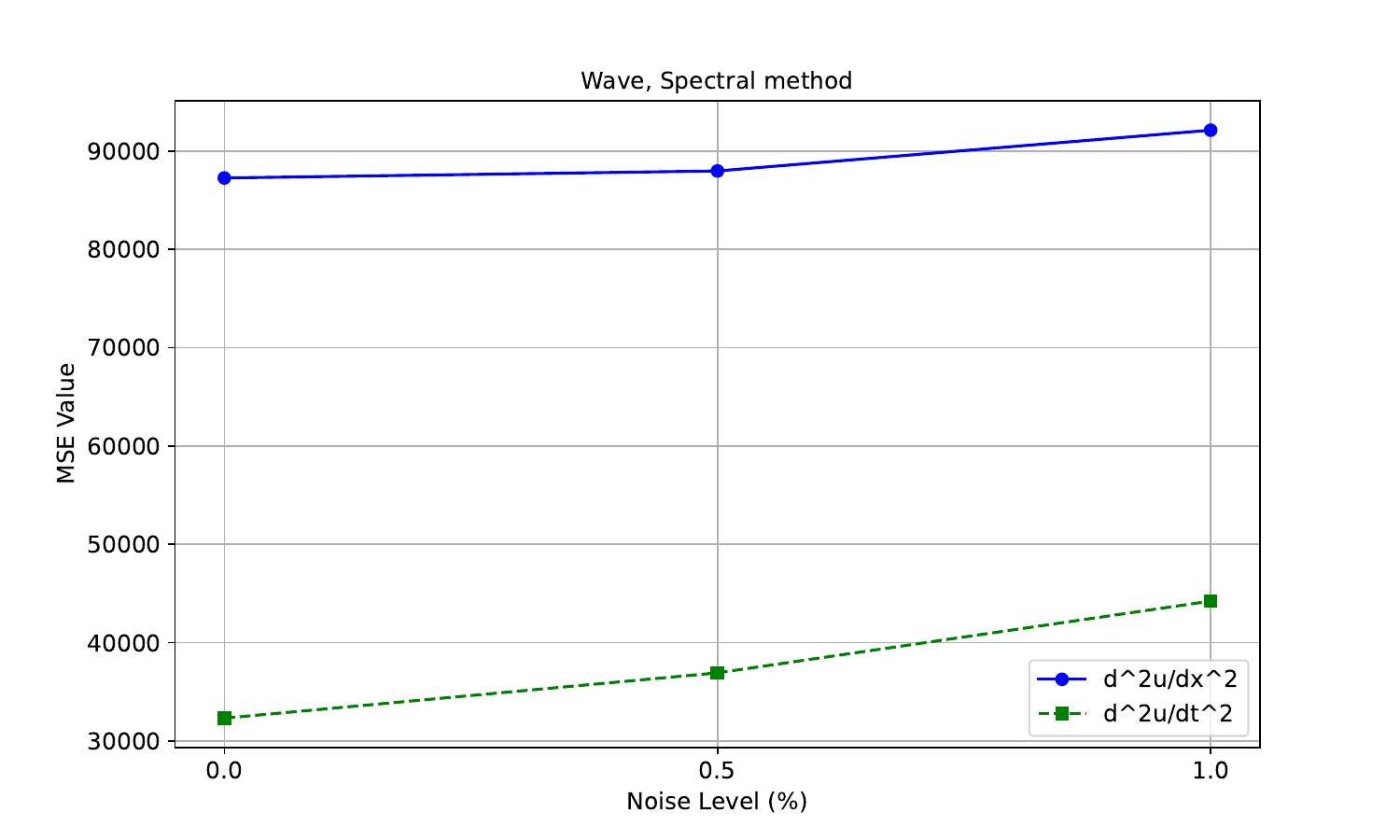}\hfill
          \includegraphics[width=.3\textwidth]{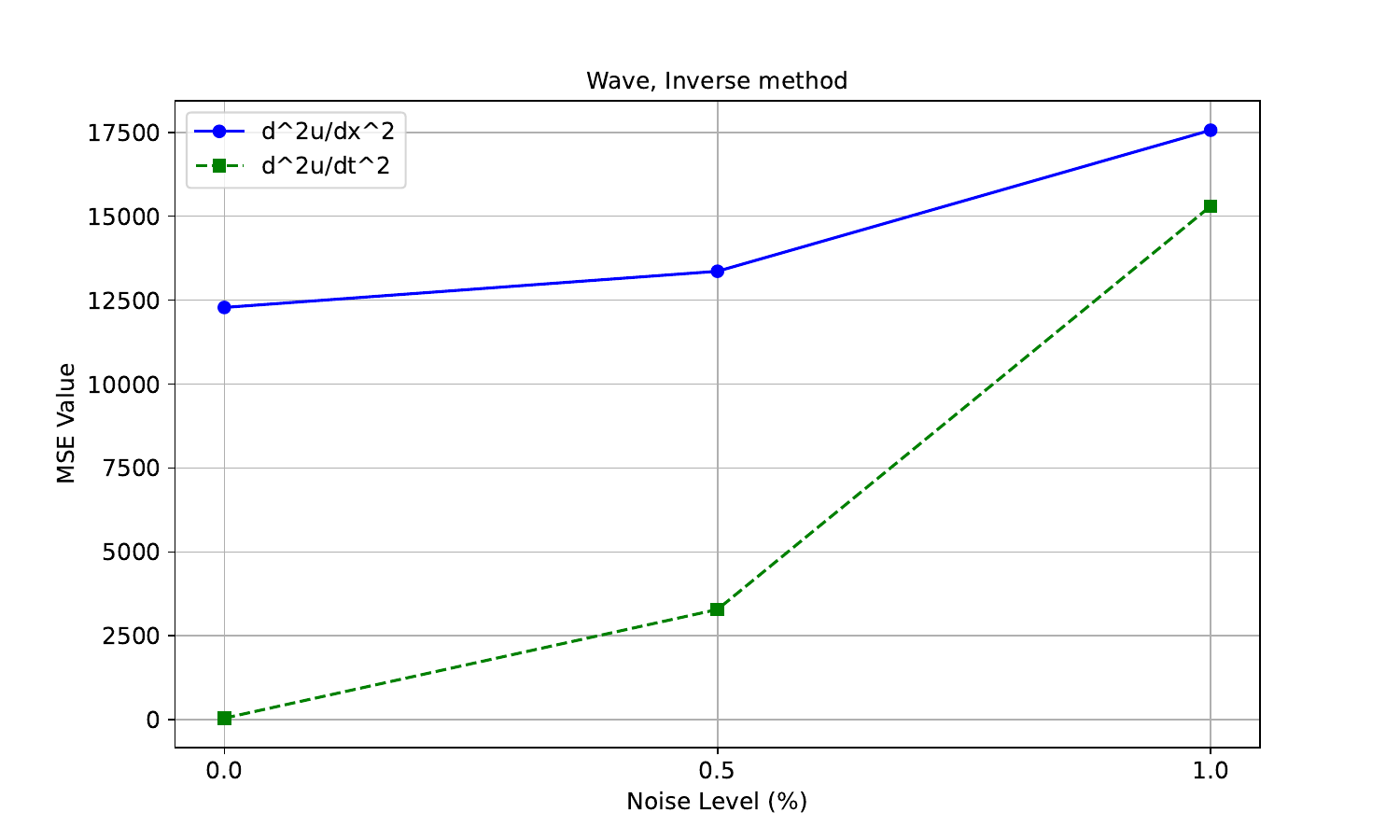}\hfill
          \includegraphics[width=.3\textwidth]{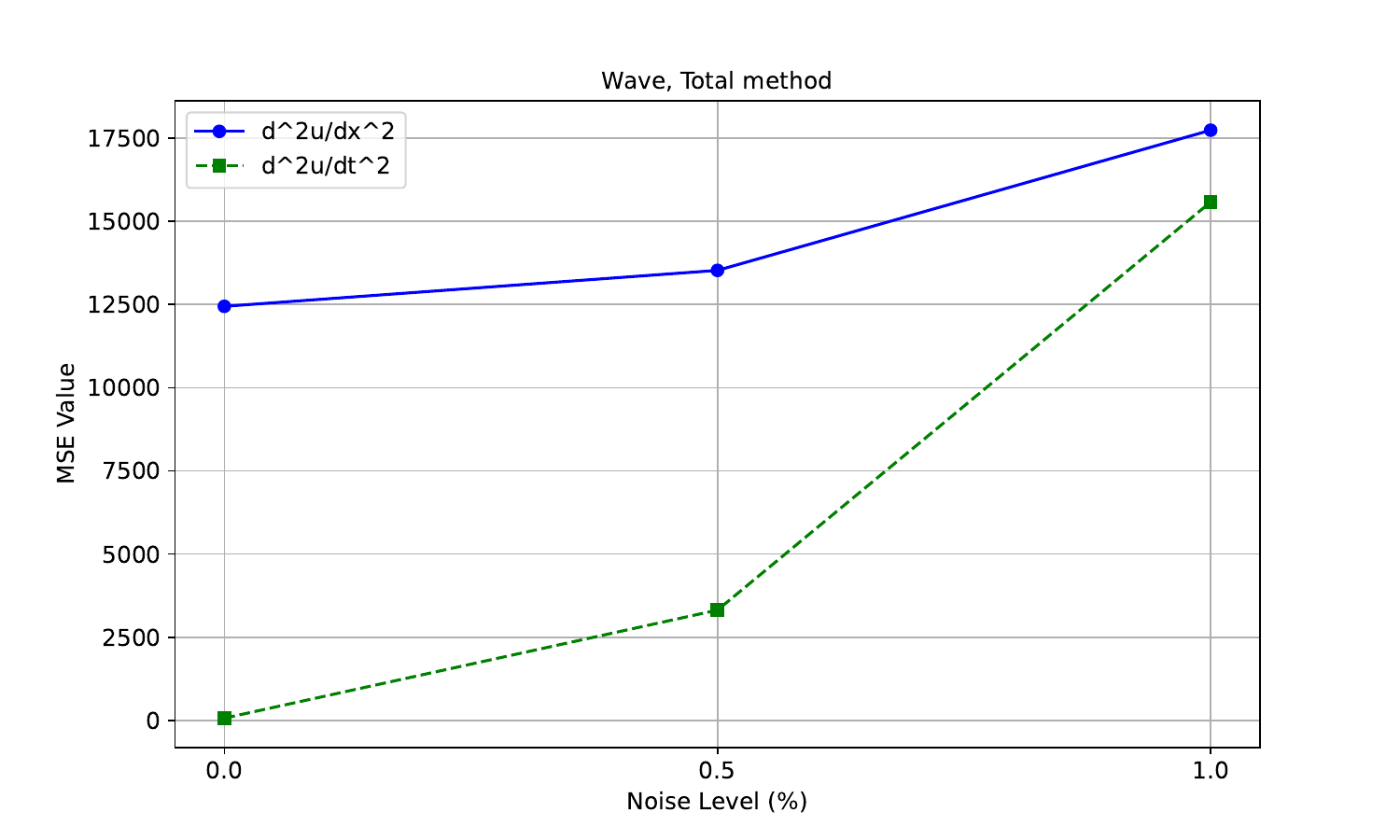}
    \end{multicols}
    \caption{Differentiation errors (MSE) for Wave equation with different noise level}
    %\label{fig:all}
\end{figure*}

\begin{table}[ht!]
\caption{Differentiation errors, noise = 0\%}
%\label{tab:my-table}
\centering
\begin{tabular}{|c|ccc|}
\hline
\multicolumn{1}{|l|}{} & \multicolumn{3}{l|}{Burgers equation}                                                                                                                       \\ \hline
Methods/Terms          & \multicolumn{1}{c|}{du/dt}                                       & \multicolumn{1}{c|}{d\textasciicircum{}2u/dx\textasciicircum{}2} & du/dx                 \\ \hline
Adaptive               & \multicolumn{1}{c|}{0.0009}                                      & \multicolumn{1}{c|}{0.05902}                                  & 0.0053               \\ \hline
Polynomial             & \multicolumn{1}{c|}{0.0012}                                      & \multicolumn{1}{c|}{1.1339}                                      & 0.0724                \\ \hline
Spectral               & \multicolumn{1}{c|}{7.0597}                                      & \multicolumn{1}{c|}{14.0607}                                     & 0.0757                \\ \hline
Inverse                & \multicolumn{1}{c|}{0.7638}                                      & \multicolumn{1}{c|}{153.0405}                                    & 4.6508                \\ \hline
Total\_var                  & \multicolumn{1}{c|}{0.6760}                                      & \multicolumn{1}{c|}{154.9538}                                    & 5.3229                \\ \hline
\multicolumn{1}{|l|}{} & \multicolumn{1}{l|}{KdV equation}                                & \multicolumn{1}{l|}{}                                            & \multicolumn{1}{l|}{} \\
\hline
\end{tabular}
\end{table}

\begin{table}[ht!]
\caption{Differentiation errors, noise = 0.5\%}
%\label{tab:my-table}
\centering
\begin{tabular}{|c|ccc|}
\hline
\multicolumn{1}{|l|}{} & \multicolumn{3}{l|}{Burgers equation}                                                                                                                       \\ \hline
Methods/Terms          & \multicolumn{1}{c|}{du/dt}                                       & \multicolumn{1}{c|}{d\textasciicircum{}2u/dx\textasciicircum{}2} & du/dx                 \\ \hline
Adaptive               & \multicolumn{1}{c|}{0.0048}                                       & \multicolumn{1}{c|}{0.6084}                                  & 0.0053               \\ \hline
Polynomial             & \multicolumn{1}{c|}{0.0234}                                      & \multicolumn{1}{c|}{1.7973}                                      & 0.0737                \\ \hline
Spectral               & \multicolumn{1}{c|}{7.0293}                                      & \multicolumn{1}{c|}{14.1771}                                     & 0.0761                \\ \hline
Inverse                & \multicolumn{1}{c|}{0.8013}                                      & \multicolumn{1}{c|}{153.7448}                                    & 4.6542                \\ \hline
Total\_var                  & \multicolumn{1}{c|}{0.7139}                                      & \multicolumn{1}{c|}{155.6498}                                    & 5.3271                \\ \hline
\multicolumn{1}{|l|}{} & \multicolumn{1}{l|}{KdV equation}                                & \multicolumn{1}{l|}{}                                            & \multicolumn{1}{l|}{} \\
\hline
\end{tabular}
\end{table}

\begin{table}[ht!]
\caption{Differentiation errors, noise = 1\%}
%\label{tab:my-table}
\centering
\begin{tabular}{|c|ccc|}
\hline
\multicolumn{1}{|l|}{} & \multicolumn{3}{l|}{Burgers equation}                                                                                                                       \\ \hline
Methods/Terms          & \multicolumn{1}{c|}{du/dt}                                       & \multicolumn{1}{c|}{d\textasciicircum{}2u/dx\textasciicircum{}2} & du/dx                 \\ \hline
Adaptive               & \multicolumn{1}{c|}{0.1311}                                      & \multicolumn{1}{c|}{1.9259}                                  & 0.0121               \\ \hline
Polynomial             & \multicolumn{1}{c|}{0.088}                                       & \multicolumn{1}{c|}{3.5219}                                      & 0.0761                \\ \hline
Spectral               & \multicolumn{1}{c|}{7.1244}                                      & \multicolumn{1}{c|}{14.7814}                                     & 0.0783                \\ \hline
Inverse                & \multicolumn{1}{c|}{0.8933}                                      & \multicolumn{1}{c|}{154.1399}                                    & 4.6562                \\ \hline
Total\_var                  & \multicolumn{1}{c|}{0.8085}                                      & \multicolumn{1}{c|}{156.0558}                                    & 5.3286                \\ \hline
\multicolumn{1}{|l|}{} & \multicolumn{1}{l|}{KdV equation}                                & \multicolumn{1}{l|}{}                                            & \multicolumn{1}{l|}{} \\
\hline
\end{tabular}
\end{table}

\begin{table}[ht!]
\caption{Differentiation errors without boundaries, noise = 0\%}
%\label{tab:my-table}
\centering
\begin{tabular}{|c|ccc|}
\hline
\multicolumn{1}{|l|}{} & \multicolumn{3}{l|}{Burgers equation}                                                                                                                       \\ \hline
Methods/Terms          & \multicolumn{1}{c|}{du/dt}                                       & \multicolumn{1}{c|}{d\textasciicircum{}2u/dx\textasciicircum{}2} & du/dx                 \\ \hline
Adaptive               & \multicolumn{1}{c|}{0.0008}                                      & \multicolumn{1}{c|}{0.7898}                                  & 0.0061               \\ \hline
Polynomial             & \multicolumn{1}{c|}{0.001}                                       & \multicolumn{1}{c|}{1.3151}                                      & 0.082                \\ \hline
Spectral               & \multicolumn{1}{c|}{0.2112}                                      & \multicolumn{1}{c|}{6.3855}                                     & 0.0156                \\ \hline
Inverse                & \multicolumn{1}{c|}{0.8751}                                      & \multicolumn{1}{c|}{206.6064}                                    & 5.4783                \\ \hline
Total\_var                  & \multicolumn{1}{c|}{0.7988}                                      & \multicolumn{1}{c|}{208.9618}                                    & 6.2428                \\ \hline
\multicolumn{1}{|l|}{} & \multicolumn{1}{l|}{KdV equation}                                & \multicolumn{1}{l|}{}                                            & \multicolumn{1}{l|}{} \\
\hline
\end{tabular}
\end{table}

\begin{table}[ht!]
\caption{Differentiation errors without boundaries, noise = 0.5\%}
%\label{tab:my-table}
\centering
\begin{tabular}{|c|ccc|}
\hline
\multicolumn{1}{|l|}{} & \multicolumn{3}{l|}{Burgers equation}                                                                                                                       \\ \hline
Methods/Terms          & \multicolumn{1}{c|}{du/dt}                                       & \multicolumn{1}{c|}{d\textasciicircum{}2u/dx\textasciicircum{}2} & du/dx                 \\ \hline
Adaptive               & \multicolumn{1}{c|}{0.0009}                                       & \multicolumn{1}{c|}{0.7936}                                  & 0.0061               \\ \hline
Polynomial             & \multicolumn{1}{c|}{0.0206}                                      & \multicolumn{1}{c|}{1.3864}                                      & 0.0832                \\ \hline
Spectral               & \multicolumn{1}{c|}{0.2275}                                      & \multicolumn{1}{c|}{6.4897}                                     & 0.0162                \\ \hline
Inverse                & \multicolumn{1}{c|}{0.9018}                                      & \multicolumn{1}{c|}{207.6153}                                    & 5.4904                \\ \hline
Total\_var                  & \multicolumn{1}{c|}{0.8239}                                      & \multicolumn{1}{c|}{209.9772}                                    & 6.2568                \\ \hline
\multicolumn{1}{|l|}{} & \multicolumn{1}{l|}{KdV equation}                                & \multicolumn{1}{l|}{}                                            & \multicolumn{1}{l|}{} \\
\hline
\end{tabular}
\end{table}

\begin{table}[ht!]
\caption{Differentiation errors without boundaries, noise = 1\%}
%\label{tab:my-table}
\centering
\begin{tabular}{|c|ccc|}
\hline
\multicolumn{1}{|l|}{} & \multicolumn{3}{l|}{Burgers equation}                                                                                                                       \\ \hline
Methods/Terms          & \multicolumn{1}{c|}{du/dt}                                       & \multicolumn{1}{c|}{d\textasciicircum{}2u/dx\textasciicircum{}2} & du/dx                 \\ \hline
Adaptive               & \multicolumn{1}{c|}{0.1373}                                      & \multicolumn{1}{c|}{2.3416}                                  & 0.0146               \\ \hline
Polynomial             & \multicolumn{1}{c|}{0.0474}                                       & \multicolumn{1}{c|}{2.5502}                                      & 0.0868                \\ \hline
Spectral               & \multicolumn{1}{c|}{0.2652}                                      & \multicolumn{1}{c|}{7.2152}                                     & 0.02                \\ \hline
Inverse                & \multicolumn{1}{c|}{0.9020}                                      & \multicolumn{1}{c|}{207.6163}                                    & 5.4917                \\ \hline
Total\_var                  & \multicolumn{1}{c|}{0.8236}                                      & \multicolumn{1}{c|}{209.9705}                                    & 6.2560                \\ \hline
\multicolumn{1}{|l|}{} & \multicolumn{1}{l|}{KdV equation}                                & \multicolumn{1}{l|}{}                                            & \multicolumn{1}{l|}{} \\
\hline
\end{tabular}
\end{table}

\begin{table}[h!]
\caption{Means of diff errors without boundaries, noise=0\%}
\label{tab:integral0}
\centering
\begin{tabular}{|l|l|l|l|}
\hline
Methods & D1. error & D2. error   & D3. error      \\ \hline
Gradient       & 0.0002   & 0.0027 & 0.0050 \\ \hline
Adaptive       & 0.0016    & 0.2023 & 0.0121  \\ \hline
Polynomial     & 0.0193     & 0.3344 & 0.0350   \\ \hline
Spectral       & 0.0683  & 11.7856 & 14.1910 \\ \hline
Inverse        & 1.5583  &52.5442 & 0.4601 \\ \hline
Total\_var          & 1.6218  & 52.9117 & 0.3273 \\ \hline
\end{tabular}
\end{table}

\begin{table}[h!]
\caption{Means of diff errors without boundaries, noise=0.5\%}
\label{tab:integral05}
\centering
\begin{tabular}{|l|l|l|l|}
\hline
Methods & D1. error & D2. error   & D3.error      \\ \hline
Gradient       & 0.0017    & 0.0024 & 0.017 \\ \hline
Adaptive       & 0.0025  & 0.2079 & 0.0221 \\ \hline
Polynomial     & 0.0236  & 0.6421  & 0.0366  \\ \hline
Spectral       & 0.0716 & 13.0997 & 14.1966 \\ \hline
Inverse        & 1.5655  & 71.5591 & 0.4617 \\ \hline
Total\_var          & 1.6292   & 54.4074 & 0.3275 \\ \hline
\end{tabular}
\end{table}

\begin{table}[h!]
\caption{Means of diff errors without boundaries, noise=1\%}
\label{tab:integral1}
\centering
\begin{tabular}{|l|l|l|l|}
\hline
Methods & D1. error & D2. errors   & D3. errors      \\ \hline
Gradient       & 0.0065     & 0.0036 & 0.0502 \\ \hline
Adaptive       & 0.0341  & 0.6116 & 0.0432  \\ \hline
Polynomial     & 0.0302  & 1.9630 & 0.039 \\ \hline
Spectral       & 0.0800 & 17.8008 & 14.2271 \\ \hline
Inverse        & 1.5676 & 76.2041 & 0.4931 \\ \hline
Total\_var          & 1.6296  & 58.7889 & 0.3275 \\ \hline
\end{tabular}
\end{table}

\begin{table}[ht!]
\caption{Differentiation errors on boundaries, noise = 0\%}
%\label{tab:my-table}
\centering
\begin{tabular}{|c|ccc|}
\hline
\multicolumn{1}{|l|}{} & \multicolumn{3}{l|}{Burgers equation}                                                                                                                       \\ \hline
Methods/Terms          & \multicolumn{1}{c|}{du/dt}                                       & \multicolumn{1}{c|}{d\textasciicircum{}2u/dx\textasciicircum{}2} & du/dx                 \\ \hline
Adaptive               & \multicolumn{1}{c|}{0.0010}                                      & \multicolumn{1}{c|}{0.4113}                                  & 0.0047               \\ \hline
Polynomial             & \multicolumn{1}{c|}{0.0014}                                       & \multicolumn{1}{c|}{0.9714}                                      & 0.0639                \\ \hline
Spectral               & \multicolumn{1}{c|}{13.1979}                                      & \multicolumn{1}{c|}{20.9399}                                     & 0.1296                \\ \hline
Inverse                & \multicolumn{1}{c|}{0.6641}                                      & \multicolumn{1}{c|}{105.0296}                                    & 3.9091                \\ \hline
Total\_var                  & \multicolumn{1}{c|}{0.5659}                                      & \multicolumn{1}{c|}{106.5467}                                    & 4.4983                \\ \hline
\multicolumn{1}{|l|}{} & \multicolumn{1}{l|}{KdV equation}                                & \multicolumn{1}{l|}{}                                            & \multicolumn{1}{l|}{} \\
\hline
\end{tabular}
\end{table}

\begin{table}[ht!]
\caption{Differentiation errors on boundaries, noise = 0.5\%}
%\label{tab:my-table}
\centering
\begin{tabular}{|c|ccc|}
\hline
\multicolumn{1}{|l|}{} & \multicolumn{3}{l|}{Burgers equation}                                                                                                                       \\ \hline
Methods/Terms          & \multicolumn{1}{c|}{du/dt}                                       & \multicolumn{1}{c|}{d\textasciicircum{}2u/dx\textasciicircum{}2} & du/dx                 \\ \hline
Adaptive               & \multicolumn{1}{c|}{0.0082}                                       & \multicolumn{1}{c|}{0.4425}                                  & 0.0046              \\ \hline
Polynomial             & \multicolumn{1}{c|}{0.0259}                                      & \multicolumn{1}{c|}{2.1655}                                      & 0.0651                \\ \hline
Spectral               & \multicolumn{1}{c|}{13.1257}                                      & \multicolumn{1}{c|}{21.0673}                                     & 0.1299                \\ \hline
Inverse                & \multicolumn{1}{c|}{0.7112}                                      & \multicolumn{1}{c|}{105.4608}                                    & 3.9048                \\ \hline
Total\_var                  & \multicolumn{1}{c|}{0.6153}                                      & \multicolumn{1}{c|}{106.9564}                                    & 4.4938                \\ \hline
\multicolumn{1}{|l|}{} & \multicolumn{1}{l|}{KdV equation}                                & \multicolumn{1}{l|}{}                                            & \multicolumn{1}{l|}{} \\
\hline
\end{tabular}
\end{table}

\begin{table}[ht!]
\caption{Differentiation errors on boundaries, noise = 1\%}
%\label{tab:my-table}
\centering
\begin{tabular}{|c|ccc|}
\hline
\multicolumn{1}{|l|}{} & \multicolumn{3}{l|}{Burgers equation}                                                                                                                       \\ \hline
Methods/Terms          & \multicolumn{1}{c|}{du/dt}                                       & \multicolumn{1}{c|}{d\textasciicircum{}2u/dx\textasciicircum{}2} & du/dx                 \\ \hline
Adaptive               & \multicolumn{1}{c|}{0.0241}                                      & \multicolumn{1}{c|}{0.5434}                                  & 0.0048               \\ \hline
Polynomial             & \multicolumn{1}{c|}{0.0740}                                       & \multicolumn{1}{c|}{5.2326}                                      & 0.0660                \\ \hline
Spectral               & \multicolumn{1}{c|}{13.2611}                                      & \multicolumn{1}{c|}{21.4309}                                     & 0.1311                \\ \hline
Inverse                & \multicolumn{1}{c|}{0.8143}                                      & \multicolumn{1}{c|}{106.2819}                                    & 3.9179                \\ \hline
Total\_var                  & \multicolumn{1}{c|}{0.7134}                                      & \multicolumn{1}{c|}{107.7962}                                    & 4.5084                \\ \hline
\multicolumn{1}{|l|}{} & \multicolumn{1}{l|}{KdV equation}                                & \multicolumn{1}{l|}{}                                            & \multicolumn{1}{l|}{} \\
\hline
\end{tabular}
\end{table}

\begin{table}[h!]
\caption{Means of diff errors on boundaries, noise=0\%}
\label{tab:integral0}
\centering
\begin{tabular}{|l|l|l|l|}
\hline
Methods & D1. error & D2. error   & D3. error      \\ \hline
Gradient       & 0.0017    & 0.0024 & 0.0030 \\ \hline
Adaptive       & 0.0058  & 0.2585 & 0.0058 \\ \hline
Polynomial     & 0.0216  & 0.2594  & 0.0168  \\ \hline
Spectral       & 2.9927 & 110.2729 & 14.2714 \\ \hline
Inverse        & 1.1916  & 27.5245 & 0.1948 \\ \hline
Total\_var          & 1.2506   & 27.9763 & 0.2002 \\ \hline
\end{tabular}
\end{table}

\begin{table}[h!]
\caption{Means of diff errors on boundaries, noise=0.5\%}
\label{tab:integral05}
\centering
\begin{tabular}{|l|l|l|l|}
\hline
Methods & D1. error & D2. error   & D3.error      \\ \hline
Gradient       & 0.0023    & 0.0345 & 0.0076 \\ \hline
Adaptive       & 0.0076  & 4.584 & 0.0088 \\ \hline
Polynomial     & 0.0272  & 30.7636  & 0.0620  \\ \hline
Spectral       & 2.9789 & 124.8664 & 14.2823 \\ \hline
Inverse        & 1.1999  & 1702.7400 & 0.1951 \\ \hline
Total\_var          & 1.2605   & 39.3739 & 0.2001 \\ \hline
\end{tabular}
\end{table}

\begin{table}[h!]
\caption{Means of diff errors on boundaries, noise=1\%}
\label{tab:integral1}
\centering
\begin{tabular}{|l|l|l|l|}
\hline
Methods & D1. error & D2. errors   & D3. errors      \\ \hline
Gradient       & 0.0120     & 0.0429 & 0.0315 \\ \hline
Adaptive       & 0.0141  & 12.3909 & 0.0248  \\ \hline
Polynomial     & 0.0414  & 118.7723 & 0.1662 \\ \hline
Spectral       & 3.7583 & 148.1988 & 14.2965 \\ \hline
Inverse        & 1.2277 & 44900.2712 & 0.2346 \\ \hline
Total\_var          & 1.2824  & 62.7069 & 0.2002 \\ \hline
\end{tabular}
\end{table}

\end{document}